\documentclass[a4paper,fleqn]{cas-sc}

\usepackage{graphicx}
\usepackage{hyperref}
\usepackage{amssymb}
\usepackage{amsmath}
\usepackage{subfigure}

\usepackage{makecell}

\usepackage{pifont}
\newcommand{\xmark}{\ding{55}}

\usepackage{dcolumn}
\usepackage{array,multirow}
\usepackage[dvipsnames]{xcolor}

\usepackage[numbers]{natbib}

\def\tsc#1{\csdef{#1}{\textsc{\lowercase{#1}}\xspace}}
\tsc{WGM}
\tsc{QE}
\tsc{EP}
\tsc{PMS}
\tsc{BEC}
\tsc{DE}

\begin{document}
\let\WriteBookmarks\relax
\def\floatpagepagefraction{1}
\def\textpagefraction{.001}

\shorttitle{Physics-Informed PointNet}

\shortauthors{A. Kashefi \& T. Mukerji}

\title [mode = title]{Physics-informed PointNet: A deep learning solver for steady-state incompressible flows and thermal fields on multiple sets of irregular geometries}  




%

\author[1]{Ali Kashefi}
[orcid=0000-0003-0014-9051]

\cormark[1]


\ead{kashefi@stanford.edu}





\address[1]{Department of Civil and Environmental Engineering, Stanford University, Stanford, 94305, CA, USA}





\author%
[2]
{Tapan Mukerji}[orcid=0000-0003-1711-1850]
\ead{mukerji@stanford.edu}


\address[2]{Department of Energy Resources Engineering, Stanford University, Stanford, 94305, CA, USA}

\cortext[cor1]{Corresponding author}



\begin{abstract}
We present a novel physics-informed deep learning framework for solving steady-state incompressible flow on multiple sets of irregular geometries by incorporating two main elements: using a point-cloud based neural network to capture geometric features of computational domains, and using the mean squared residuals of the governing partial differential equations, boundary conditions, and sparse observations as the loss function of the network to capture the physics. While the solution of the continuity and Navier-Stokes equations is a function of the geometry of the computational domain, current versions of physics-informed neural networks have no mechanism to express this functionally in their outputs, and thus are restricted to obtain the solutions only for one computational domain with each training procedure. Using the proposed framework, three new facilities become available. First, the governing equations are solvable on a set of computational domains containing irregular geometries with high variations with respect to each other but requiring training only once. Second, after training the introduced framework on the set, it is now able to predict the solutions on domains with unseen geometries from seen and unseen categories as well. The former and the latter both lead to savings in computational costs. Finally, all the advantages of the point-cloud based neural network for irregular geometries, already used for supervised learning, are transferred to the proposed physics-informed framework. The effectiveness of our framework is shown through the method of manufactured solutions and thermally-driven flow for forward and inverse problems.
\end{abstract}




\maketitle

\section{Introduction and motivation \label{Sect1}}

To design a neural network for predicting the solution of coupled partial differential equations (PDEs) governing physical phenomena of interest, generally speaking, there are two common deep learning methodologies: ``supervised models'' and ``physics-informed models.'' In supervised models (see e.g., Refs. \cite{sekar2019fast,xiao2018novel,thuerey2019deep,kashefi2021point,kim2020prediction}), a neural network is trained over a set of labeled data as pairs of input and outputs of the network. The loss function in this approach is usually the mean squared or mean absolute errors, measuring the difference between the solutions of PDEs predicted by the neural network and the ground truth. The ground truth solutions could be the exact analytical solutions of PDEs if applicable, or solutions obtained using high-fidelity numerical solvers and reliable experiments in labs. On the other hand, physics-informed models (see e.g., Refs. \cite{raissi2019physics,meng2020ppinn,jagtap2020conservative,pang2019fpinns,zhang2020learning}) fall in the category of unsupervised or weakly-supervised deep learning. In physics-informed models, the loss function is defined as the mean squared residuals of the governing PDEs along with the mean squared errors of the associated initial (in transient problems) and boundary conditions for forward problems as well as the mean squared errors of sparse observations for inverse problems. In this approach, time and space domains are modeled as ``symbolic tensors'' and the spatial and temporal gradient operators of PDEs are computed using the available technology of ``automatic differentiation'' in TensorFlow \cite{tensorflow2015-whitepaper} or other deep learning platforms. In automatic differentiation, the derivative of a variable of interest (as the network output) with respect to a temporal or spatial coordinate (as the network input) is mathematically calculated by the chain rule, meaning all other parameters of the network are involved in the derivative. Neural networks of physics-informed models were first introduced by \citet{raissi2019physics} and have been widely used in extensive computational physics areas such as compressible flows \cite{mao2020physics}, incompressible flows \cite{lou2021physics,jin2021nsfnets}, heat transfer problems \cite{cai2021physics,wang2021reconstruction}, solid mechanics \cite{haghighat2021physics,rao2021physics}, porous media \cite{almajid2022prediction}, chemical kinetics \cite{ji2021stiff}, etc. One may see Refs. \cite{karniadakis2021physics,cai2022physics,cuomo2022scientific} for a review of recent articles in this area. It is worth noting that physics-informed models are useable for both solving the forward and inverse problems \cite{raissi2019physics}. In this article, we follow \citet{lou2021physics} for the definition of forward and inverse problems in the scope of physics-informed models. Accordingly, for forward problems, governing PDEs and corresponding boundary conditions are known; however, for inverse problems, governing PDEs as well as sparse observations of interior points of computational domains are known, whereas all or some corresponding boundary conditions are unknown. As another note, while strictly speaking any multi-dimensional array of numbers is not a tensor unless it follows the tensor coordinate transformation rules, following the software engineering point of view, we frequently use the term ``tensor'' in this article, which refers to the definition of multi-dimensional arrays as in TensorFlow \cite{tensorflow2015-whitepaper}.

There are two main advantages for physics-informed models compared to supervised models. For forward problems, there is no need for labeled data. Note that generating labeled data usually requires numerical simulations or experimental measurements in labs, which by itself might still enforce high costs in terms of time and computational budget. For this reason, physics-informed models substantially lessen such costs by relieving the labeled data requirements. For inverse problems, unknowns of a scientific problem such as parameters, boundary conditions, and full solutions of the corresponding PDEs can be discovered using ``sparse'' measurements. Contrarily, supervised models cannot carry out such features as they are fundamentally work based on plentiful measurements. Notwithstanding these two advantages, physics-informed models come with a few limitations.

First, physics-informed models necessitate higher memory of Graphics Processing Unit (GPU) compared to supervised models. This is because presence of a more complicated loss function in physics-informed models ends in a more complicated loss function gradient, producing large tensors through the back-propagation process by the chain rule. Second, neural network structures based on physics-informed models are significantly slower in training convergence compared to supervised-based networks, specifically if a first order iterative optimization algorithm is used (e.g., approximately 80000 iterations was required in Ref. \cite{laubscher2021simulation}, 11000 in Ref. \cite{wang2021reconstruction}, and 80000 in Ref. \cite{cai2021physics}). Third, physics-informed models are currently slower than conventional numerical solvers for forward problems (see e.g., Sect. 3.1 of Ref. \cite{wang2021reconstruction}, Sect. 5 of Ref. \cite{lou2021physics}, and Sect. 1.2 of Ref. \cite{mao2021deepm} for a full discussion). Fourth, obtaining solutions of complex physical phenomena such as turbulent flows is yet a considerable challenge for physics-informed neural network models (see e.g., Ref. \cite{jin2021nsfnets}). Fifth, in the current versions of physics-informed neural networks, the associated loss function treats each spatial point of a computational domain individually and evaluates the residuals of PDEs of interest at that specific point independent of the rest of points in that computational domain. In other words, these physics-informed neural networks are not informed of the geometry of the computational domain constructed by the input spatial points; even though the solution of PDEs indeed depends very much on the domain geometry. Therefore, a physics-informed neural network can be trained exclusively on one geometry. It means that for any new geometry, one must retrain the physics-informed network. As a consequence, physics-informed neural networks are practically unusable for accelerating computational physics to investigate a wide range of geometric parameters in industrial designs. In this article, we concentrate on the latest issue and propose a novel and robust methodology to resolve it and address relevant concerns.

Hypothetically, there is no mechanism in the above cited physics-informed neural networks to capture geometric characteristics of the computational domain of interest. This is mainly due to their architecture, which is a fully connected layer with a few  hidden layers (see e.g., Fig. 2 of Ref. \cite{raissi2019deepVortex}, Fig. 3 of Ref. \cite{haghighat2021physics}, Fig. 2 of Ref. \cite{wang2021reconstruction}, Fig. 2 of Ref. \cite{laubscher2021simulation}, and Fig. 2 of Ref. \cite{xu2021explore}). To alleviate the aforementioned limitation, a neural network is desirable that takes care of both the spatial coordinate of its input spatial points and simultaneously the geometric properties of the computational space to which this specific input spatial point belongs. Potentially, two options are available. The first one is a neural network designed according to a sequence of convolutional encoders and decoders (see e.g., Refs. \cite{bhatnagar2019prediction,guo2016convolutional,thuerey2019deep}), namely Convolutional Neural Networks (CNNs). In 2016, \citet{guo2016convolutional} for the first time designed an end-to-end CNN for prediction of steady flows in two and three dimensional spaces around bluff bodies. The second option is PointNet \cite{qi2017pointnet}. In 2021, \citet{kashefi2021point} for the first time used a PointNet-based deep learning strategy \cite{qi2017pointnet}, though in a supervised framework, to predict the velocity and pressure fields of incompressible flows around a cylinder with various cross-sectional shapes. Regardless of supervised or physics-informed models, there are several key advantages for PointNet \cite{qi2017pointnet} over CNNs. These advantages have been discussed in details in Ref. \cite{kashefi2021point}. Here, we first briefly review them and then, we list a few other specific benefits of PointNet \cite{qi2017pointnet} in comparison with CNNs for physics-informed models.

Using the PointNet \cite{qi2017pointnet} approach, the geometry of either the target computational domain or objects inside the domain is represented without any pixelation technique, in contrast with CNN-based approaches. This aspect avoids introducing artifacts to the domain geometry (see e.g., Fig. 1 of Ref. \cite{kashefi2021point}). Additionally, because the geometry of the domain is precisely captured by the network, the corresponding output is sensitive to even minor changes from one geometry to another (e.g., a small variation in the angle of attack of an airfoil). Moreover, depending on where a more accurate result or a smoother geometry representation is needed in the domain, the spatial distribution of points can vary adaptively from fine to coarse scales (see e.g., Fig. 19 of Ref. \cite{kashefi2021point}), leading to a reduction in the computational effort. Finally, the spatial size of the computational domain can vary from one to another over all the geometries of interest and does not have to be fixed (see e.g., Fig. 11 of Ref. \cite{kashefi2021point}).

Apart from the above mentioned advantages, PointNet \cite{qi2017pointnet} delivers three particular advantages just for physics-informed models. First, the input of PointNet \cite{qi2017pointnet} is the spatial coordinates of a set of points establishing a point cloud. Thus, the associated spatial derivatives in the loss function of physics-informed models are ``explicitly'' computed with respect to the network input. Nevertheless, the primary input of CNNs is a binary image, for example, with 0 and 1, representing respectively solid object and fluid flow spaces. Hence, each pixel of the image is required to be labeled with a spatial coordinate in an ``implicit'' manner, imposing extra difficulties for loss function implementation. Second, labeling pixels near objects with irregular geometries inside the domain is not straightforward in CNNs, specifically when the boundary passes through a pixel. If the pixel is labeled as a point located on the object boundary, that pixel contributes to the boundary condition part of the loss function; on the other hand, if the pixel is labeled as an interior point of the computational domain, that point contributes to the PDE components of the loss function. Contrarily to CNNs, the boundary and interior points of a computational domain are explicitly separated. Third, because the boundary of irregular objects can be ``smoothly'' represented in a PointNet \cite{qi2017pointnet} configuration, the spatial gradients of variables of interest in the physic-informed model loss function also smoothly change across the boundary, leading to more physically accurate outcomes and faster convergence in the training procedure.

\begin{figure*}
\centering
\includegraphics[width=0.95\textwidth]{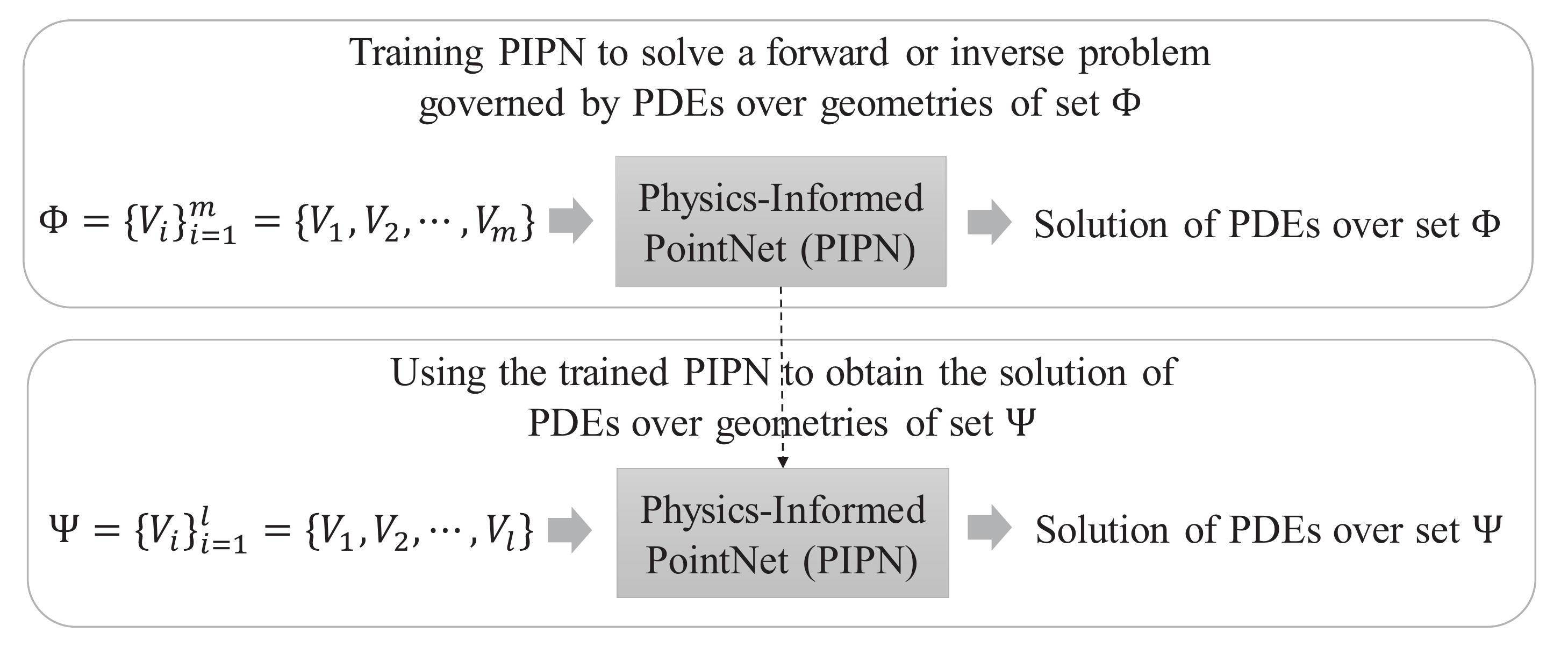}
\caption{Physics-Informed PointNet (PIPN) methodology; PIPN first solves a forward (without labeled data) or inverse (with sparse scattered labeled data) problem over geometries of the set $\Phi=\{V_i\}_{i=1}^m$. In this way, PIPN is trained over the set $\Phi=\{V_i\}_{i=1}^m$. Afterwards, the trained PIPN predicts the problem solution over geometries of the set $\Psi=\{V_i\}_{i=1}^l$. The set $\Psi=\{V_i\}_{i=1}^l$ contains domains with unseen geometries from seen and unseen categories with reference to the set $\Phi=\{V_i\}_{i=1}^m$.}
\label{Fig1}
\end{figure*}

Taking advantage of the segmentation branch of PointNet \cite{qi2017pointnet} into account, we propose a simple and elegant Physics-Informed PointNet (PIPN) model for solving PDEs. Distinctively, the PIPN tactic provides users with two novel capabilities. First, PIPN predicts the solutions of target PDEs on a set of multiple computational domains (rather than a single domain) with irregular geometries. Second, the PIPN trained on that set is able to predict the solutions on a new set of computational domains with unseen geometries from seen and unseen categories (during the training procedure). This scenario is illustrated in Fig. \ref{Fig1}. None of these two remarkable properties are accessible in the extant versions of physics-informed models.

An alternative approach in physics-informed models is not to use the technology of automatic differentiation. Instead, spatial derivatives of PDEs are discretized using a finite difference method and then the associated finite difference stencil is represented via a convolution operation (e.g., in a CNN) with non-trainable filters (see e.g., Appendix B of Ref. \cite{gao2021phygeonet} and Fig. 3 of Ref. \cite{zhao2021physicsHeat}). For the first time, \citet{gao2021phygeonet} incorporated this discretization technique (see e.g., Refs. \cite{zhu2019physicsUncer,long2018pde,long2019pde}) into a physics-informed model and introduced PhyGeoNet \cite{gao2021phygeonet} to solve parameterized steady-state PDEs without labeled data. However, PhyGeoNet \cite{gao2021phygeonet} and its later versions \cite{ren2022phycrnet,gao2021super} have three main shortcomings. First, once a finite difference scheme is chosen, all its limitations such as its accuracy order and its issues near boundaries of a computational domain for high order methods are enforced to the PhyGeoNet \cite{gao2021phygeonet} framework (see also Fig. 5 and Sect. 3.2 of Ref. \cite{zhao2021physicsHeat} for a visual illustration). Second, finite difference methods and consequently PhyGeoNet \cite{gao2021phygeonet} typically work in Cartesian grids. To obviate this shortcoming; however, \citet{gao2021phygeonet} proposed a solution: using forward/inverse elliptic coordinate transformations between irregular physical domains and a rectangular reference domain (see Sect. 2.2.1 of Ref. \cite{gao2021phygeonet}). We believe that this solution is highly demanding because for any new irregular geometry, a considerable amount of offline effort is required for the mapping procedure even before training PhyGeoNet \cite{gao2021phygeonet}. Moreover, for highly irregular non-parameterizable geometries such forward/inverse coordinate transformations may not be available. Third, although \citet{gao2021phygeonet} claimed that PhyGeoNet \cite{gao2021phygeonet} was trainable on a set of irregular geometries (see Sect. 3.2.2 of Ref. \cite{gao2021phygeonet}), in practice they examined PhyGeoNet \cite{gao2021phygeonet} on a set of geometries, all parameterized with a single scalar (see Eq. 19, Fig. 12, and Fig. 13 of Ref. \cite{gao2021phygeonet}). The efficiency of PhyGeoNet \cite{gao2021phygeonet} over a set of non-parameterizable geometries as well as geometries with high variations with respect to each other is still to be proven. According to \citet{gao2021phygeonet}, PhyGeoNet faces challenges for handling geometries with more than five $C_0$ continuous boundaries. This article demonstrates that PIPN does not suffer from any of these limitations.

The efficiency of PIPN and accuracy of its predictions are assessed by solving PDEs of conservation of mass, momentum, and energy of incompressible flows through two representative test cases: the method of manufactured solutions in a set of domains with irregular geometries and natural convection induced by a hot inner cylinder with various shapes located inside of a cold outer square cylinder. Furthermore, we pioneer investigations of the influence of pressure boundary conditions on the accuracy of the velocity and pressure fields predicted by the physics-informed model. Moreover, a comparison between implementation of the Navier-Stokes equations in the conservative and non-conservative forms is made in terms of error analysis and computational cost. In addition, different techniques for implementing body forces of the Navier-Stokes equations in the TensorFlow \cite{tensorflow2015-whitepaper} software are compared. Generalizability of PIPN is tested by predicting the velocity, pressure, and temperature fields for unseen computational domains from unseen geometry categories. Computational expenses of PIPN and regular physics-informed neural networks are compared.

The rest of this research article is structured as follows. Governing equations of mass, momentum, and energy balance for incompressible flow are provided in Sect. \ref{Sect2}. We describe the methodology and key concepts of PIPN in Sect. \ref{Sect31}. The loss function of PIPN is illustrated in Sect. \ref{Sect32}. The architecture of PIPN is explained in detail in Sect. \ref{Sect33}. We evaluate the performance of PIPN using the method of manufactured solutions in Sect. \ref{Sect41}. Results for the natural convection in a square enclosure with a cylinder are presented as the second test case in Sect. \ref{Sect42}. Summary of the work and notes for extensions of the PIPN configuration are given in Sect. \ref{Sect5}.

\section{Governing equations of interest \label{Sect2}}

The governing equations of conservation of mass, momentum, and energy for an incompressible steady flow of a Newtonian fluid in two dimensional spaces are respectively given by
\begin{eqnarray}
\label{Eq1}
\nabla \cdot \textbf{\textit{u}}=0 \textrm{ in } V,
\end{eqnarray}
\begin{eqnarray}
\label{Eq2}
\rho \big(\textbf{\textit{u}}\cdot \nabla \big)\textbf{\textit{u}} -\mu \Delta \textbf{\textit{u}} + \nabla p=\textbf{\textit{f}} \textrm{ in } V,
\end{eqnarray}
\begin{eqnarray}
\label{Eq3}
\rho \big(\textbf{\textit{u}}\cdot \nabla \big)T -\frac{\kappa}{c_p} \Delta T  =0 \textrm{ in } V,
\end{eqnarray}
where $\textbf{\textit{u}}$ is the velocity vector with components $u$ and $v$ in the $x$ and $y$ directions, respectively. $p$ indicates the pressure of the fluid and $T$ shows the temperature of the fluid. $\textbf{\textit{f}}$ is the external body force. We denote the $x$ and $y$ components of $\textbf{\textit{f}}$ respectively by $f^x$ and $f^y$. The fluid density and the dynamic viscosity are denoted by $\rho$ and $\mu$, respectively. $\kappa$ is the thermal conductivity and $c_p$ stands for the specific heat at a constant pressure. For the first test case (Sect. \ref{Sect41}), we exclusively deal with the continuity and Navier-Stokes equations (Eqs. \ref{Eq1}--\ref{Eq2}), while the energy equation (Eq. \ref{Eq3}) is involved in the second test case (Sect. \ref{Sect42}) as well. Note that energy dissipation is ignored in Eq. \ref{Eq3}. Our goal is to obtain the solutions of the PDEs (Eqs. \ref{Eq1}--\ref{Eq3}) in a set of non-trivial geometries of the fluid domain $V$. $V$ is a non-simply connected space expressed as
\begin{equation}
\label{Eq4}
V:=H-W,
\end{equation}
where $H$ is a square space with the side length of $L$. We characterize the space $W$ for each test case investigated separately in Sect. \ref{Sect41} and Sect. \ref{Sect42}.  

\section{Physics-informed PointNet (PIPN) \label{Sect3}}
\subsection{Methodology \label{Sect31}}

In this subsection, we express the key ideas of PIPN in a general and abstract manner. After that, details of PIPN are explained in Sect. \ref{Sect32} and Sect. \ref{Sect33}. Fundamentally, the PIPN skeleton is the combination of two major components: a neural network designed based on the PointNet \cite{qi2017pointnet} mechanism, and an associated loss function accorded to the physics-informed models. 

As discussed in Sect. \ref{Sect1}, the goal is to first train PIPN for obtaining the solution of PDEs (Eqs. \ref{Eq1}--\ref{Eq3}) over a set of computational spaces $\Phi=\{V_i\}_{i=1}^m$ in a forward or inverse problem; and second to use the trained PIPN for predicting the solution over a new set of computational spaces $\Psi=\{V_i\}_{i=1}^l$. Figure \ref{Fig1} visualizes this concept. We characterize each $V_i$ with a point cloud $\mathcal{X}_i$ containing $N$ points defined as $\mathcal{X}_i=\{\mathbf{x}_j \in \mathbb{R}^d\}_{j = 1}^N$ where $d$ indicates the spatial dimension. Consider a mapping from each $\mathcal{X}_i$ to the corresponding PDEs solutions $\mathcal{Y}_i$ defined as $\mathcal{Y}_i=\{\mathbf{y}_j \in \mathbb{R}^{n_{\text{PDE}}}\}_{j = 1}^N$, where $n_{\text{PDE}}$ is the number of uknown fields in the coupled PDEs. For any pair of $\mathcal{X}_i$ and $\mathcal{Y}_i$, the output $\mathbf{y}_j$ in $\mathcal{Y}_i$ corresponds to the input point $\mathbf{x}_j$ in $\mathcal{X}_i$. The output $\mathbf{y}_j$ depends on both the corresponding spatial coordinate $\mathbf{x}_j$ in $\mathcal{X}_i$ and the geometry of domain $V_i$ constructed by the point cloud $\mathcal{X}_i$. Mathematically, it is written as
\begin{equation}
\label{Eq5}
\textbf{y}_j = f\left(\textbf{x}_j, g\left(\mathcal{X}_i \right)\right),  \textrm{ with } \textbf{x}_j \textrm{ in } \mathcal{X}_i;\textrm{ }\textbf{y}_j \textrm{ in } \mathcal{Y}_i;\textrm{ }1 \leq i \leq m;\textrm{ }1 \leq j \leq N,
\end{equation}
where $f$ is the mapping and $g$ is a function encoding the geometric feature of $\mathcal{X}_i$. In PIPN, both the mapping $f$ and encoder $g$ are operated by a single neural network. Given the desired aim of PIPN, two features are mandatory for the neural network. First, it must be able to extract the geometric feature of the point cloud $\mathcal{X}$. Second, it should be capable to handle unstructured and unordered set of points constructing the point cloud $\mathcal{X}$. We demonstrate how PointNet \cite{qi2017pointnet} provides these two features in Sect. \ref{Sect33}. To determine the network parameters, an optimization problem is solved by gradient descent to minimize a physics-informed loss function. For forward problems, the PIPN loss function is the mean squared residuals of the governing PDEs (Eqs. \ref{Eq1}--\ref{Eq3}) evaluated at interior points of a point cloud $\mathcal{X}$, together with the mean squared error of labeled boundary conditions of PDEs (Eqs. \ref{Eq1}--\ref{Eq3}) evaluated at boundary points of the point cloud. For inverse problems, the mean squared error of sparse scattered labeled data at observation points is added to the PIPN loss function as well. To obtain the residuals of PDEs, the derivative of the network output ($\mathcal{Y}$) with respect to the network input ($\mathcal{X}$) are computed by means of automatic differentiation of TensorFlow \cite{tensorflow2015-whitepaper}. In other words, auto-differentiation of TensorFlow \cite{tensorflow2015-whitepaper} uses the chain rule to calculate variations in residuals of the PDEs (Eqs. \ref{Eq1}--\ref{Eq3}) and thereby perturbations of the input point locations through the backpropagation procedure. Note that because the current work is focused on steady state incompressible flows, there is no term associated with temporal differentiation in the PIPN loss function.

\begin{figure*}
\centering
\includegraphics[width=0.98\textwidth]{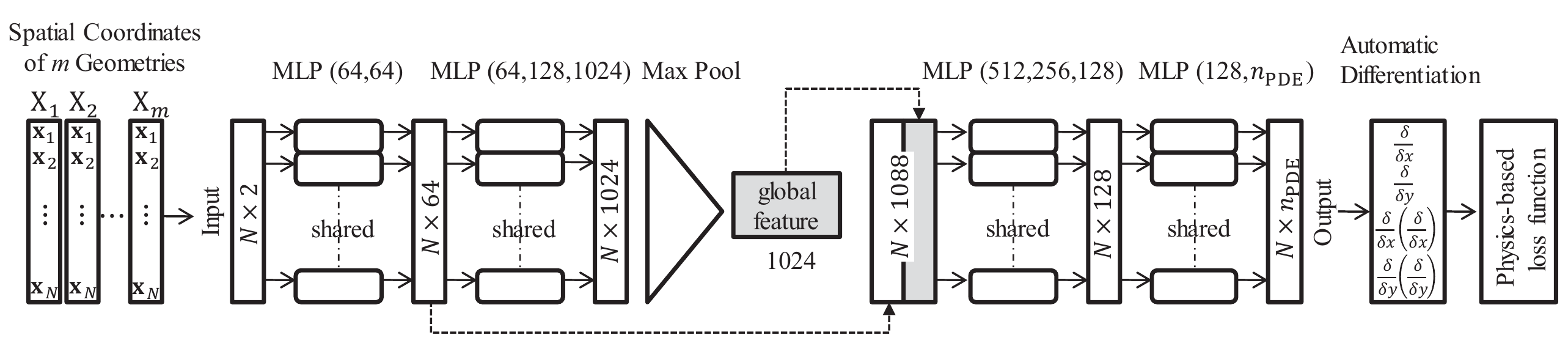}
\caption{\textcolor{blue}{Architecture of Physics-Informed PointNet (PIPN); MLPs with the labels of $(A_1,A_2)$ and $(A_1,A_2,A_3)$ are defined in Sect. \ref{Sect33}. $n_{\textrm{PDE}}$ indicates the number of desired fields to be predicted. PIPN computes spatial derivatives (e.g, $\frac{\delta \Tilde{u}}{\delta x}$, $\frac{\delta \Tilde{p} }{\delta y}$, $\frac{\delta}{\delta x}(\frac{\delta \Tilde{v}}{\delta x})$, etc.) using automatic differentiation to consturct the loss function based on PDEs describing the physics (e.g., see Eqs. \ref{Eq6}--\ref{Eq9})}}
\label{Fig2}
\end{figure*}

\subsection{Loss function \label{Sect32}}
PIPN potentially works for three dimensional problems; however, we restrict our studies to two dimensional cases in the current article and set $d=2$. Hence, the input vector $\mathbf{x}$ is specified as $\mathbf{x}=(x,y)$. The velocity ($u$ and $v$) and pressure ($p$) fields are the outputs of interest in the first test case (see Sect. \ref{Sect41}) and thus we set $n_{\text{PDE}}=3$, while in the thermally-driven flow problem (see Sect. \ref{Sect42}), we seek the solution of the temperature ($T$) field as well and hence, we set $n_{\text{PDE}}=4$. Consequently, the output vector $\mathbf{y}$ is specified as $\mathbf{y}=(u,v,p)$ and $\mathbf{y}=(u,v,p,T)$ respectively for the first (Sect. \ref{Sect41}) and second (Sect. \ref{Sect42}) test cases. For any $V_i \in \Phi$ and its corresponding pair of $\mathcal{X}_i$ and $\mathcal{Y}_i$, the residuals of the continuity equation $\big(r_i^{\text{continuity}}\big)$, momentum conservation equations in the $x-$ direction $\big(r_i^{\text{momentum}_x}\big)$ and in the $y-$directions $\big(r_i^{\text{momentum}_y}\big)$, energy conservation equation $\big(r_i^{\text{energy}}\big)$, along with the residuals of the Dirichlet boundary conditions of the velocity $\big(r_i^{\text{velocity}_{\text{BC}}}\big)$, pressure $\big(r_i^{\text{pressure}_{\text{BC}}}\big)$, and temperature $\big(r_i^{\text{temperature}_{\text{outer-BC}}}\big)$ and sparse observations of the velocity $\big(r_i^{\text{velocity}_{\text{obs}}}\big)$, pressure $\big(r_i^{\text{pressure}_{\text{obs}}}\big)$, and temperature $\big(r_i^{\text{temperature}_{\text{obs}}}\big)$ are respectively defined as follows:
\begin{eqnarray}
\label{Eq6}
r_i^{\text{continuity}} = \frac{1}{M_1} \sum_{k=1}^{M_1} \left (\frac{\delta \Tilde{u}_k}{\delta x_k} + \frac{\delta \Tilde{v}_k}{\delta y_k}  \right)^2,
\end{eqnarray}
\begin{eqnarray}
\label{Eq7}
r_i^{\text{momentum}_x} = \frac{1}{M_1} \sum_{k=1}^{M_1} \left ( \rho \left(\Tilde{u}_k \frac{\delta \Tilde{u}_k}{\delta x_k} + \Tilde{v}_k \frac{\delta \Tilde{u}_k}{\delta y_k} \right) +  \frac{\delta \Tilde{p}_k}{\delta x_k}-\mu \left(\frac{\delta}{\delta x_k} \left(\frac{\delta \Tilde{u}_k}{\delta x_k} \right) + \frac{\delta}{\delta y_k}\left(\frac{\delta \Tilde{u}_k}{\delta y_k} \right) \right)  - f_k^x \right)^2,
\end{eqnarray}
\begin{eqnarray}
\label{Eq8}
r_i^{\text{momentum}_y} = \frac{1}{M_1} \sum_{k=1}^{M_1} \left ( \rho \left(\Tilde{u}_k \frac{\delta \Tilde{v}_k}{\delta x_k} + \Tilde{v}_k \frac{\delta \Tilde{v}_k}{\delta y_k} \right) +  \frac{\delta \Tilde{p}_k}{\delta y_k}-\mu \left(\frac{\delta}{\delta x_k} \left(\frac{\delta \Tilde{v}_k}{\delta x_k} \right) + \frac{\delta}{\delta y_k}\left(\frac{\delta \Tilde{v}_k}{\delta y_k} \right) \right)  - f_k^y \right)^2,
\end{eqnarray}
\begin{eqnarray}
\label{Eq9}
r_i^{\text{energy}} = \frac{1}{M_1} \sum_{k=1}^{M_1} \left ( \rho \left(\Tilde{u}_k \frac{\delta \widetilde{T}_k}{\delta x_k} + \Tilde{v}_k \frac{\delta \widetilde{T}_k}{\delta y_k} \right) -\frac{\kappa}{c_p} \left(\frac{\delta}{\delta x_k} \left(\frac{\delta \widetilde{T}_k}{\delta x_k} \right) + \frac{\delta}{\delta y_k}\left(\frac{\delta \widetilde{T}_k}{\delta y_k} \right) \right) \right)^2,
\end{eqnarray}
\begin{eqnarray}
\label{Eq10}
r_i^{\text{velocity}_{\text{BC}}} = \frac{1}{M_2} \sum_{k=1}^{M_2} \left (\left(\Tilde{u}_k - u_k \right)^2 + \left(\Tilde{v}_k - v_k \right)^2  \right),
\end{eqnarray}
\begin{eqnarray}
\label{Eq11}
r_i^{\text{pressure}_{\text{BC}}} = \frac{1}{M_2} \sum_{k=1}^{M_2} \left (\Tilde{p}_k - p_k \right)^2,
\end{eqnarray}
\begin{eqnarray}
\label{Eq12}
r_i^{\text{temperature}_{\text{outer-BC}}} = \frac{1}{M_3} \sum_{k=1}^{M_3} \left (\widetilde{T}_k - T_k \right)^2,
\end{eqnarray}
\begin{eqnarray}
\label{Eq13}
r_i^{\text{velocity}_{\text{obs}}} = \frac{1}{M_4} \sum_{k=1}^{M_4} \left (\left(\Tilde{u}_k - u_k \right)^2 + \left(\Tilde{v}_k - v_k \right)^2  \right),
\end{eqnarray}
\begin{eqnarray}
\label{Eq14}
r_i^{\text{pressure}_{\text{obs}}} = \frac{1}{M_5} \sum_{k=1}^{M_5} \left (\Tilde{p}_k - p_k \right)^2,
\end{eqnarray}
\begin{eqnarray}
\label{Eq15}
r_i^{\text{temperature}_{\text{obs}}} = \frac{1}{M_5} \sum_{k=1}^{M_5} \left (\widetilde{T}_k - T_k \right)^2,
\end{eqnarray}
where $M_1$ is the number of interior points of $\mathcal{X}_i$. $M_2$ denotes the number of points located on the inner and outer boundaries of $\mathcal{X}_i$, while $M_3$ exclusively indicates the number of points placed on the outer boundaries of $\mathcal{X}_i$. $M_4$ is the number of sensors located in the computational domain to measure the velocity values, sparsely. $M_5$ is similarly defined for the temperature and pressure. Obviously, $M_1+M_2=N$. In this study, we use a fixed value for $M_1$, $M_2$, $M_3$, $M_4$, and $M_5$ and over all $\mathcal{X}_i$. Note that while $M_1$, $M_2$, $M_3$, $M_4$, and $M_5$ could conceptually vary from one point cloud to another, $N$ has to be fixed over all $\mathcal{X}_i$. The predicted solutions obtained by PIPN are shown by $(\Tilde{u},\Tilde{v},\Tilde{p},\widetilde{T})$, whereas the ground truth obtained by a numerical solver, lab experiment, or analytical solutions are indicated by $(u,v,p,T)$. For each test case (see Sect. \ref{Sect41} and Sect. \ref{Sect42}), the final PIPN loss function is specified as a combination of the residuals introduced above summed over all $V_i \in \Phi$ ($1 \leq i \leq m$). Note that we indicate the automatic differentiation operator by $\delta$ rather than $\partial$, due to different mathematical definitions of each of these two operators. Moreover the second order derivative (e.g., with respect to the $x$ component) is denoted by $\frac{\delta}{\delta x} (\frac{\delta}{\delta x})$ rather than $\frac{\delta ^2}{\delta x^2}$, since only the first order gradient operator exists in the TensorFlow \cite{tensorflow2015-whitepaper} software. \textcolor{blue}{Note that because the network outputs (i.e., $\textbf{y}_j$) are a function of geometric features of point clouds (i.e., $g\left(\mathcal{X}_i \right)$) according to Eq. \ref{Eq5}, the spatial derivatives of the network outputs (i.e., $\textbf{y}_j$) via automatic differentiation in Eqs. \ref{Eq6}--\ref{Eq9} become a function of these geometric features as well.} To solve the optimization problem, the Adam optimizer \cite{kingma2014adam} with hyperparameters of $\beta_1=0.9$, $\beta_2=0.999$, and $\hat{\epsilon}=10^{-6}$ is used. The mathematical definition of $\beta_1$, $\beta_2$, and $\hat{\epsilon}$ can be found in Ref. \cite{kingma2014adam}.

\subsection{Neural network architecture \label{Sect33}}

Before describing the details of the neural network architecture, we briefly review notations for two main components of regular neural networks: Multilayer Perceptron (MLP) and Fully Connected (FC) layer. In simple words, several sequential FC layers establish an MLP component. We denote by $(A_1,A_2)$ an MLP component with two layers with the size of $A_1$ and $A_2$. An MLP component with three layers is similarly noted by $(A_1,A_2,A_3)$. Each FC layer contains a weight matrix $\mathbf{W}$ and a bias vector $\mathbf{b}$. The number of rows in the matrix $\mathbf{W}$ identifies the size of the corresponding FC layer. Two sequential FC layers are connected together through the following mathematical formulation:
\begin{equation}
\label{Eq16}
\mathbf{a}_i = \sigma \big(\mathbf{W}_i\mathbf{a}_{i-1} + \mathbf{b}_i \big),
\end{equation}
where $\mathbf{a}_i$ and $\mathbf{a}_{i-1}$ are respectively the input and output of $i$th FC and $i-1$th FC layers. $\mathbf{W}_i$ and $\mathbf{b}_i$ are similarly defined. Furthermore, $\sigma$ is a nonlinear activation function and is applied to each vector component elementwise. One may refer to Ref. \cite{goodfellow2016deep} for further explanations of MLP and FC layers. We next explain the proposed neural network structure.

The neural network is mainly designed based on the segmentation component of PointNet \cite{qi2017pointnet} and two main branches constitute its architecture, as exhibited in Fig. \ref{Fig2}. Through the first branch, the geometric feature of the input set $\mathcal{X}$ is encoded in a latent global feature with a vector of size 1024. Afterwards, the second branch decodes the latent global feature to the output set $\mathcal{Y}$. As discussed in Sect. \ref{Sect31}, PointNet \cite{qi2017pointnet} extracts the geometric feature of set $\mathcal{X}$, while the obtained feature is invariant with respect to ordering over the input set $\mathcal{X}$. PointNet \cite{qi2017pointnet} handles this procedure using two mathematical concepts. First, PointNet \cite{qi2017pointnet} uses a ``shared'' function (here called $h$) over all members of each input set $\mathcal{X}$ (i.e., $\mathbf{x}_j$). Second, to encode the geometric feature of each input set $\mathcal{X}$, PointNet \cite{qi2017pointnet} utilizes a permutation invariant function such as maximum, minimum, average, or summation operators. The original version of PointNet \cite{qi2017pointnet} proposed by \citet{qi2017pointnet} uses the ``max'' function and we implement it here as well. Thus, after applying the two MLPs of the first branch, the latent global feature can be approximated as

\begin{equation}
\label{Eq17}
g(\mathcal{X}) \approx \max \left(h(\mathbf{x}_1), \dots, h(\mathbf{x}_N)\right).
\end{equation}
Note that we introduced the function $g$ in Eq. \ref{Eq5} in Sect. \ref{Sect31}. As mentioned earlier, $h$ is a shared function representing all the mathematical operations carried out on each $\mathbf{x}_j$ of the set $\mathcal{X}$ through the two MLPs in the first branch. \textcolor{blue}{One may refer to Ref. \cite{lin2013network} for a detailed description of shared MLPs from computer science perspectives.} The input of the second branch is a vector of size 1088, which is the consequence of concatenating the global feature vector of size 1024 and the intermediate feature vector of size 64 taken from the first branch, as shown in Fig. \ref{Fig2}. In the next stage, two MLPs with shared functions respectively in the form of (512, 256, 128) and (128, $n_\text{PDE}$) operates on the second branch input to predict the solution (i.e., the output set $\mathcal{Y}$) of the desired PDEs. Note that we remove Transform Nets (T-Nets) from the original version of PointNet \cite{qi2017pointnet} because we obtain high efficiency even without implementing them. One may refer to Ref. \cite{qi2017pointnet} for further elaboration of PointNet \cite{qi2017pointnet}.
\\
The hyperbolic tangent activation function defined as
\begin{equation}
\label{Eq18}
\sigma(\gamma) = \frac{e^{2\gamma}-1}{ e^{2\gamma}+1},
\end{equation}
is used for all the layers. After each FC layer, we use batch normalization \cite{ioffe2015batch}.

To close this subsection, we address a few points. First, since our PDEs of interest involve second order derivatives of the velocity and temperature fields (see Eqs. \ref{Eq2}--\ref{Eq3}), it is critical to use an activation function such that its second order derivative is well-defined. For example, using the Rectfied Linear Unit (ReLU) activation function expressed as
\begin{equation}
\label{Eq19}
\sigma(\gamma) = \max(0, \gamma). 
\end{equation}
leads to divergence of the training procedure of PIPN. Note that ReLU successfully performs in the work by \citet{kashefi2021point}, when they used PointNet \cite{qi2017pointnet} in a supervised model for deep learning of velocity and pressure fields of incompressible flows. Second, the choice of activation functions in the last layer depends on our prior knowledge about the problem solutions. For instance, if the solution of PDEs is in the range of $[0, 1]$, the sigmoid activation function defined 
\begin{equation}
\label{Eq20}
\sigma(\gamma) = \frac{1}{1 +e^{-\gamma}},
\end{equation}
could be a good candidate. However, if prior knowledge is not available, an option is not to enforce any activation function in the last layer keeping possible outcomes of PIPN (i.e., numerical values of $\mathbf{y}_j$) unbounded. Third, the PDE solutions of benchmark problems considered in Sect. \ref{Sect41} and Sect. \ref{Sect42} are a function of the geometry of the space $W$ (see Eq. \ref{Eq4}), this is while the input set $\mathcal{X}$ ``implicitly'' represents $W$ through its null space. In other words, PIPN predicts the solutions of PDEs in the active space of $\mathcal{X}$, by learning the geometry of null space of $W$. The idea of implicit shape representation has been discussed further in Sect. II.C of Ref. \cite{kashefi2021point}. Fourth, to generate point clouds $\mathcal{X}_i$, we first discretize the space of $V_i$ using the Gmsh \cite{geuzaine2009gmsh} application, and then take the finite-element grid vertices of unstructured triangular meshes to construct $\mathcal{X}_i$. Note that although the point distributions might locally affect the accuracy of physics-informed neural networks (see e.g., Fig. 2 of Ref. \cite{mao2020physics}), the PIPN methodology is independent of point generation schemes in general. Fifth, network training is performed on a TESLA P40 graphics card with the memory clock rate of 1.531 GHz and 24 Gigabytes of RAM.

\begin{table}[h]
\caption{Description of geometries of the set $\Phi=\{V_i\}_{i=1}^{26}$ discussed in Sect. \ref{Sect411} through Sect. \ref{Sect414}; $\Omega$ indicates counterclockwise rigid rotation (in radian) of space $W$ with respect to its geometric center with reference to the current pose of schematic figures.}
\centering
\begin{tabular}{c c c c}
\toprule
\vtop{\hbox{\strut Shape of $W$}\hbox{\strut (see Eq. \ref{Eq4})}} & \vtop{\hbox{\strut Schematic}\hbox{\strut figure}} & \vtop{\hbox{\strut Geometric }\hbox{\strut description}} & Number of data \\
\midrule
Circle & \includegraphics[width=0.07\linewidth]{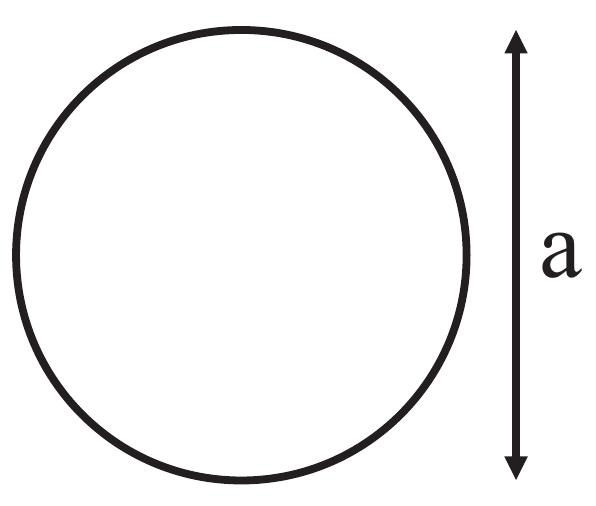} & $a=0.8\pi \textrm{ m}$ & $1$ \\
\midrule
Semi circle &
\includegraphics[width=0.07\linewidth]{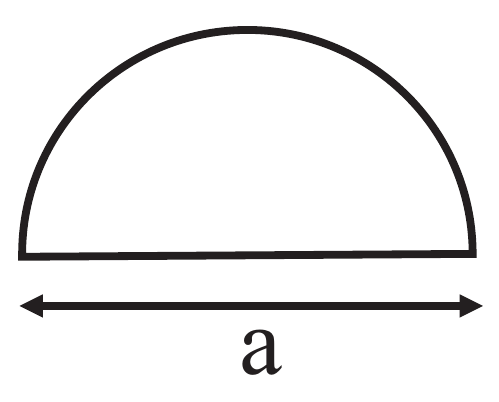} & $a=0.8\pi \textrm{ m}$ with $\Omega=0, \pi$ & $2$ \\
\midrule
Three-quarter sector of a circle &
\includegraphics[width=0.075\linewidth]{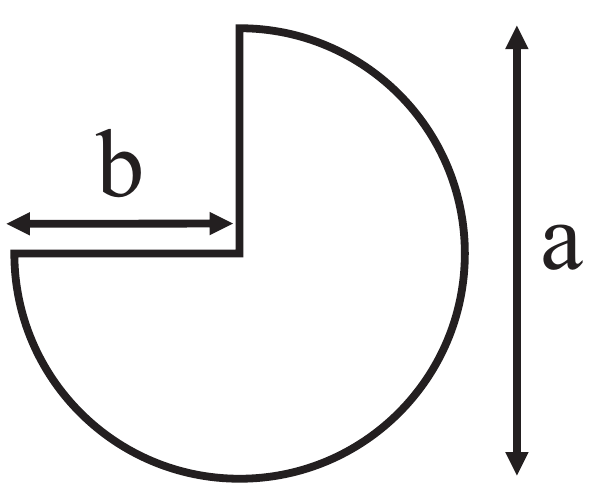}
 & $a=0.8\pi \textrm{ m, } b=0.4\pi \textrm{ m}$ with $\Omega=0, \pi$ & $2$ \\
\midrule
Equilateral triangle & \includegraphics[width=0.07\linewidth]{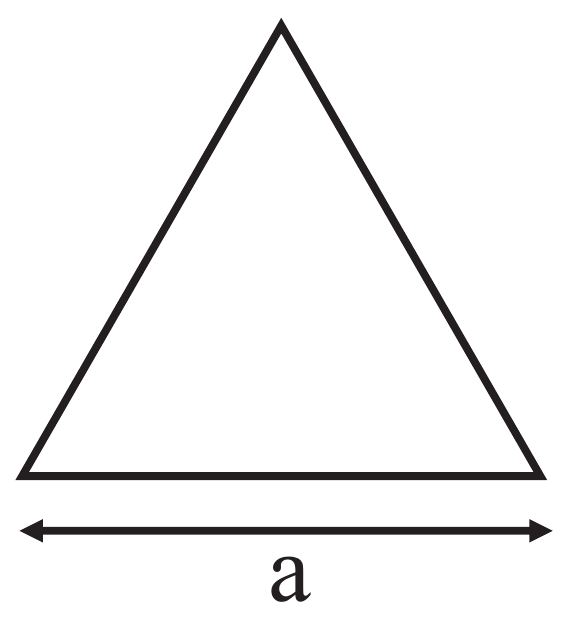} & $a=0.8 \pi \textrm{ m}$ with $\Omega=\frac{\pi}{6}, \frac{\pi}{3}, \frac{\pi}{2}$ & $3$ \\
\midrule
Equilateral hexagon & \includegraphics[width=0.07\linewidth]{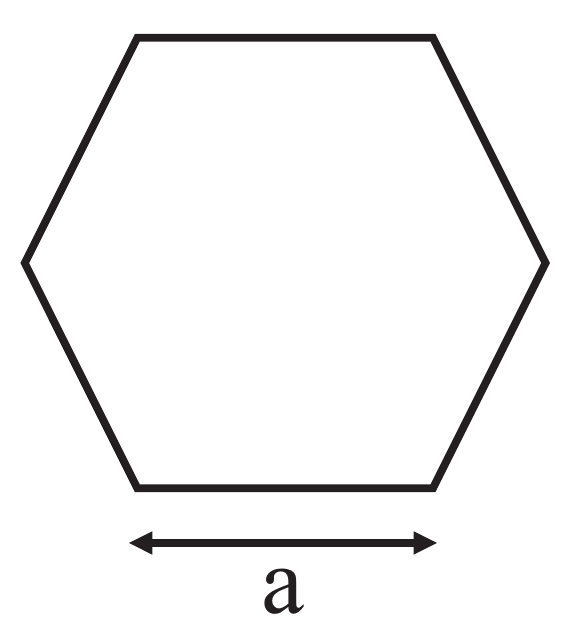} & $a=0.8\pi \frac{\sqrt{3}}{3} \textrm{ m}$ with $\Omega=0, \frac{\pi}{6}$ & $2$ \\
\midrule
Equilateral octagon & \includegraphics[width=0.073\linewidth]{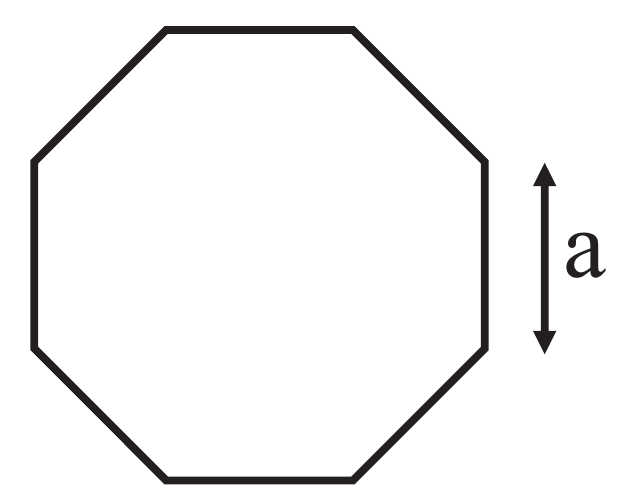} & $a=0.8\pi(\sqrt{2}-1) \textrm{ m}$ with $\Omega=0, \frac{\pi}{8}$ & $2$ \\
\midrule
Trapezoid & \includegraphics[width=0.073\linewidth]{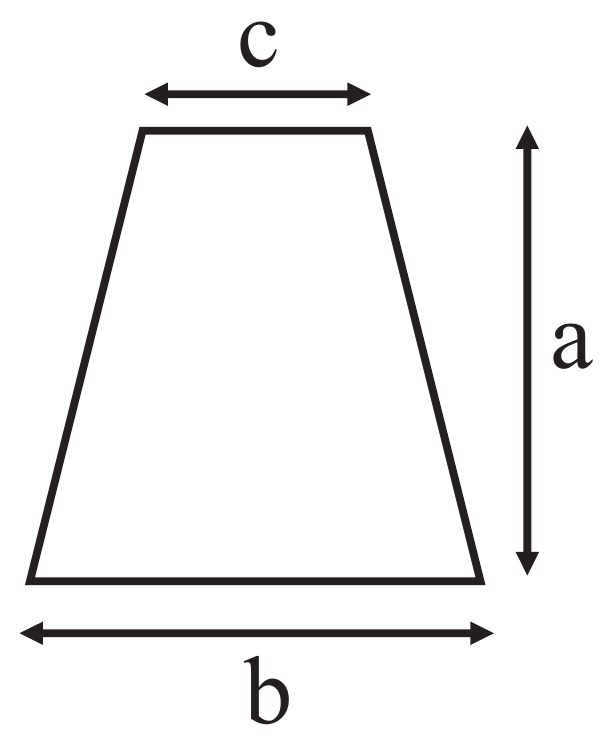} &
\makecell{$a=0.8\pi \textrm{ m, } b=\frac{14}{15}\pi \textrm{ m, }$ \\
$c=0.4\pi \textrm{ m}$ with $\Omega=0, \pi$} & $2$ \\
\midrule
Square & \includegraphics[width=0.07\linewidth]{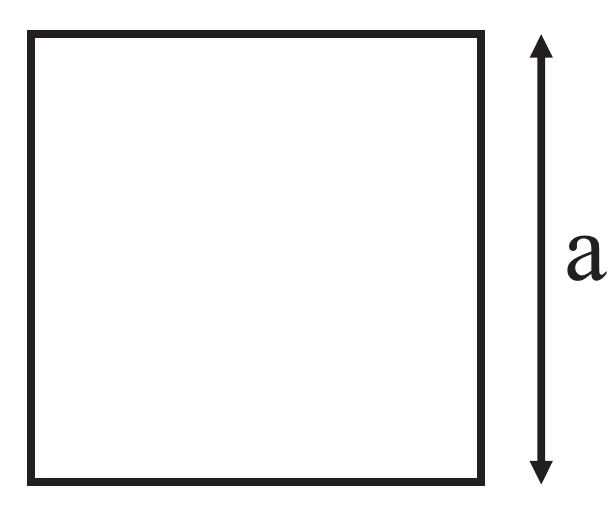} & $a=0.8\pi \textrm{ m}$ with $\Omega=0, \frac{\pi}{4}$ & $2$ \\
\midrule
Symmetrical star & \includegraphics[width=0.073\linewidth]{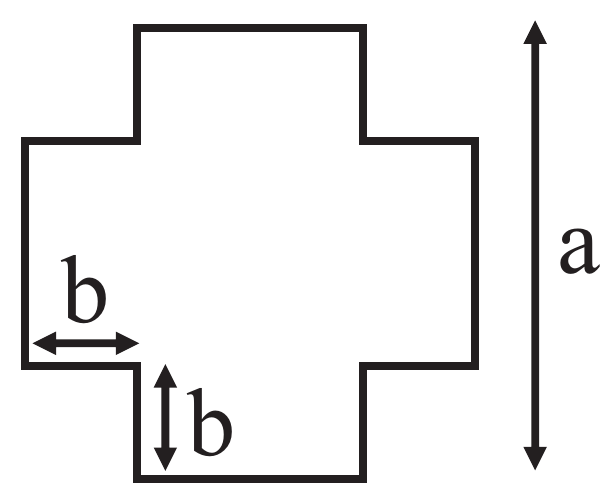} & $a=0.8 \pi \textrm{ m, } b = 0.16 \pi \textrm{ m}$ with $\Omega=0, \frac{\pi}{4}$ & $2$ \\
\midrule
Ellipse & \includegraphics[width=0.08\linewidth]{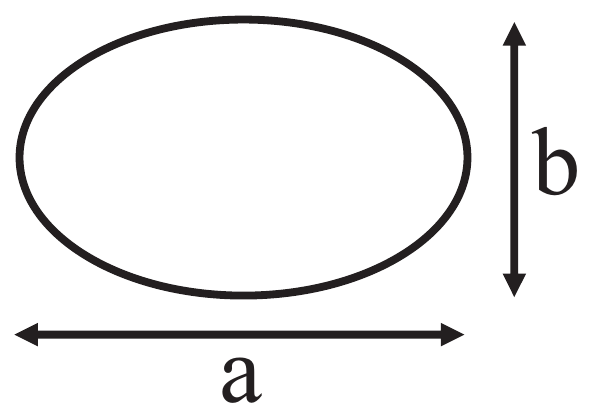} & 
\makecell{ $a=1.04 \pi \textrm{ m, } b=0.4 \pi  \textrm{ m}$ with $\Omega=\frac{\pi}{6}, \frac{5\pi}{6}$ \\ 
and $a=0.96 \pi \textrm{ m, } b=0.4 \pi  \textrm{ m}$ with $\Omega=0, \frac{\pi}{2}$} & $4$ \\
\midrule
Rectangle & \includegraphics[width=0.08\linewidth]{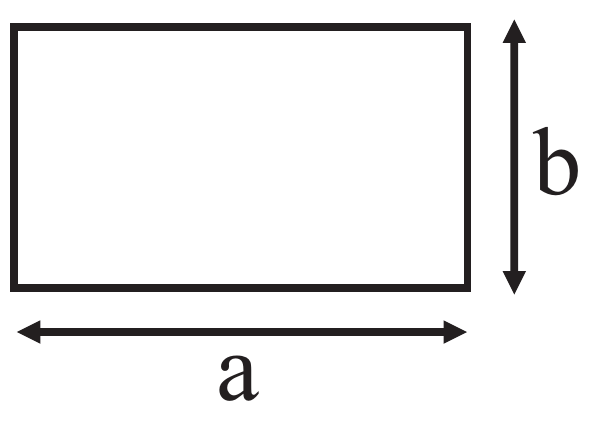} &
\makecell{ $a=1.04 \pi \textrm{ m, } b=0.4 \pi  \textrm{ m}$ with $\Omega=\frac{\pi}{3}, \frac{2\pi}{3}$ \\ and $a=0.96 \pi \textrm{ m, } b=0.4 \pi  \textrm{ m}$ with $\Omega=\frac{\pi}{6}, \frac{5\pi}{6}$} & $4$ \\
\bottomrule
\end{tabular}
\label{Table1}
\end{table}

\section{Results and discussion \label{Sect4}}
\subsection{Method of manufactured solutions in non-trivial geometries \label{Sect41}}

The concern of this subsection is to assess the efficiency of PIPN for a forward problem. We set up a machine-learning experiment by means of the method of manufactured solutions. The method of manufactured solutions is a widely-used scheme for the purpose of code verification (see e.g., Refs. \cite{shunn2012verification,veeraragavan2016use,brady2012code}) and algorithm examination (see e.g., Refs. \cite{ecca2009evaluation,waltz2013verification,vedovoto2011application}). To utilize this strategy, we consider a divergence free velocity field given by
\begin{eqnarray}
\label{Eq21}
u(x,y)=\cos (x) \sin (y),
\end{eqnarray}
\begin{eqnarray}
\label{Eq22}
v(x,y)= - \cos (y) \sin (x),
\end{eqnarray}
with an arbitrary pressure field given by
\begin{equation}
\label{Eq23}    
p(x,y)=-\frac{\rho}{4} \big(\cos (2x) +\cos (2y)\big).
\end{equation}
To satisfy the Navier-Stokes equations (Eq. \ref{Eq2}), the forcing terms read
\begin{eqnarray}
\label{Eq24}
f^x=2 \mu \cos (x) \sin (y),
\end{eqnarray}
\begin{eqnarray}
\label{Eq25}
f^y=-2 \mu \cos (y) \sin (x).
\end{eqnarray}
A density of $\rho=1 \text{ kg/m}^3$ and viscosity of $\mu=0.01 \text{ Pa}\cdot  \text{s}$ are set. The side length ($L$) of $H$ is set to $2\pi$ m and $H:=[\pi \text{ m},3\pi \text{ m}]\times[\pi \text{ m},3\pi \text{ m}]$ (see Eq. \ref{Eq4}). The Dirichlet boundary conditions are described by the manufactured solutions (Eqs. \ref{Eq21}--\ref{Eq23}).

\begin{figure*}[htbp]
\centering
\includegraphics[width=0.95\textwidth]{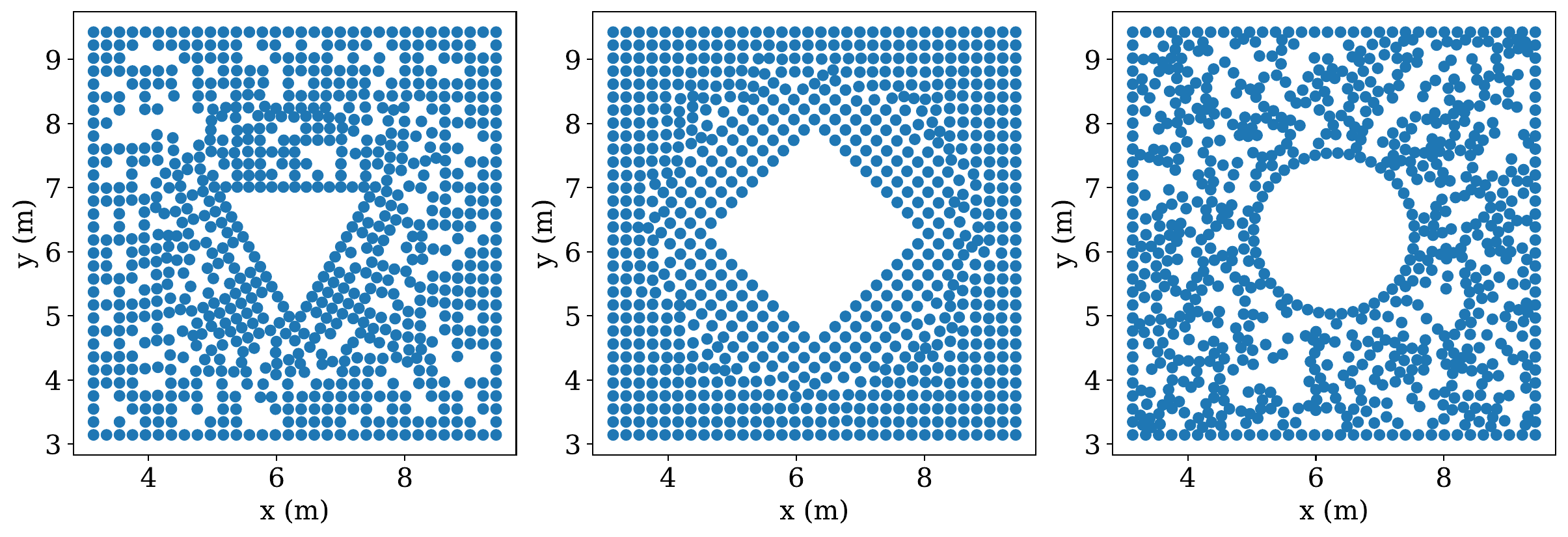}
\includegraphics[width=0.95\textwidth]{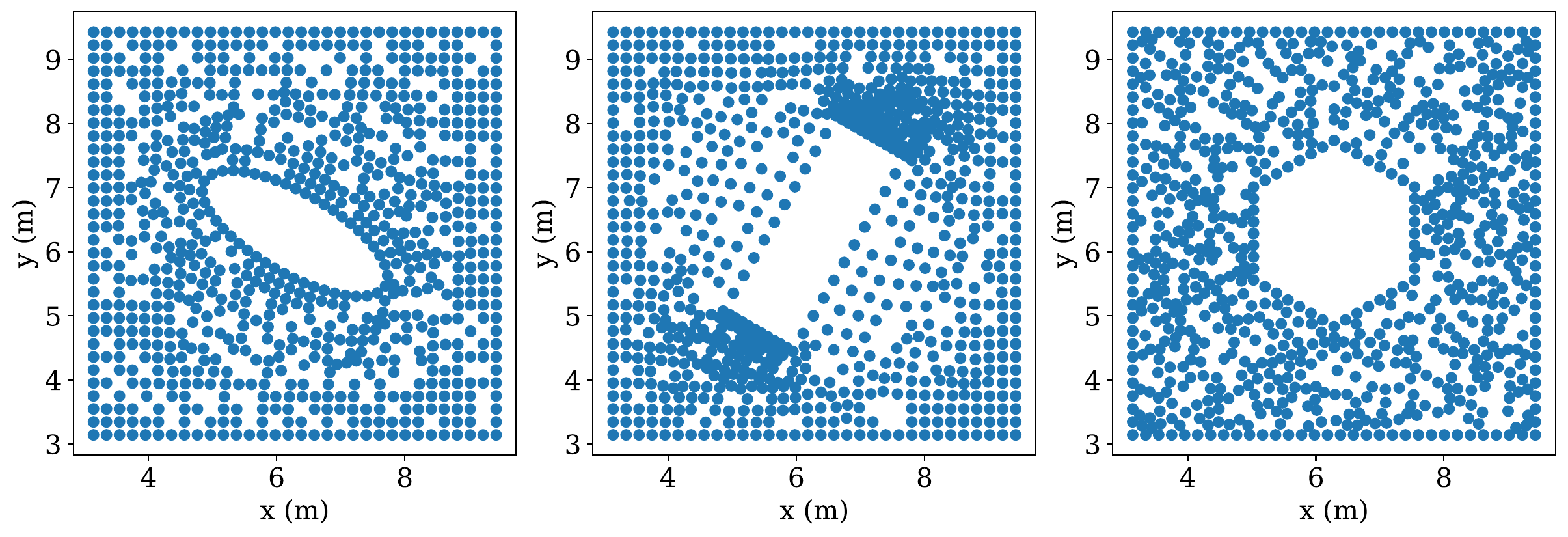}
\includegraphics[width=0.95\textwidth]{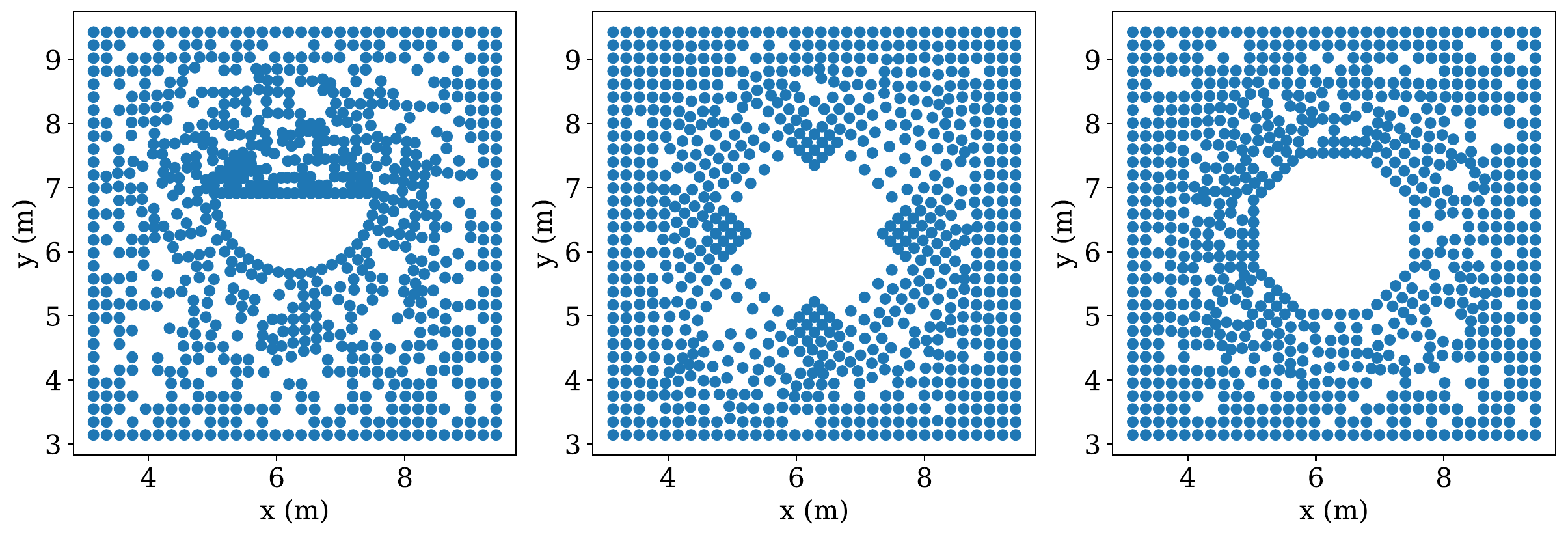}
\includegraphics[width=0.95\textwidth]{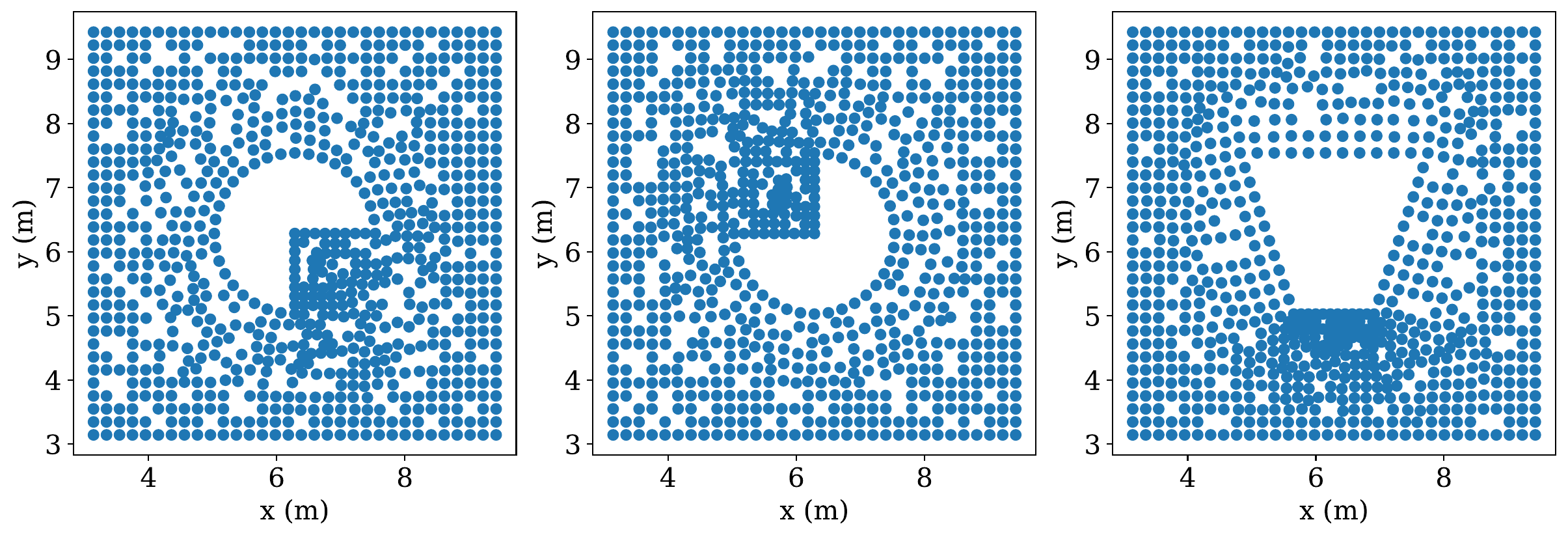}
\caption{A few examples of point clouds representing computational domains of the set $\Phi=\{V_i\}_{i=1}^{26}$ considered for the method of manufactured solutions in Sect. \ref{Sect41}}
\label{Fig3}
\end{figure*}

We establish the set $\Phi=\{V_i\}_{i=1}^{26}$ by generating 26 geometries $(m=26)$ described in Table \ref{Table1}. A few examples of generated point clouds are depicted in Fig. \ref{Fig3}. In the generated point clouds, $N= 835$, $M_1= 667$, and $M_2=168$ are set. As tabulated in Table \ref{Table1} and shown in Fig. \ref{Fig3}, the set $\Phi=\{V_i\}_{i=1}^{26}$ contains geometries of $W_i$ with basic shapes such as circle, semi-circle, three-quarter sector of a circle, square, equilateral triangle, equilateral hexagon, equilateral octagon, rectangle, trapezoid, ellipse, and symmetrical star; representing a set with high geometric variation with respect to each other.

We define the forward problem as follows: given the velocity and pressure Dirichlet boundary condition on all the boundaries; find the full velocity and pressure fields in the inquiry points. Accordingly, the loss function of this problem is determined as
\begin{equation}
\label{Eq26}
\mathcal{J}=\frac{1}{m} \sum_{i=1}^m \left( \lambda_1 r_i^{\text{continuity}} + \lambda_2 r_i^{\text{momentum}_x} + \lambda_3 r_i^{\text{momentum}_y} + \lambda_4 r_i^{\text{velocity}_{\text{BC}}} + \lambda_5 r_i^{\text{pressure}_{\text{BC}}} \right ),
\end{equation}
where $\lambda_1, \dots, \lambda_5$ are the associated weights of each residuals and their units are the inverse of the units of their corresponding residuals. In the experiment conducted here, we set the weights equal to 1 with the appropriate units. In this way, the loss function ($\mathcal{J}$) is dimensionless. The batch size of $\mathcal{B}=13$ is chosen and the set of geometries is shuffled at each iteration (i.e., epoch). A constant learning rate of $\alpha=0.001$ is selected and the training procedure is executed until the following criterion is satisfied:
\begin{equation}
\label{Eq27}
\mathcal{J}  \leq 4 \times 10^{-5}.
\end{equation}

At the end of this subsection several notes are listed. First, since the exact solutions are available, computed errors are purely due to the PIPN framework. However, if the PIPN solutions were compared to those computed by a computational fluid dynamics (CFD) solver, unavoidable numerical errors caused by that solver would negatively affect the error analysis of the results obtained by PIPN. Second, the velocity and pressure boundary conditions of this case are not constant and vary point by point along the boundaries, creating extra challenges for PIPN. Third, the proposed manufactured solutions for steady-state incompressible flows in two dimensions could be useful for other researchers to verify their own physics-informed machine learning framework. Fourth, the manufactured solutions offered in Eqs. \ref{Eq21}--\ref{Eq23} represent a quasi Taylor-Green vortex \cite{taylor1937mechanism} mechanism with the difference that the proposed velocity and pressure fields in Eqs. \ref{Eq21}--\ref{Eq23} only contain the spatial components of the exact solution of the Taylor-Green vortex \cite{taylor1937mechanism} problem in a two-dimensional space, while the exponential decaying temporal term does not appear.

\subsubsection{General analysis \label{Sect411}}
\begin{figure*}
\centering
\includegraphics[width=0.95\textwidth]{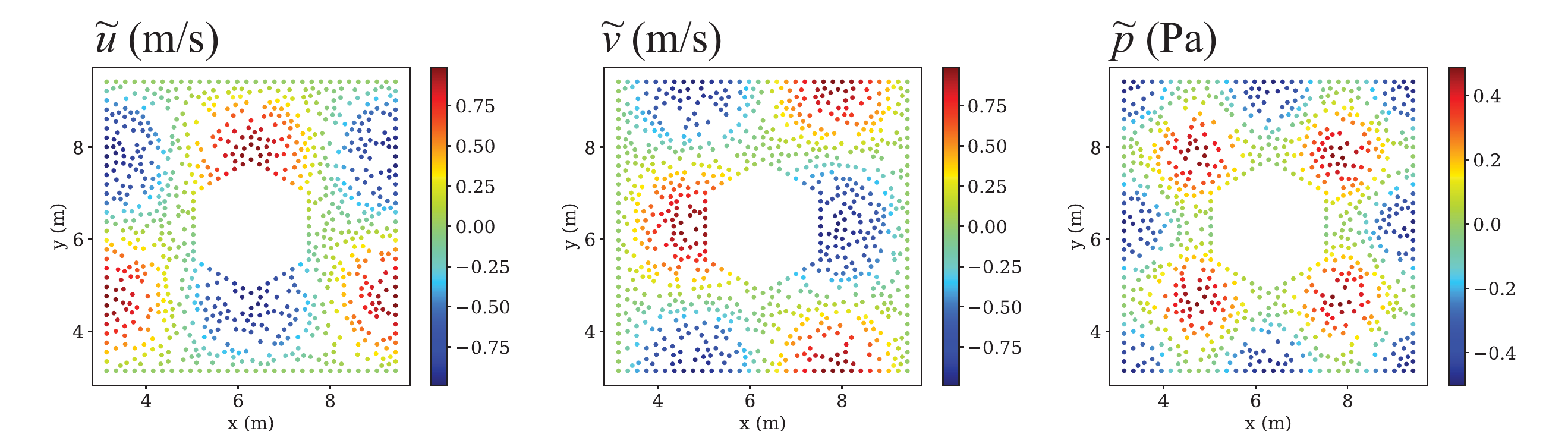}
\includegraphics[width=0.95\textwidth]{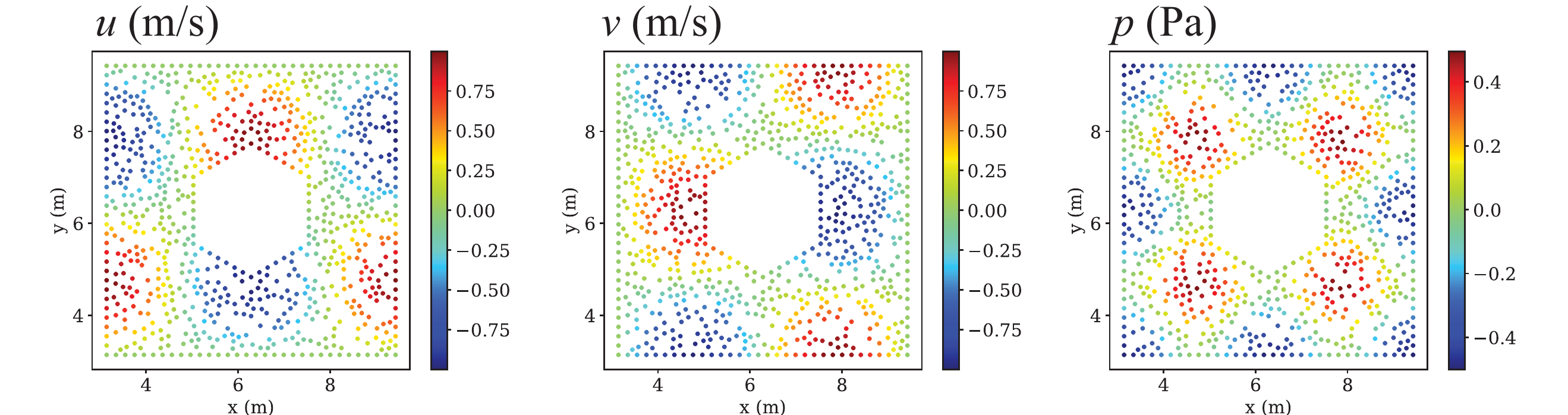}
\includegraphics[width=0.95\textwidth]{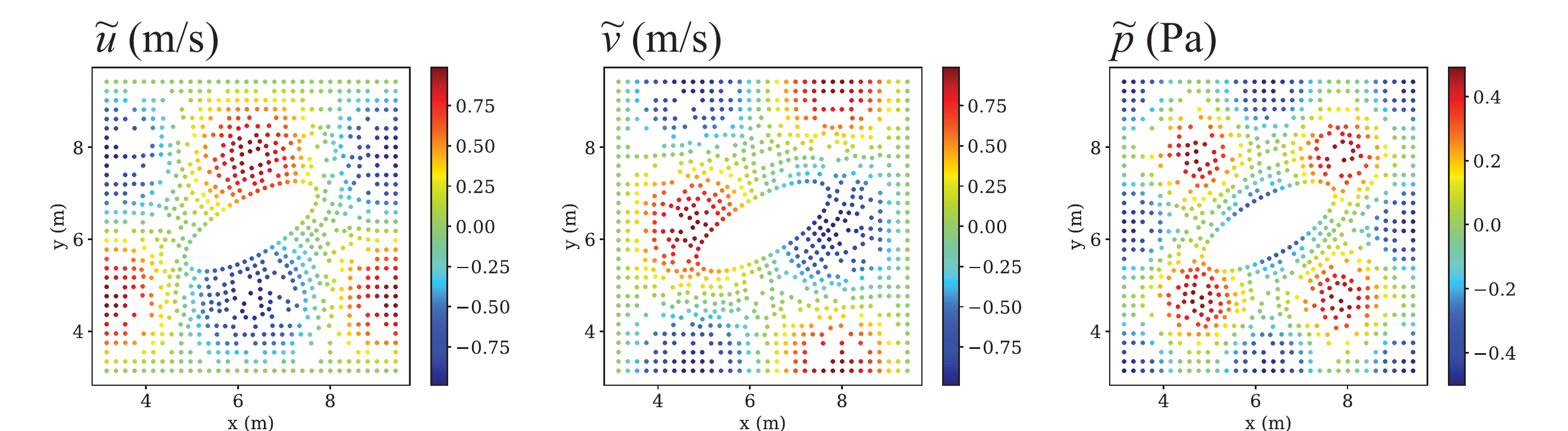}
\includegraphics[width=0.95\textwidth]{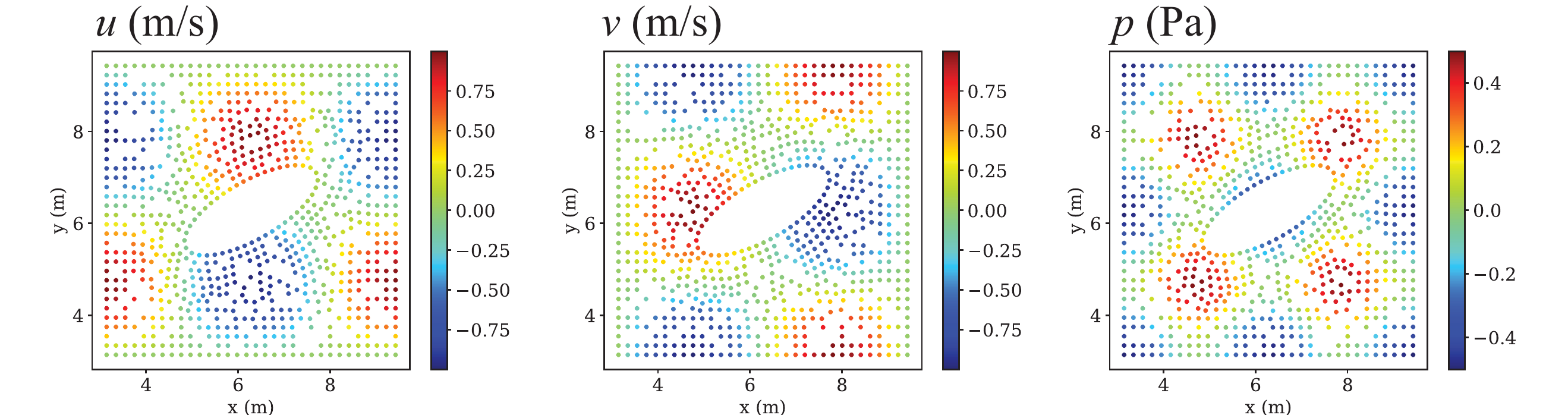}
\caption{The first group of examples taken from the set $\Phi=\{V_i\}_{i=1}^{26}$, comparing the analytical solutions (Eqs. \ref{Eq21}--\ref{Eq23}) to the PIPN predictions for the velocity and pressure fields for the method of manufactured solutions (Sect. \ref{Sect411})}
\label{Fig4}
\end{figure*}

\begin{figure*}
\centering
\includegraphics[width=0.95\textwidth]{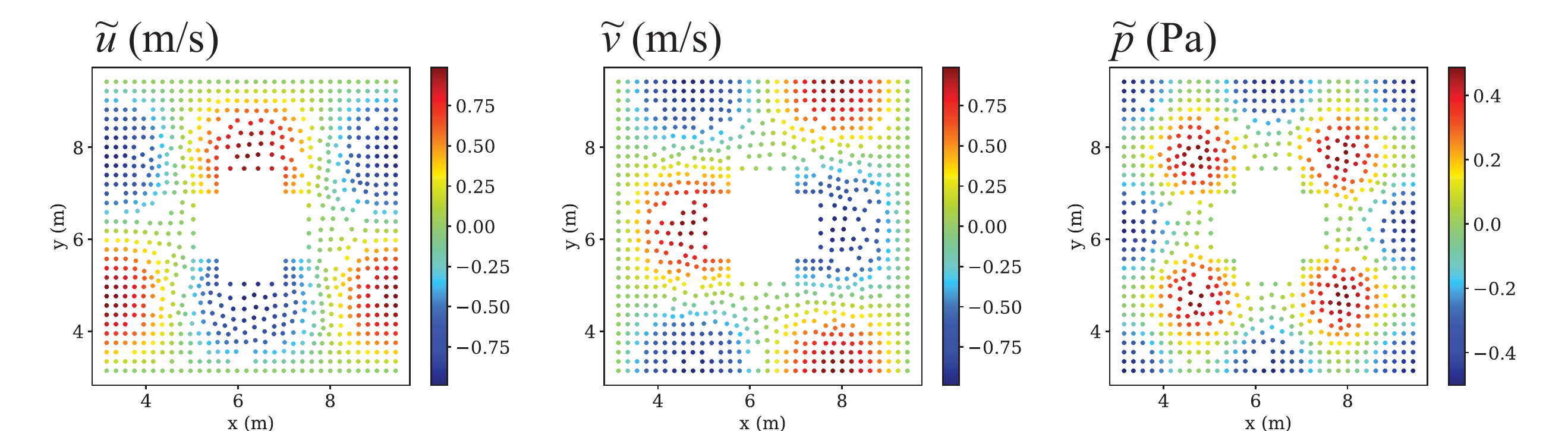}
\includegraphics[width=0.95\textwidth]{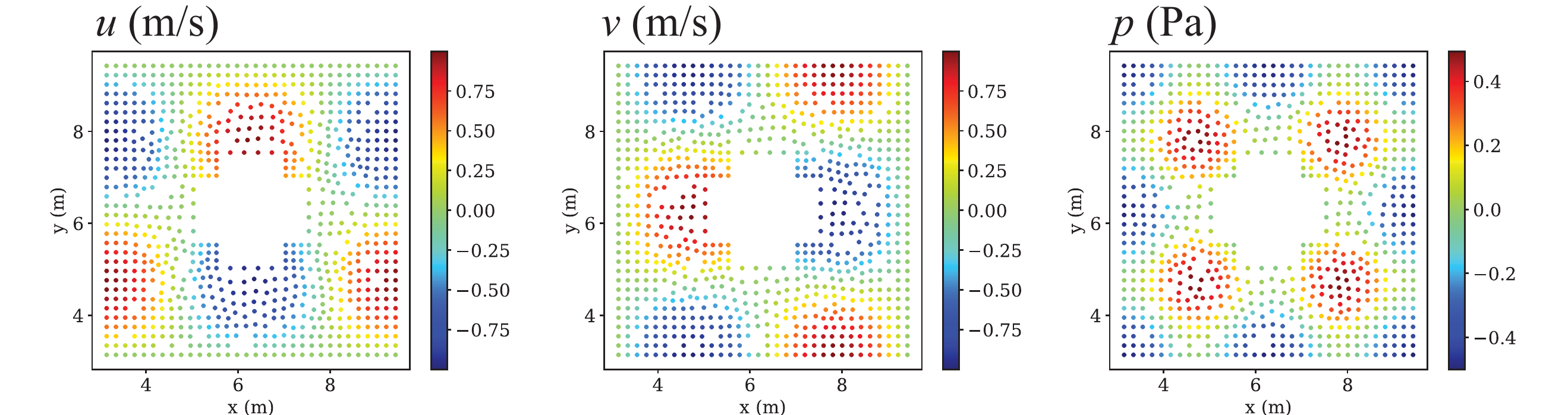}
\includegraphics[width=0.95\textwidth]{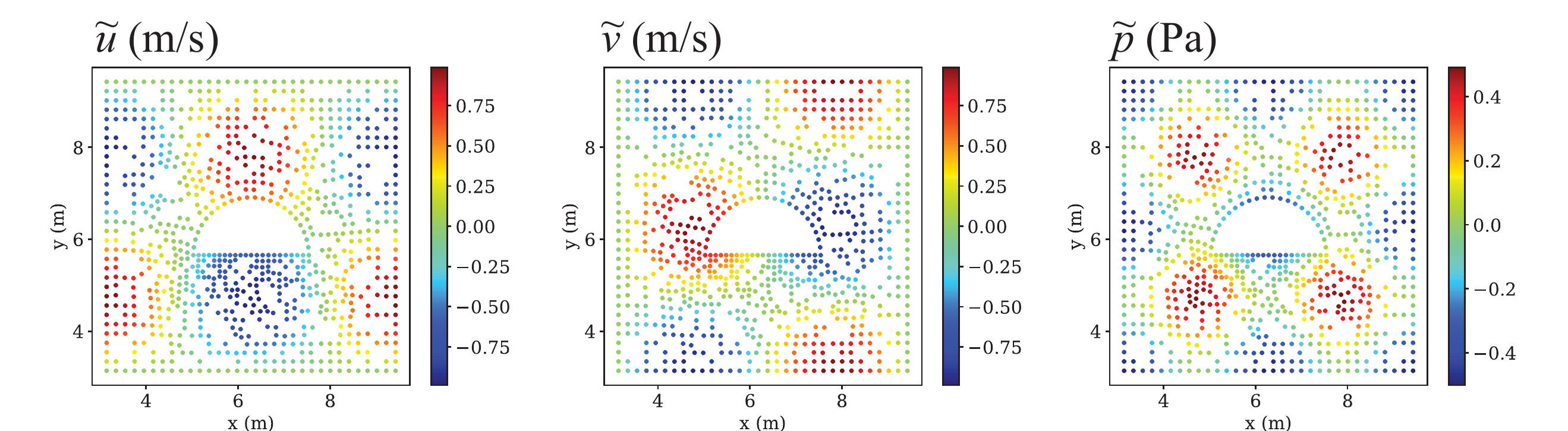}
\includegraphics[width=0.95\textwidth]{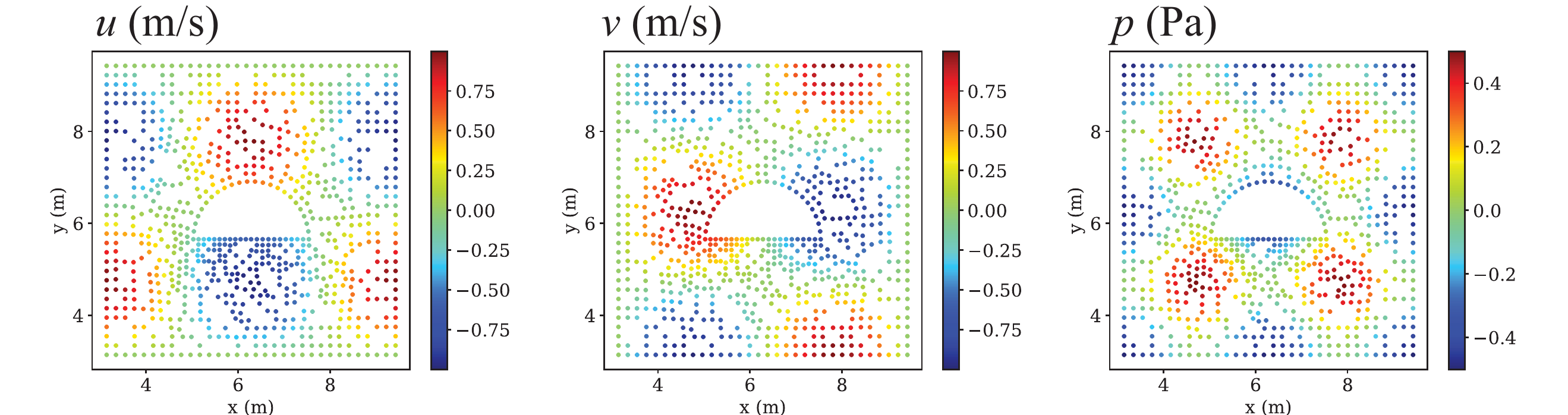}
\caption{The second group of examples taken from the set $\Phi=\{V_i\}_{i=1}^{26}$, comparing the analytical solutions (Eqs. \ref{Eq21}--\ref{Eq23}) to the PIPN predictions for the velocity and pressure fields for the method of manufactured solutions (Sect. \ref{Sect411})}
\label{Fig5}
\end{figure*}

\begin{figure*}
\centering
\includegraphics[width=0.95\textwidth]{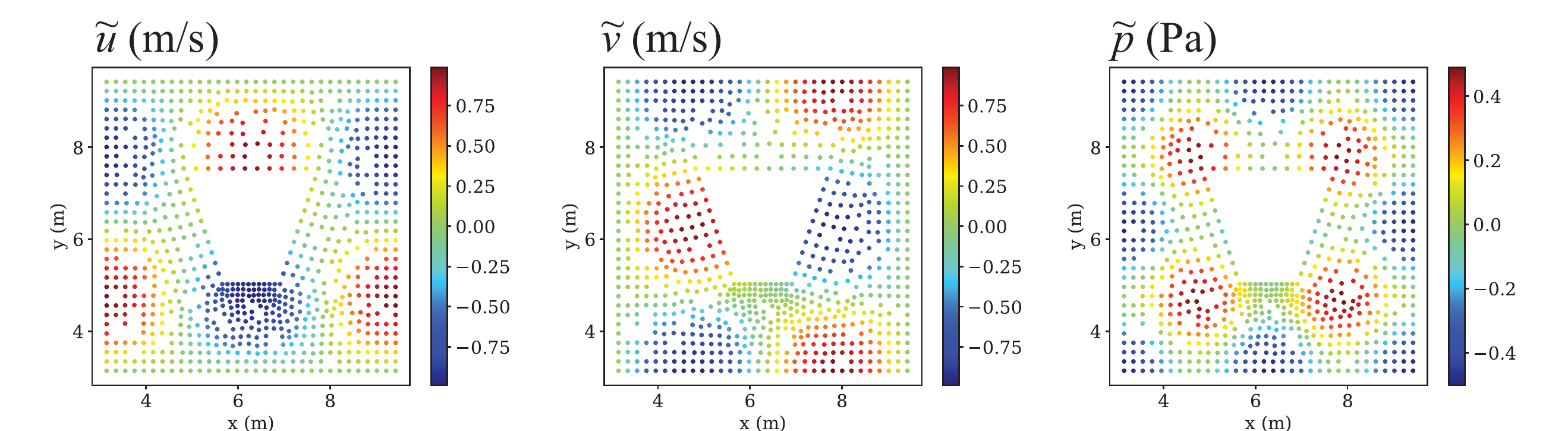}
\includegraphics[width=0.95\textwidth]{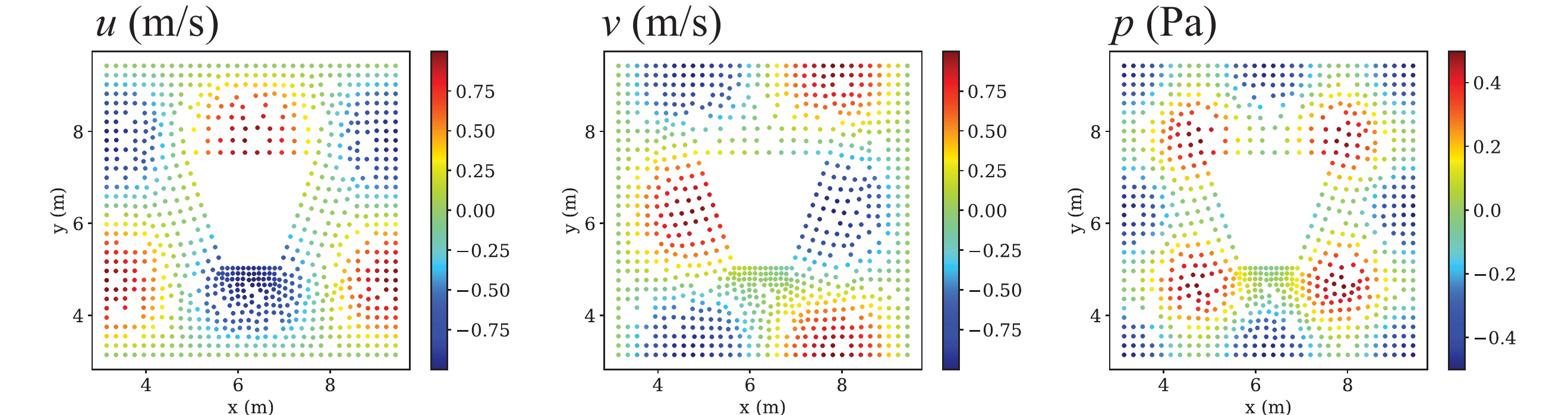}
\includegraphics[width=0.95\textwidth]{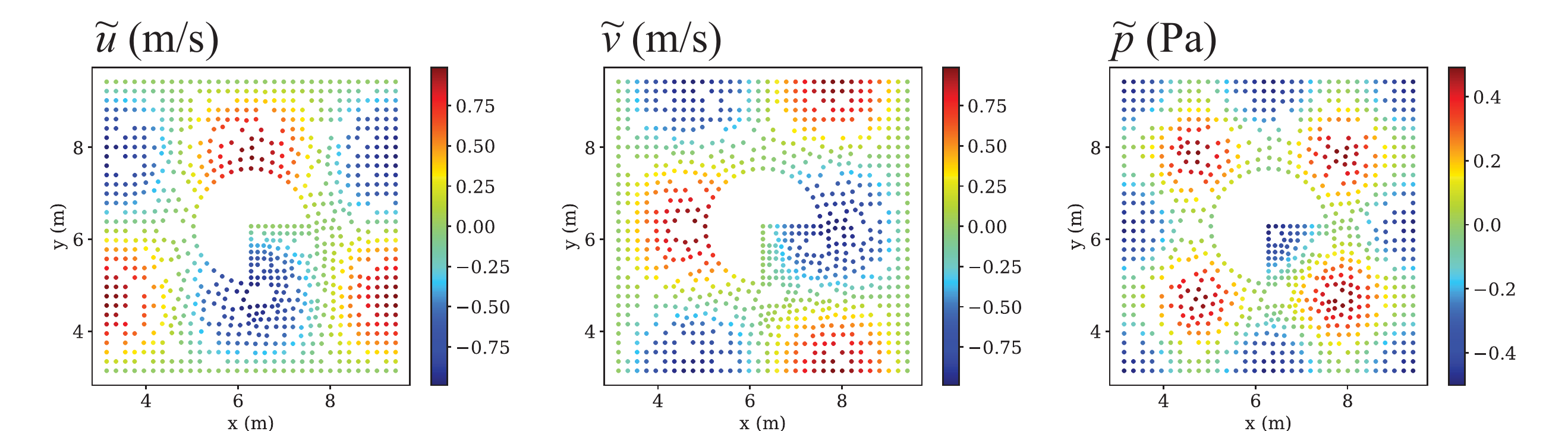}
\includegraphics[width=0.95\textwidth]{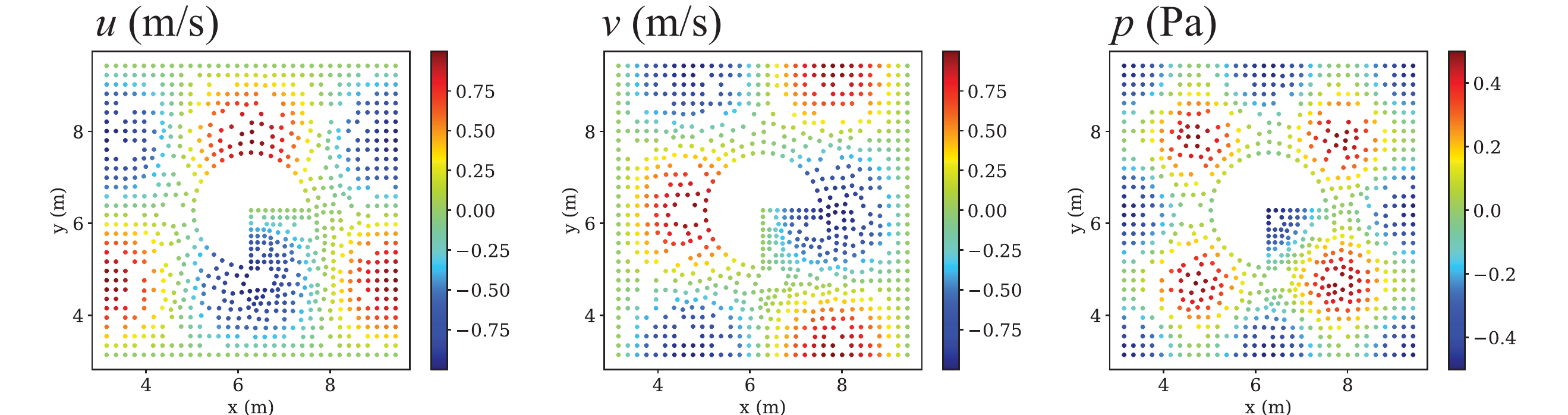}
\caption{The third group of examples taken from the set $\Phi=\{V_i\}_{i=1}^{26}$, comparing the analytical solutions (Eqs. \ref{Eq21}--\ref{Eq23}) to the PIPN predictions for the velocity and pressure fields for the method of manufactured solutions (Sect. \ref{Sect411})}
\label{Fig6}
\end{figure*}

The second column of Table \ref{Table2} tabulates the average, maximum, and minimum relative pointwise errors ($L^2$ norm) of the velocity and pressure fields predicted by PIPN over all 26 geometries of the set $\Phi=\{V_i\}_{i=1}^{26}$. Moreover, a graphical comparison between the velocity and pressure fields computed by PIPN and by the manufactured solutions (Eqs. \ref{Eq21}--\ref{Eq23}) is made in Figs. \ref{Fig4}--\ref{Fig6} for a few examples selected from the set $\Phi=\{V_i\}_{i=1}^{26}$. In general, the data listed in Table \ref{Table2} and graphical results in Figs. \ref{Fig4}--\ref{Fig6} illustrate a successful performance of PIPN in solving the steady-state Navier-Stokes and continuity equations over the set $\Phi=\{V_i\}_{i=1}^{26}$, containing geometries with various shapes, orientations, and spatial point distribution densities. Allocated data in Table \ref{Table3} demonstrates that 19141 iterations and consequently 51673 s (approximately 15 hours) are required for satisfying the convergence criterion (Eq. \ref{Eq27}). Similar to extant physics-informed neural network, PIPN also imposes significantly higher computational cost compared to supervised models as discussed in Sect. \ref{Sect1}.

\begin{figure*}
\centering
\includegraphics[width=0.95\textwidth]{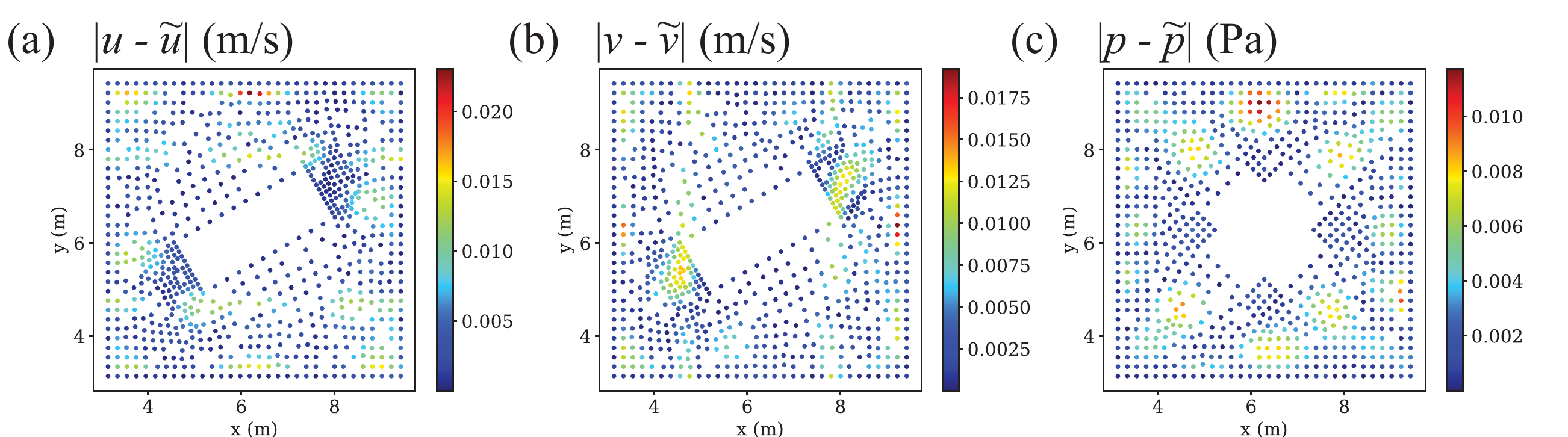}
\includegraphics[width=0.95\textwidth]{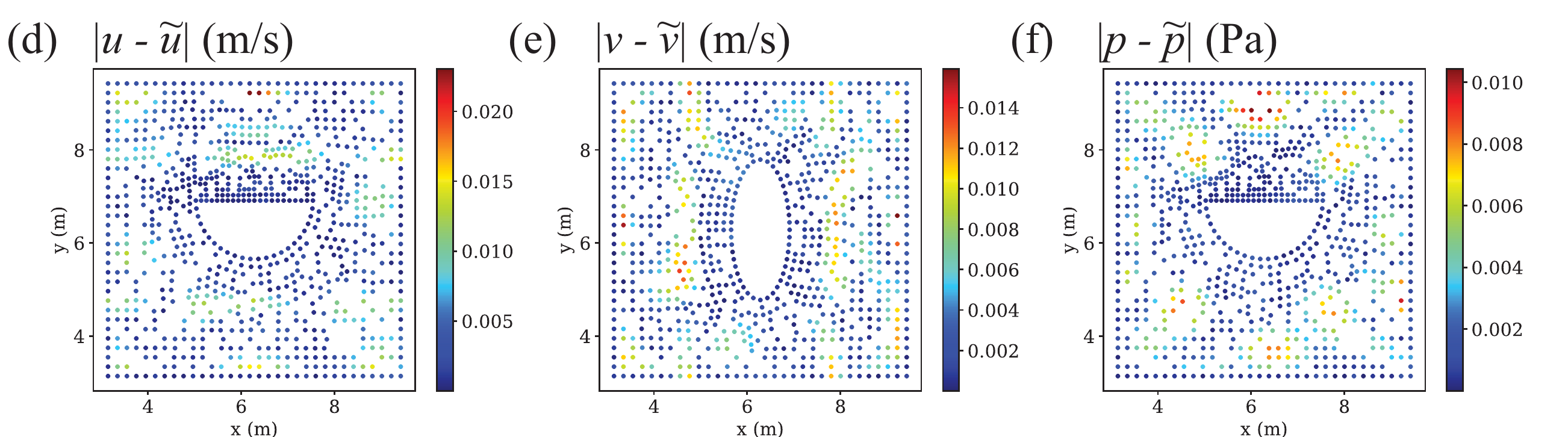}
\caption{Distribution of absolute pointwise error when the relative pointwise error ($L^2$ norm) becomes (\textbf{a}) maximum for $\Tilde{u}$, (\textbf{b}) maximum for $\Tilde{v}$, (\textbf{c}) maximum for $\Tilde{p}$, (\textbf{d}) minimum for $\Tilde{u}$, (\textbf{e}) minimum for $\Tilde{v}$, and (\textbf{f}) minimum for $\Tilde{p}$ for the method of manufactured solutions in the set $\Phi=\{V_i\}_{i=1}^{26}$ (Sect. \ref{Sect411})}
\label{Fig7}
\end{figure*}

Absolute error distributions of the predicted velocity and pressure fields over geometries with the maximum and minimum relative pointwise error ($L^2$ norm) are presented in Fig. \ref{Fig7}. Based on the results shown in Fig. \ref{Fig7}, the maximum local errors are observable both on the boundary and interior points of the computational domains. Note that the interior points are governed by the corresponding continuity and linear momentum equations and the boundary points are governed by the pressure and velocity boundary terms in the loss function (see Eqs. \ref{Eq6}--\ref{Eq8} and Eqs. \ref{Eq10}--\ref{Eq11}). In the current setting, all of these terms are equally weighted (i.e., $\lambda_i=1$ with $1 \leq i \leq 5$), and thus the numerical optimizer put equal efforts for minimizing them. One may adjust the weights based on a desired application to minimize local errors either on boundaries or interiors nodes. Alternatively, one may take care of a partial domain, for example, only points located on inner boundaries. The PIPN framework provides these flexibilities for users. As a general comment, incorporating later versions of PointNet \cite{qi2017pointnet} such as PointNet++ \cite{qi2017pointnet++} or KpConv \cite{kpconv} either with supervised or physics-informed models could potentially improve the issue of local errors as PointNet++ \cite{qi2017pointnet++} and KpConv \cite{kpconv} pays more attention to local features of geometries compared to PointNet \cite{qi2017pointnet}.

As shown in Fig. \ref{Fig7}a--\ref{Fig7}c, maximum errors of the velocity and pressure fields happen in geometries with rectangle and symmetrical star as $W$ (see Eq. \ref{Eq4}). The reason is relevant to the point cloud distributions in these geometries. As can be observed in Fig. \ref{Fig7}a--\ref{Fig7}c, there is a high concentration of points on the short side of the rectangle and symmetrical star, whereas a sparser point set exists on the longer side. Moreover, recall that we use the idea of implicit shape representation as discussed in Sect. \ref{Sect33}. Hence, such point cloud distributions impair the implicit representation of the rectangles and symmetrical star and cause relatively higher errors compared to other geometries of the set $\Phi=\{V_i\}_{i=1}^{26}$. 

\begin{table}[htbp]
\caption{Error analysis of the velocity and pressure fields for the method of manufactured solutions for the set $\Phi=\{V_i\}_{i=1}^{26}$ discussed in Sect. \ref{Sect411} through Sect. \ref{Sect414}. The $L^2$ norm over the entire domain ($V$) is indicated by $||\cdots||_V$.}
\centering
\begin{tabular}{l l l l l}
\toprule
Subsection & Sect. \ref{Sect411} & Sect. \ref{Sect412} & Sect. \ref{Sect413} & Sect. \ref{Sect414} \\
\midrule
No pressure boundary condition & \xmark & \checkmark & \xmark  & \xmark \\
\midrule
Conservative implementation of the momentum balance & \xmark  & \xmark  & \checkmark & \xmark \\
\midrule
Symbolic body force implementation & \xmark & \xmark  & \xmark  & \checkmark\\
\midrule
Average $||\tilde{u}-u||_V/||u||_V$ & $9.79855\textrm{E}-3$ & $1.33735\textrm{E}-2$ & $1.06916\textrm{E}-2$ & $9.74648\textrm{E}-3$\\
Maximum $||\tilde{u}-u||_V/||u||_V$ & $1.06049\textrm{E}-2$ & $1.55837\textrm{E}-2$ & $1.17249\textrm{E}-2$ & $1.09788\textrm{E}-2$ \\
Minimum $||\tilde{u}-u||_V/||u||_V$ & $9.11389\textrm{E}-3$ & $1.27436\textrm{E}-2$ & $9.92275\textrm{E}-3$ & $9.00381\textrm{E}-3$ \\
\midrule
Average $||\tilde{v}-v||_V/||v||_V$ & $9.55093\textrm{E}-3$ & $1.37429\textrm{E}-2$ & $1.02523\textrm{E}-2$ & $9.60376\textrm{E}-3$ \\
Maximum $||\tilde{v}-v||_V/||v||_V$ & $1.03773\textrm{E}-2$ & $1.49985\textrm{E}-2$ & $1.10241\textrm{E}-2$ & $1.06318\textrm{E}-2$ \\
Minimum $||\tilde{v}-v||_V/||v||_V$ & $8.89982\textrm{E}-3$ & $1.24088\textrm{E}-2$ & $9.46344\textrm{E}-3$ & $8.82407\textrm{E}-3$ \\
\midrule
Average $||\tilde{p}-p||_V/||p||_V$ & $1.27507\textrm{E}-2$ & $6.34154\textrm{E}-1$ & $1.37225\textrm{E}-2$ & $1.32666\textrm{E}-2$\\
Maximum $||\tilde{p}-p||_V/||p||_V$ & $1.35610\textrm{E}-2$ & $6.67422\textrm{E}-1$ & $1.42983\textrm{E}-2$ & $1.36537\textrm{E}-2$\\
Minimum $||\tilde{p}-p||_V/||p||_V$ & $1.19376\textrm{E}-2$ & $5.96424\textrm{E}-1$ & $1.27864\textrm{E}-2$ & $1.27184\textrm{E}-2$\\
\bottomrule
\end{tabular}
\label{Table2}
\end{table}

\subsubsection{Effect of pressure boundary condition \label{Sect412}}

\begin{figure*}[htbp]
\centering
\includegraphics[width=0.95\textwidth]{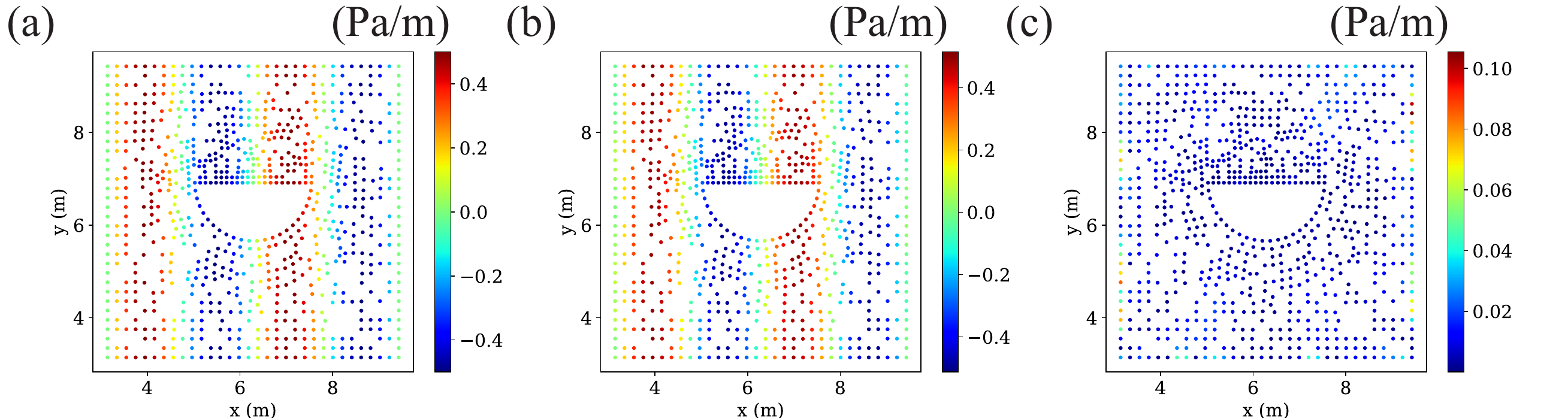}
\includegraphics[width=0.95\textwidth]{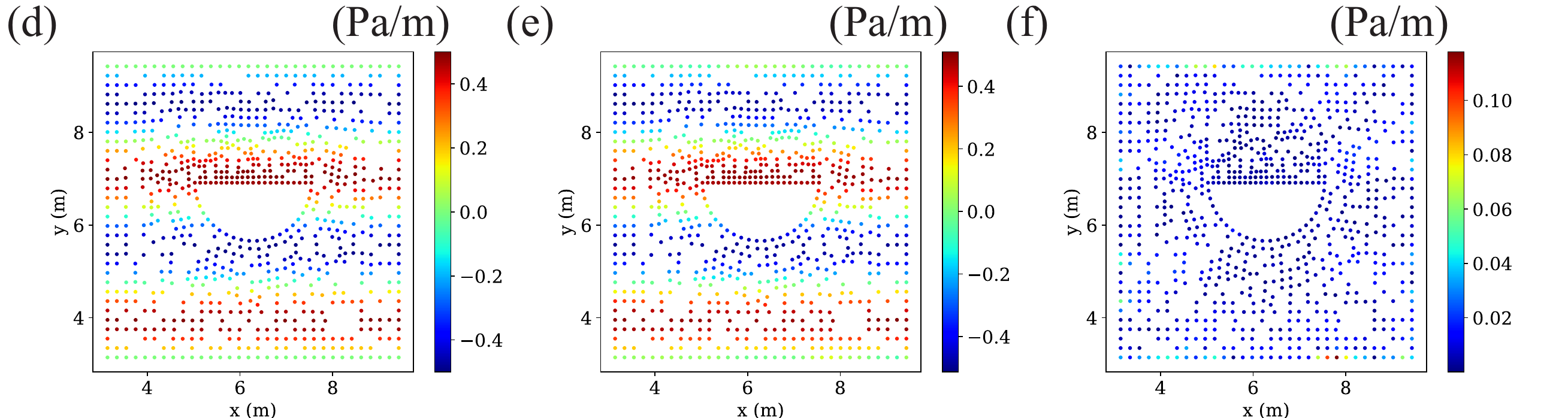}
\caption{A comparison between the pressure gradient field computed by the exact solution (Eq. \ref{Eq23}) and the pressure gradient field predicted by PIPN when the pressure boundary condition is not involved in its loss function (Eq. \ref{Eq30}) for the geometry with the maximum relative error ($L^2$ norm) of the pressure in the set $\Phi=\{V_i\}_{i=1}^{26}$ (Sect. \ref{Sect412}); (\textbf{a}) exact pressure gradient in the $x$ direction $\frac{\partial p}{ \partial x}$, (\textbf{b}) predicted pressure gradient in the $x$ direction $\frac{\delta \Tilde{p}}{\delta x}$, (\textbf{c}) absolute pointwise error of the pressure gradient in the $x$ direction $\big|\frac{\partial p}{\partial x}-\frac{\delta \Tilde{p}}{\delta x}\big|$, (\textbf{d}) exact pressure gradient in the $y$ direction $\frac{\partial p}{\partial y}$, (\textbf{e}) predicted pressure gradient in the $y$ direction $\frac{\delta \Tilde{p}}{\delta y}$, (\textbf{f}) absolute pointwise error of the pressure gradient in the $y$ direction $\big|\frac{ \partial p}{\partial y}-\frac{\delta \Tilde{p}}{\delta y}\big|$}
\label{Fig8}
\end{figure*}

Pressure is an implicit variable in the system of Eqs. \ref{Eq1}--\ref{Eq2} and is required to be determined such that the incompressibility constraint of the velocity field is satisfied \cite{timmermans1996approximate}. The influence of pressure boundary condition on the accuracy of numerical solvers of incompressible flows has been widely discussed in the literature (see e.g., Refs. \cite{gresho1987pressure,orszag1986boundary,hosseini2011pressure}). In this subsection, we investigate the effect of pressure boundary condition on the performance of PIPN. If we denote the boundary of the computational domain ($V$) by $\Gamma$, then $\Gamma$ consists of two non-overlapping subsets of Dirichlet ($\Gamma_D$) and Neumann ($\Gamma_N$) boundaries (see e.g., Eq. 1 in Ref. \cite{jobelin2006finite} and Eqs. 3--4 of Ref. \cite{timmermans1996approximate}) expressed as
\begin{eqnarray}
\label{Eq28}
\textbf{\textit{u}}=\textbf{\textit{u}}_{\Gamma_D} \textrm{ on } \Gamma_D,
\end{eqnarray}
\begin{eqnarray}
\label{Eq29}
-p\textbf{\textit{n}}+\mu \nabla \textbf{\textit{u}} \cdot \textbf{\textit{n}}=\textbf{\textit{t}}_{\Gamma_N} \textrm{ on } \Gamma_N,
\end{eqnarray}
where $\textbf{\textit{n}}$ is the outward unit vector normal to the boundaries and $\textbf{\textit{t}}_{\Gamma_N}$ is the stress vector acting on $\Gamma_N$. Mathematically, because there is no overlap between $\Gamma_D$ and $\Gamma_N$, if $\Gamma$ is solely described by the velocity Dirichlet boundary condition (Eq. \ref{Eq28}), the continuity and Navier-Stokes equations (Eqs. \ref{Eq1}--\ref{Eq2}) are solvable without specifying any boundary condition for the pressure quantity (see e.g., Sect. 2 of Ref. \cite{timmermans1996approximate}, Sect. 1 of Ref. \cite{shen1992error}, Sect. 1 of Ref. \cite{hosseini2011pressure} for a full discussion of this topic). \textcolor{red}{Moreover, note that the pressure (see Eq. \ref{Eq23}) and consequently the pressure gradient spatially vary over the boundary points of the geometries of the set $\Phi=\{V_i\}_{i=1}^{26}$. Hence, the presence of $r_i^{\text{pressure}_{\text{BC}}}$ in the loss function $\mathcal{J}$ (see Eq. \ref{Eq26}) provides valuable information to PIPN and thus contributes to a more accurate prediction of the pressure gradient and subsequently the velocity fields.} With the pressure boundary condition residual dropped from the loss function (see Eq. \ref{Eq26}), $\mathcal{J}$ is adjusted as
\begin{equation}
\label{Eq30}
\mathcal{J}=\frac{1}{m} \sum_{i=1}^m \left( \lambda_1 r_i^{\text{continuity}} + \lambda_2 r_i^{\text{momentum}_x} + \lambda_3 r_i^{\text{momentum}_y} + \lambda_4 r_i^{\text{velocity}_{\text{BC}}} \right ).
\end{equation}
Based on the relative pointwise error ($L^2$ norm) analysis listed in Table \ref{Table2}, the average relative error of the predicted velocity fields in the $x$ direction increases by 36.484\%, in the $y$ direction by 43.890\%, and the average relative error in the predicted pressure fields increases by 4873.483\%, compared to when the pressure boundary condition is enforced in the loss function. While the relative $L^2$ norm of the velocity field error still remains within reasonable accuracy (order of $10^{-2}$), the relative error order of the pressure field is approximately magnified by a factor of 10. This is simply because no pressure boundary condition is imposed to the loss function (Eq. \ref{Eq30}). Although PIPN does not preserve the accuracy of the pressure field in the absence of the pressure boundary term in the loss function, the pressure gradient predictions $\big(\frac{\delta \tilde{p}}{\delta x} \textrm{ and } \frac{\delta \tilde{p}}{\delta y}\big)$ retain a high level of accuracy, as exhibited in Fig. \ref{Fig8} and reported in Table \ref{Table4}. Since only the pressure gradient appears in the Navier-Stokes equations (not the pressure variable itself), a tremendous decay in the accuracy of the pressure field does not significantly negatively affect the velocity field predictions. Note that we validate $\frac{\delta \tilde{p}}{\delta x}$ and $\frac{\delta \tilde{p}}{\delta y}$ respectively with reference to $\frac{\partial p}{\partial x}$ and $\frac{\partial p}{\partial y}$ in Fig. \ref{Fig8} and Table \ref{Table4}, though $\frac{\delta}{\delta x}$ and $\frac{\partial }{\partial x}$ (or similarly $\frac{\delta }{\delta y}$ and $\frac{\partial}{\partial y}$) are not mathematically equivalent as discussed in Sect. \ref{Sect32}. However, $\frac{\partial p}{\partial x}$ and $\frac{\partial p}{\partial y}$ with $p$ as the analytical solution (see Eq. \ref{Eq23}) are the best available options for performing these error analysis.

As a last point in this subsection, we address the computational cost recorded in Table \ref{Table3}. As can be realized from the third column of Table \ref{Table3}, omitting the pressure boundary term from the loss function ends in substantially relaxing the optimization problem and requiring 14036 fewer seconds (approximately 4 fewer hours) for convergence. A similar conclusion is made in the PointNet supervised model by \citet{kashefi2021point}, where inviscid flows (allowed to slip on surfaces) had a faster convergence than viscous flow (requiring no slip condition on surfaces).

\begin{table}[h]
\caption{Computational cost for the convergence (see Eq. \ref{Eq26}) of the method of manufactured solutions for the set $\Phi=\{V_i\}_{i=1}^{26}$ discussed in Sect. \ref{Sect411} through Sect. \ref{Sect414}}
\centering
\begin{tabular}{l l l l l}
\toprule
Subsection & Sect. \ref{Sect411} & Sect. \ref{Sect412} & Sect. \ref{Sect413} & Sect. \ref{Sect414} \\
\midrule
No pressure boundary condition & \xmark  & \checkmark & \xmark  & \xmark \\
\midrule
Conservative implementation of the momentum balance & \xmark  & \xmark  & \checkmark & \xmark \\
\midrule
Symbolic body force implementation & \xmark  & \xmark  & \xmark & \checkmark\\
\midrule
Number of iterations for convergence & $19141$ & $14330$ & $21710$ & $21841$ \\
Wall time for convergence (s)  & $51673$ & $37637$ & $57760$ & $59704$ \\
Average wall time per iteration (s) & $2.6995$ & $2.6264$ & $2.6605$ & $2.7335$ \\
\bottomrule
\end{tabular}
\label{Table3}
\end{table}

\begin{table}[htbp]
\caption{Error analysis of the pressure gradient for the set $\Phi=\{V_i\}_{i=1}^{26}$ when the pressure boundary condition is denied in the PIPN loss function (Sect. \ref{Sect412}). $||\cdots||_V$ shows the $L^2$ norm over the entire domain ($V$).}
\centering
\begin{tabular}{l l l}
\toprule
& $\big|\big|\frac{\delta \tilde{p}}{\delta x}-\frac{\partial p}{\partial x}\big|\big|_V\big/\big|\big|\frac{\partial p}{\partial x}\big|\big|_V$ & $\big|\big|\frac{\delta \tilde{p}}{\delta y}-\frac{\partial p}{\partial y}\big|\big|_V\big/\big|\big|\frac{\partial p}{\partial y}\big|\big|_V$ \\
\midrule
Average & $5.40516\textrm{E}-2$ & $7.56162\textrm{E}-2$ \\
Maximum & $5.57193\textrm{E}-2$ & $8.57065\textrm{E}-2$ \\
Minimum & $5.18154\textrm{E}-2$ & $6.08985\textrm{E}-2$ \\
\bottomrule
\end{tabular}
\label{Table4}
\end{table}


\subsubsection{Conservative vs. non-conservative implementation \label{Sect413}}
The conservative form of the steady-state Navier-Stokes equations (Eq. \ref{Eq2}) in a two dimensional space and in the presence of a non-conservative body force can be written as
\begin{eqnarray}
\label{Eq31}
\frac{\partial}{\partial x} \left(\rho u^2 - 2\mu \frac{\partial u}{\partial x} + p \right) + \frac{\partial}{\partial y} \left(\rho u v -\mu \left (\frac{\partial v}{\partial x} + \frac{\partial u}{\partial y} \right) \right) = f^x, 
\end{eqnarray}
\begin{eqnarray}
\label{Eq32}
\frac{\partial}{\partial x} \left(\rho u v -\mu \left (\frac{\partial v}{\partial x} + \frac{\partial u}{\partial y} \right) \right) + \frac{\partial}{\partial y}\left (\rho v^2 - 2\mu \frac{\partial v}{\partial y} + p \right) = f^y. 
\end{eqnarray}
In this sense, the residuals of equations of the momentum balance in the $x$-direction $\big(r_i^{\text{momentum}_x}\big)$ and $y$-direction $\big(r_i^{\text{momentum}_y}\big)$ can be rewritten as
\begin{eqnarray}
\label{Eq33}
r_i^{\text{momentum}_x} = \frac{1}{M_1} \sum_{k=1}^{M_1} \left ( \frac{\delta}{\delta x_k} \left(\rho \Tilde{u}_k^2 - 2\mu \frac{\delta \Tilde{u}_k}{\delta x_k} + \Tilde{p}_k \right) + \frac{\delta}{\delta y_k} \left(\rho \Tilde{u}_k \Tilde{v}_k -\mu \left (\frac{\delta \Tilde{v}_k}{\delta x_k} + \frac{\delta \Tilde{u}_k}{\delta y_k} \right) \right)  - f_k^x \right)^2,
\end{eqnarray}
\begin{eqnarray}
\label{Eq34}
r_i^{\text{momentum}_y} = \frac{1}{M_1} \sum_{k=1}^{M_1} \left ( \frac{\delta}{\delta x_k} \left(\rho \Tilde{u}_k \Tilde{v}_k -\mu \left (\frac{\delta \Tilde{v}_k}{\delta x_k} + \frac{\delta \Tilde{u}_k}{\delta y_k} \right) \right) + \frac{\delta}{\delta y_k}\left (\rho \Tilde{v}_k^2 - 2\mu \frac{\delta \Tilde{v}_k}{\delta y_k} + \Tilde{p}_k \right)  - f_k^y \right)^2.
\end{eqnarray}
Comparing the coding implementation of conservative (Eqs. \ref{Eq7}--\ref{Eq8}) with the non-conservative form (Eqs. \ref{Eq33}--\ref{Eq34}), first it is realized that the term of $\left(\rho \Tilde{u}_k \Tilde{v}_k -\mu \left (\frac{\delta \Tilde{v}_k}{\delta x_k} + \frac{\delta \Tilde{u}_k}{\delta y_k} \right) \right)$ is common in both $\big(r_i^{\text{momentum}_x}\big)$ and $\big(r_i^{\text{momentum}_y}\big)$ in the conservative form. Furthermore, the conservative form involves 7 automatic differentiation operations for each $V_i$, while the non-conservative form involves 10 automatic differentiation operations. \textcolor{red}{Note that the TensorFlow \cite{tensorflow2015-whitepaper} software constructs the computation graph only once at the beginning of the PIPN execution. Thus, the computational time saved per epoch might not be significant; however, for simulations requiring a large number of epochs with many geometries (i.e., very large $m$) the total saved computational time might be noticeable.} In this regard, there is a motivation to investigate and compare the computational time required between these two implementation techniques. To the best of our knowledge, we answer this question for the first time in the area of physics-informed machine learning. Note that our discussion here is extendable to when the body force is conservative as well; nevertheless, since the body forces articulated in Sect. \ref{Sect41} is not conservative (i.e., having a non-zero curl value), we limit our discussion to non-conservative external body forces.

The relative pointwise errors ($L^2$ norm) of the predicted fields and the associated computational cost are respectively reported in the fourth column of Table \ref{Table2} and Table \ref{Table3}. With reference to the non-conservative implementation (Sect. \ref{Sect411}), for the conservative implementation the average relative error of the computed velocity and pressure fields respectively increase by 9.114\% for $\tilde{u}$, 7.343\% for $\tilde{v}$, and 7.621\% for $\tilde{p}$. Moreover, the number of iterations for meeting the convergence criterion (see Eq. \ref{Eq27}) increases (51673 vs. 57760), while the computational expense per iteration remains approximately the same as the non-conservative form (2.6995 s vs. 2.6605 s). It suggests that although the conservative implementation requires a fewer number of automatic differentiation compared to the non-conservative implementation, it imposes extra computational cost to the system since it requires a larger number of iterations for convergence. The reason for a slower convergence can be explained as follows. The continuity equation (Eq. \ref{Eq1}) of incompressible flows implicitly exists in the conservative form of the Navier-stokes equations (see Eqs. \ref{Eq31}--\ref{Eq32}). It implies that more weights are added to the continuity equation in the PIPN loss function of conservation implementation (see Eqs. \ref{Eq33}--\ref{Eq34}). Moreover, imposing the incompressibility constraint is the stiffest component of the computations of incompressible flows. Hence, by considering extra weights for the continuity equation in the loss function, the numerical optimizer of PIPN is enforced to put extra efforts to satisfy the constraint. Hence, this strategy leads to a slower convergence.

\begin{figure*}
\centering
\includegraphics[width=0.95\textwidth]{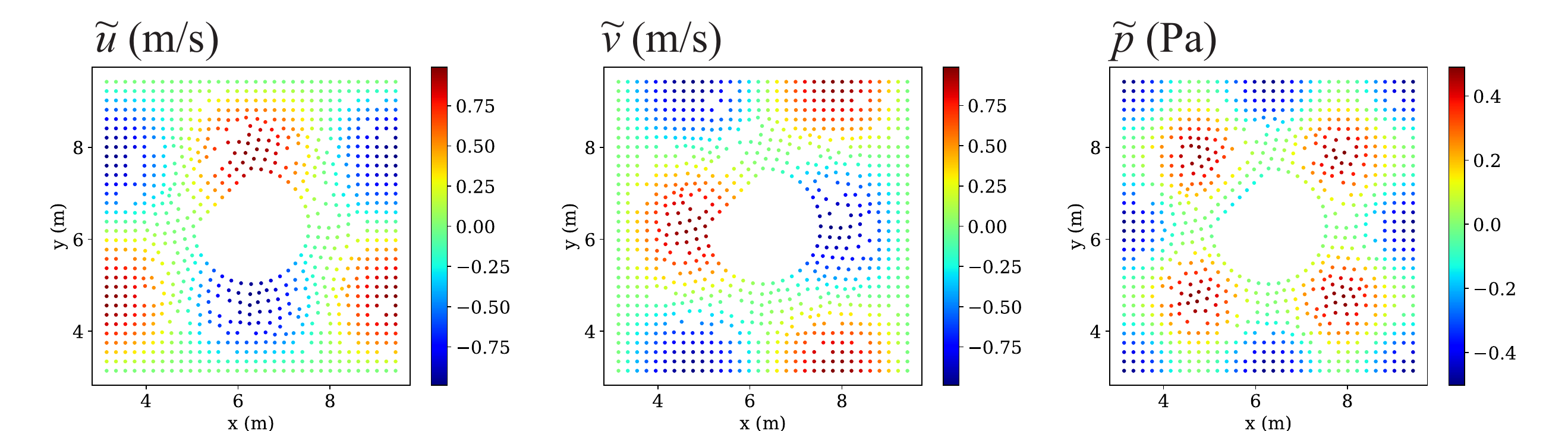}
\includegraphics[width=0.95\textwidth]{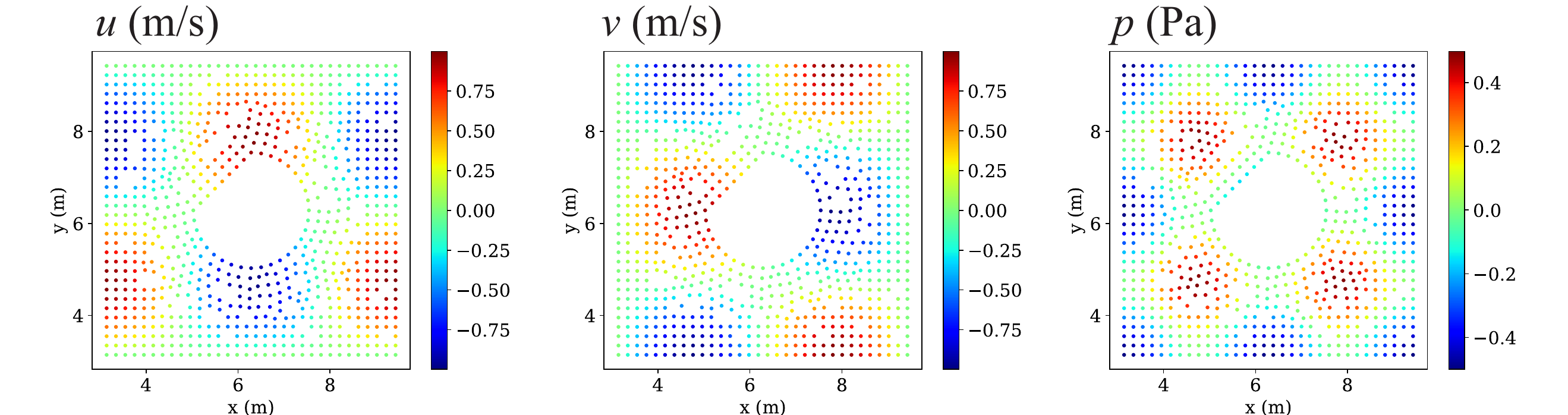}
\includegraphics[width=0.95\textwidth]{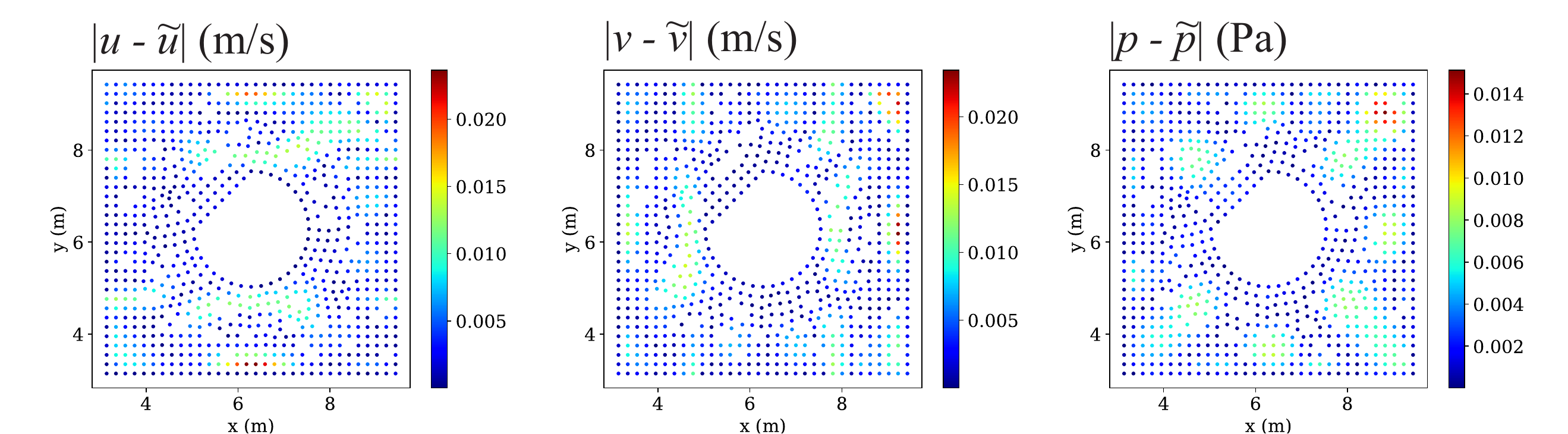}
\caption{\textcolor{red}{Comparisons between the ground truth (Eqs. \ref{Eq21}--\ref{Eq23}) and PIPN prediction for the velocity and pressure fields of a domain in the set $\Psi=\{V_i\}_{i=1}^{3}$ with a cavity created by a combination of an equilateral triangle and a three-quarter circle for testing the generalizability of PIPN (Sect. \ref{Sect415})}}
\label{Fig9}
\end{figure*}


\begin{figure*}
\centering
\includegraphics[width=0.95\textwidth]{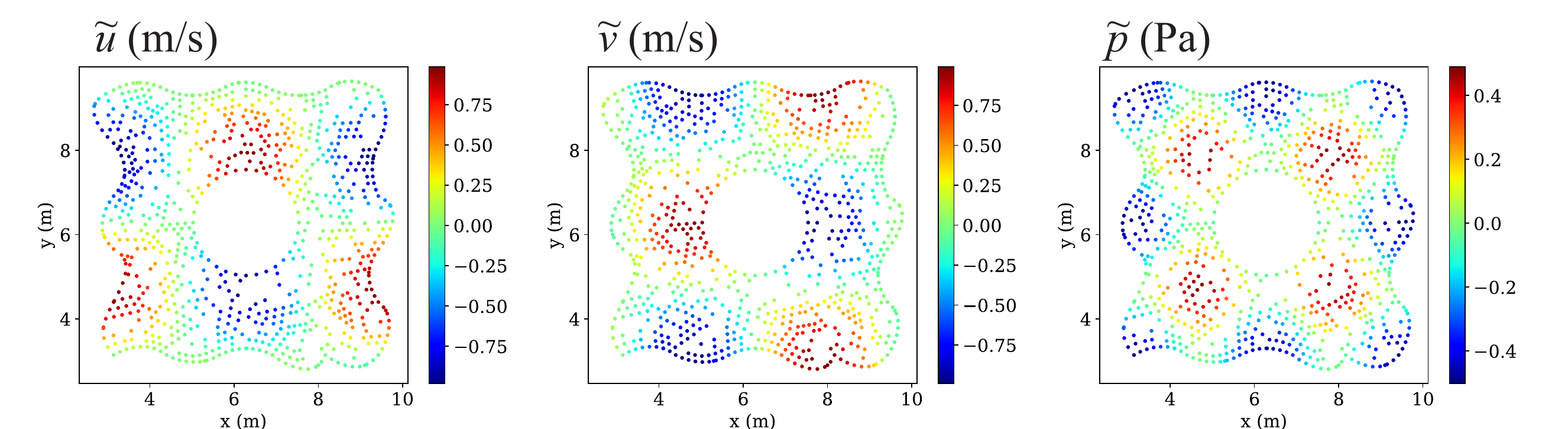}
\includegraphics[width=0.95\textwidth]{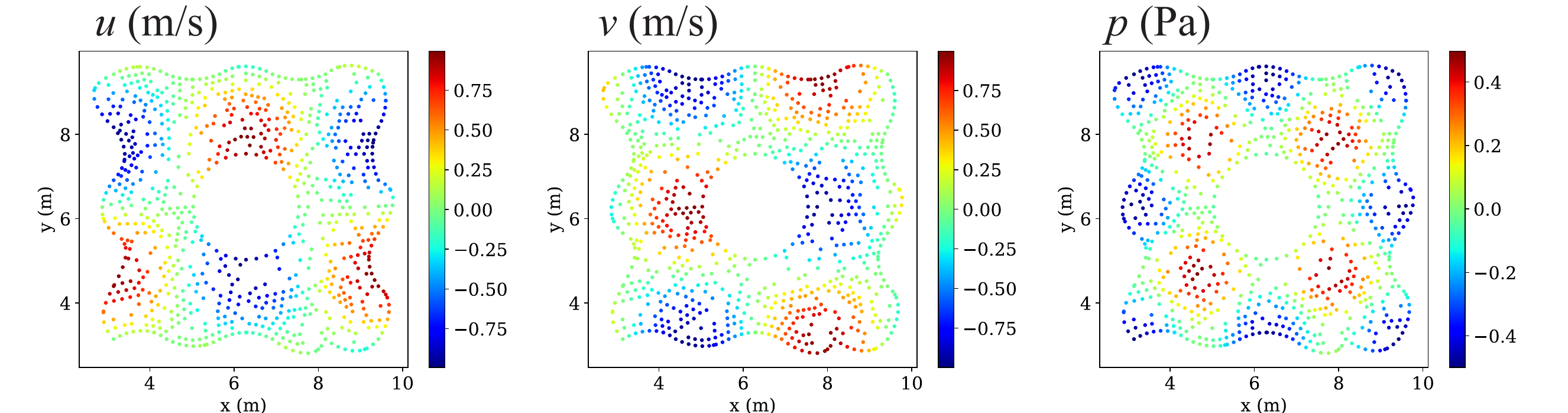}
\includegraphics[width=0.95\textwidth]{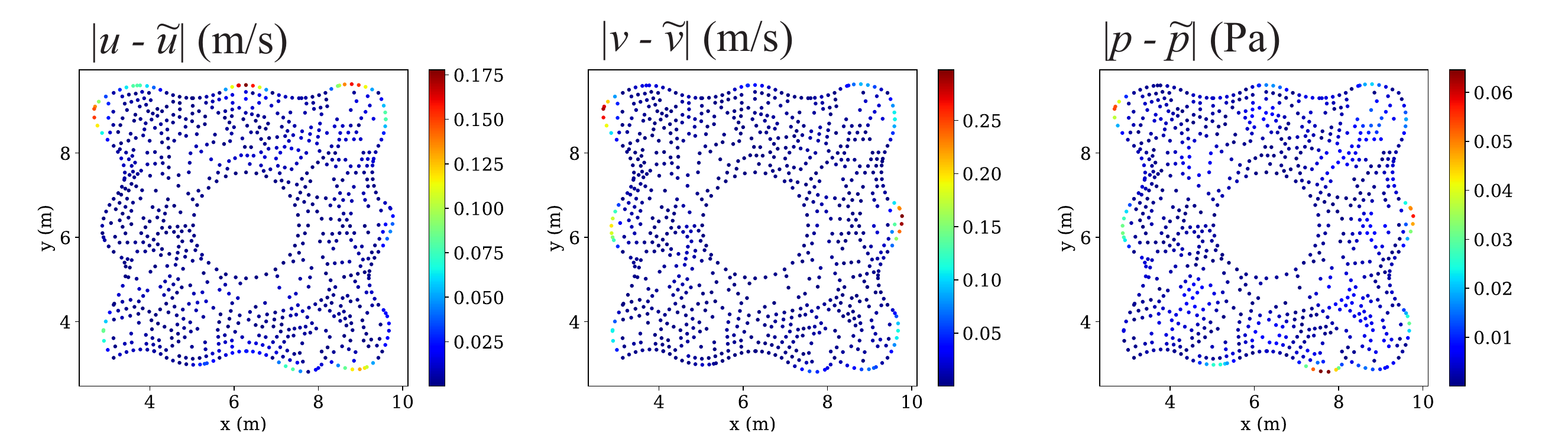}
\caption{\textcolor{red}{
Comparisons between the ground truth (Eqs. \ref{Eq21}--\ref{Eq23}) and PIPN prediction for the velocity and pressure fields of a domain in the set $\Psi=\{V_i\}_{i=1}^{3}$ with distorted outer boundaries for testing the generalizability of PIPN (Sect. \ref{Sect415})}}
\label{Fig10}
\end{figure*}


\begin{figure*}
\centering
\includegraphics[width=0.95\textwidth]{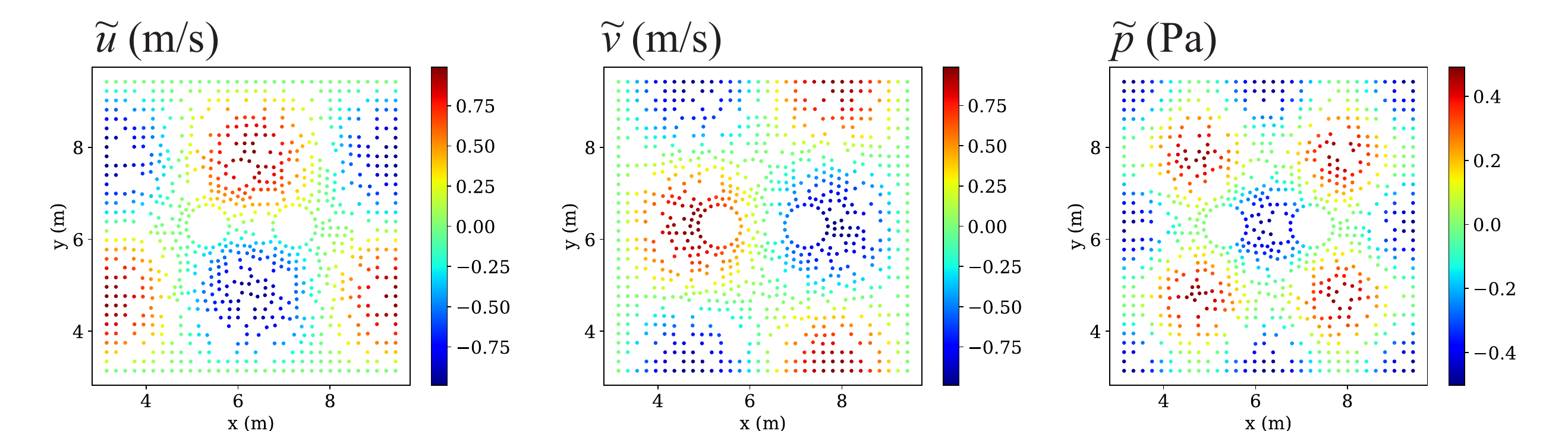}
\includegraphics[width=0.95\textwidth]{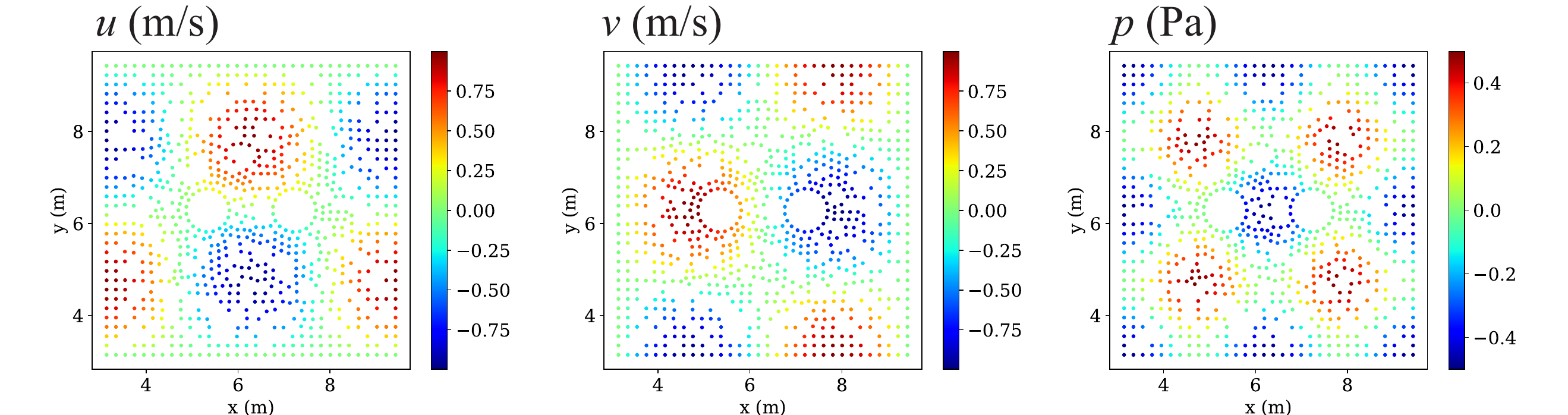}
\includegraphics[width=0.95\textwidth]{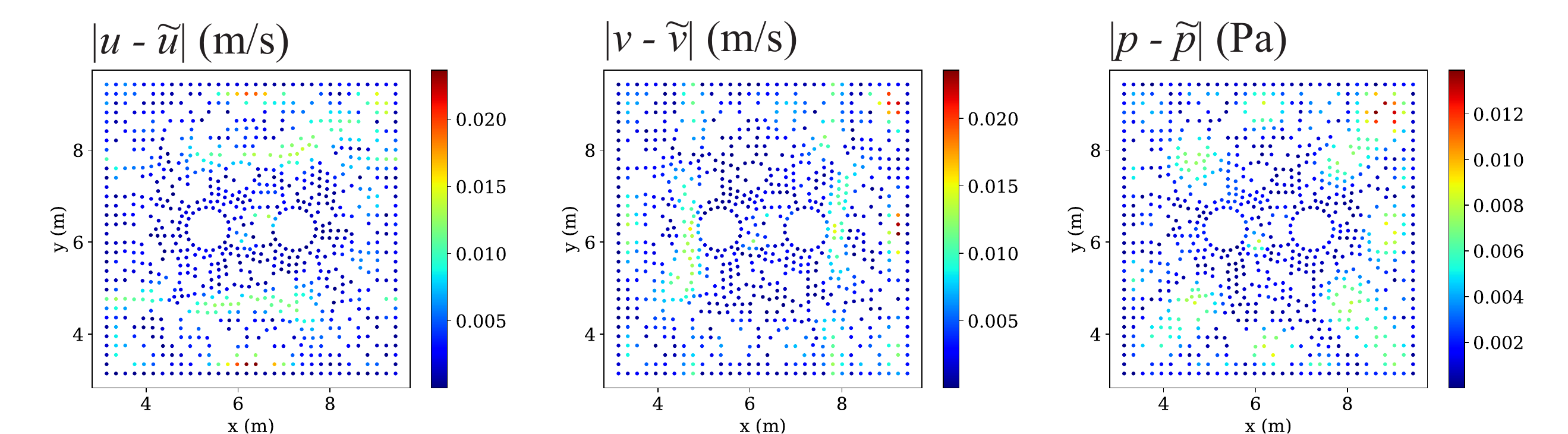}
\caption{\textcolor{red}{Comparisons between the ground truth (Eqs. \ref{Eq21}--\ref{Eq23}) and PIPN prediction for the velocity and pressure fields of a domain in the set $\Psi=\{V_i\}_{i=1}^{3}$ with two circular cavities inside for testing the generalizability of PIPN (Sect. \ref{Sect415}).}}
\label{Fig11}
\end{figure*}


\subsubsection{Body force implementation\label{Sect414}}

Because the body force terms ($f^x$ and $f^y$) are a function of the spatial domain (see Eqs. \ref{Eq24}--\ref{Eq25}), there are two strategies for their implementation in the loss function (Eq. \ref{Eq26}). The first and straightforward strategy is to implement them as ``constant'' tensors with pre-assigned values before starting the training procedure. In this strategy, the spatial derivatives of $f^x$ and $f^y$ become zero when TensorFlow \cite{tensorflow2015-whitepaper} computes the gradient of the loss function ($\mathcal{J}$). We take this approach in all subsections except the current one. The next strategy is to implement them as ``symbolic'' tensors. It is simply possible because $\mathbf{x}=(x,y)$ is defined as a symbolic tensor in PIPN. In contrast with the first implementation, the spatial derivatives of forcing terms appear in the loss function gradient during the training process in the second strategy. To explore the difference between these two, the second strategy is coded and we redo the experiment carried out in Sect. \ref{Sect411} with the same setup. The results are collected in Table \ref{Table2} and Table \ref{Table3}. Based on Table \ref{Table2} and compared to the ``constant'' body force tensors, the average relative pointwise error ($L^2$ norm) of the calculated velocity fields remains approximately unchanged (0.531\% decrease for $\tilde{u}$ and 0.553\% increase for $\tilde{v}$) and for the pressure field enlarges approximately by 4.046\%. According to Table \ref{Table3}, the second strategy is computationally more expensive requiring a larger number of iterations. Moreover, it has higher computational cost ``per iteration'' compared to the first strategy. It implies that the presence of extra terms in the loss function as a consequence of ``symbolic'' body force tensors makes the numerical optimization problem stiffer. Therefore, it is concluded that implementing the body forces as ``constant'' tensors results in a faster convergence i.e., satisfying the convergence criterion (Eq. \ref{Eq27}) with a fewer number of iterations.

\begin{table}[ht]
\caption{Error analysis of the velocity and pressure fields for testing the generalizability of PIPN using the method of manufactured solutions presented in Sect. \ref{Sect415}. The $L^2$ norm over the entire domain ($V$) is demonstrated by $||\cdots||_V$.}
\centering
\begin{tabular}{l l l l}
\toprule
& $||\tilde{u}-u||_V/||u||_V$ & $||\tilde{v}-v||_V/||v||_V$ & $||\tilde{p}-p||_V/||p||_V$ \\
\midrule
Domain with a cavity created by combination of an
\\
equilateral triangle and a three-quarter circle (Fig. \ref{Fig9}) & $1.32465\textrm{E}-2$ & $1.19450\textrm{E}-2$ & $1.62159\textrm{E}-2$ \\
Domain with distorted outer boundaries (Fig. \ref{Fig10}) & $5.24260\textrm{E}-2$ & $7.45601\textrm{E}-2$ & $3.49008\textrm{E}-2$ \\
Domain with two circular cavities inside (Fig. \ref{Fig11}) & $1.32377\textrm{E}-2$ &  $1.13976\textrm{E}-2$ & $1.47483\textrm{E}-2$ \\
\bottomrule
\end{tabular}
\label{Table5}
\end{table}


\subsubsection{Generalizability \label{Sect415}}
In this subsection, we evaluate another aspect of the PIPN algorithm: generalizability. The PIPN platform trained over defined geometries of the set $\Phi=\{V_i\}_{i=1}^{26}$ is now examined in order to predict the velocity and pressure fields over a set of never-seen geometries from unseen categories, $\Psi=\{V_i\}_{i=1}^{l}$. Practically, we inspect the generalizability of PIPN over three different domains ($l=3$). The outcomes are graphically shown in Figs. \ref{Fig9}--\ref{Fig11} and relative pointwise errors ($L^2$ norm) are tabulated in Table \ref{Table5}. Figure \ref{Fig9} exhibits a successful prediction of the velocity and pressure fields for an unseen geometry as a result of combining two seen geometries: the three-quarter circle \textcolor{red}{with the diameter of 0.8$\pi$ m} and the equilateral triangle (or alternatively the square). In fact, PIPN accomplishes digitizing geometric feature of the unseen geometry in the latent global feature (see Fig. \ref{Fig2}), which is, indeed, the result of interacting two other seen features. In Fig. \ref{Fig10}, we challenge PIPN by distorting the outer boundaries of \textcolor{red}{a computational domain with a circular cavity of the diameter of 0.8$\pi$ m}, while the network has only seen outer boundaries with straight lines. More challengingly, PIPN is asked to predict the solutions of the governing PDEs (Eqs. \ref{Eq1}--\ref{Eq2}) for points located outside of the seen spaces (i.e., $H=[\pi \text{ m},3\pi  \text{ m}]\times[\pi \text{ m},3\pi \text{ m}]$), as shown in Fig. \ref{Fig10}. By monitoring the absolute pointwise errors plotted in Fig. \ref{Fig10}, it is determined that that maximum local error also happens ``partially'' at these points (rather than interior points). It is worth noting that to smoothly represent the outer curved boundaries of this domain (see Fig. \ref{Fig10}), we increase the number of points on them compared to outer boundaries of the seen domains of the set $\Phi=\{V_i\}_{i=1}^{26}$ during training. Interestingly, PIPN is able to digest this modification and predicts accurately the velocity and pressure solutions for the most parts of the outer boundaries. All in all, an excellent level of accuracy for the desired fields is achieved except for the partial parts of the outer boundaries. In the last generalizability examination, two circular cavities \textcolor{red}{of the diameter of 0.7 m} are located at a horizontal distance \textcolor{red}{of 1.6 m} from each other, as depicted in Fig. \ref{Fig11}. PIPN has seen neither a domain with two cavities inside nor even a circular cavity with the chosen radius in this case. As summarized in Table \ref{Table5}, the relative pointwise errors ($L^2$ norm) of the velocity and pressure fields of this generalizability test are less than 2\%, indicating an acceptable performance of PIPN for predicting the solution of PDEs of interest in unseen geometries from unseen categories. Similar conclusion was given by \citet{kashefi2021point}, when their supervised PointNet \cite{qi2017pointnet} validly predicted laminar and steady incompressible flow around multiple objects, even though only single object was seen during its training (see Figs. 13--16 in Ref. \cite{kashefi2021point}).

\subsubsection{Comparison between PIPN and regular physics-informed neural networks\label{Sect416}}

In this subsection, we compare the performance of PIPN with a regular physics-informed neural network (PINN) such as those cited in Sect. \ref{Sect1} (e.g., see Refs. \cite{raissi2019physics,raissi2019deepVortex,haghighat2021physics}) in terms of computational cost and prediction accuracy. Specifically, we establish a PIPN by means of four sequential MLPs, respectively, with the size of (64, 64), (64, 64, 1024), (256, 128, 128), and (128, $N\times n_{\textrm{PDE}}$), all constructed with unshared FC layers. In this design, we apply five items for a fair comparison. First, with $n_{\textrm{PDE}}=3$, the number of trainable parameters of PINN (870129) are approximately equal to that of PIPN (892355). Second, we execute the PINN training until the value of its loss function satisfies the same convergence criterion (Eq. \ref{Eq27}) considered for PIPN. Third, similar to PIPN, the Adam optimizer \cite{kingma2014adam}, the hyperbolic tangent activation function, and batch normalization are used. Fourth, we tune the hyper-parameters of PINN to reach the highest possible performance. Fifth, the same machine and memory capacity are used for training PINN. Note that in contrast with PIPN, the PINN configuration is not able to be trained on all the geometries of $\Phi=\{V_i\}_{i=1}^{26}$ in one training practice. In fact, we need to train the PINN framework for each geometry of $\Phi=\{V_i\}_{i=1}^{26}$ individually. Table \ref{Table6} summarizes the outcomes of this comparison. Based on the information of Table \ref{Table6}, the order of accuracy of the results obtained by PINN is the same as that of the PIPN outputs. This is simply due to satisfying the same convergence criterion by the two frameworks. However, the required computational time for PINN is approximately 18 times greater than the cost consumed by PIPN. This factor is magnified by increasing the number of domains of the set $\Phi$. Additionally, to obtain the solution of velocity and pressure fields in the computational domains of $\Psi=\{V_i\}_{i=1}^{3}$, the PINN framework needs to be trained from scratch, requiring approximately 104016 seconds in total. This is while the PIPN trained over $\Phi=\{V_i\}_{i=1}^{26}$ is able to predict the desired fields on the domains of $\Psi=\{V_i\}_{i=1}^{3}$ in less than 3 seconds, as discussed in Sect. \ref{Sect415}. Tuning learning rate ($\alpha$) of PINN is another notable challenge. In contrast to PIPN, a constant learning rate of $\alpha=0.001$ does not lead to a convergence and it has to be adaptively decreased after a certain number of iterations. Moreover, the PINN optimization problem becomes unstable at some cases and there is a tendency to converge to the trivial solution of zero in the interior computational field. As a consequence of these challenges, a considerable amount of offline efforts is required by the PINN framework, even before executing the final training. This example demonstrates the advantages of the introduced PIPN over the extant PINNs from the computational cost and training procedure perspectives.   


\begin{table}[ht]
\caption{Comparison between the performance of PIPN and a regular PINN for the set $\Phi=\{V_i\}_{i=1}^{26}$ described in Sect. \ref{Sect411} and Table \ref{Table1}. Note that the PIPN framework is trained over all 26 geometries of the set $\Phi=\{V_i\}_{i=1}^{26}$ in one training set, while the regular PINN needs to be trained separately for each geometry of the set $\Phi=\{V_i\}_{i=1}^{26}$. $||\cdots||_V$ shows the $L^2$ norm over the entire domain ($V$).}
\centering
\begin{tabular}{l l l}
\toprule
Physics-informed model & PIPN & PINN \\
\midrule
Number of iterations for convergence & $19141$ & $2080133$ (in total) \\
\midrule
Wall time for convergence (s) & $51673$ & $902361$ (in total) \\
\midrule
Average $||\tilde{u}-u||_V/||u||_V$ & $9.79855\textrm{E}-3$ & $1.11438\textrm{E}-2$ \\
Average $||\tilde{v}-v||_V/||v||_V$ & $9.55093\textrm{E}-3$ & $1.06153\textrm{E}-2$ \\
Average $||\tilde{p}-p||_V/||p||_V$ & $1.27507\textrm{E}-2$ & $1.28889\textrm{E}-2$ \\
\bottomrule
\end{tabular}
\label{Table6}
\end{table}



\begin{table}
\caption{Description of the generated geometries discussed in Sect. \ref{Sect421}. $\Omega$ represents counterclockwise rigid rotation of space $W$ with respect to its geometric center.}
\centering
\begin{tabular}{l l l l}
\toprule
Shape of $W$ (see Eq. \ref{Eq4}) & Side length & $\Omega$ (variation in orientation) & Number of data \\
\midrule
Equilateral nonagon & $0.365 \times \sin{\frac{\pi}{9}} \times \csc \frac{\pi}{7}  \textrm{ m}$ & $1^\circ,2^\circ,\cdots,39^\circ,40^\circ$ & $40$ \\
Equilateral octagon & $0.8(\sqrt{2}-1)  \textrm{ m}$ & $1^\circ,2^\circ,\cdots,44^\circ,45^\circ$ & $45$ \\
Equilateral heptagon & $0.365 \textrm{ m}$ & $1^\circ,2^\circ,\cdots,49^\circ,50^\circ$ & $50$ \\
\bottomrule
\end{tabular}
\label{Table7}
\end{table}


\begin{figure*}
\centering
\includegraphics[width=0.95\textwidth]{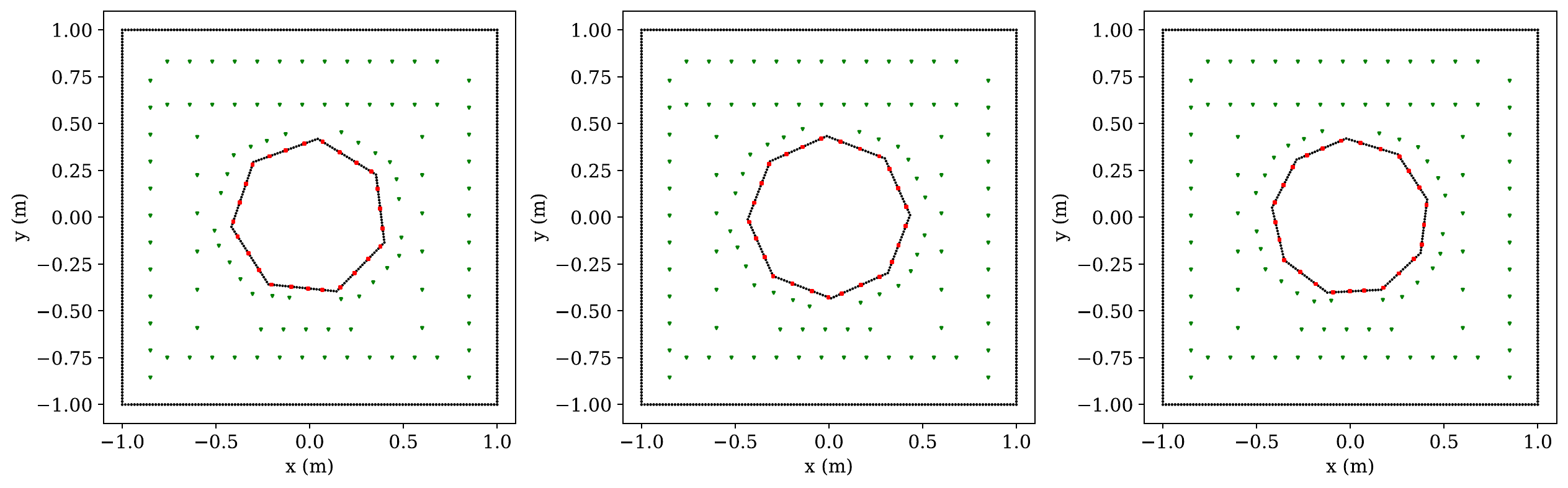}
\caption{Sensor locations for three domains from the set $\Phi=\{V_i\}_{i=1}^{108}$ (Sect. \ref{Sect421}); Green triangles show sensors measuring the velocity, pressure, and temperature values, whereas red squares demonstrate sensors measuring only the pressure and temperature values.}
\label{Fig12}
\end{figure*}


\begin{table}[htbp]
\caption{Error analysis of the velocity, pressure, and temperature fields over the set $\Phi=\{V_i\}_{i=1}^{108}$ and $\Psi=\{V_i\}_{i=1}^{27}$ with the presence of sparse pressure observations in the loss function (Eq. \ref{Eq39}) for the buoyancy induced-convection problem discussed in Sect. \ref{Sect421}. $||\cdots||_V$ shows the $L^2$ norm over the entire domain ($V$), while $||\cdots||_{\Gamma_{\textrm{inner}}}$ indicates the$ L^2$ norm over the inner surface ($\Gamma_{\textrm{inner}}$).}
\centering
\begin{tabular}{l l l}
\toprule
 & Over the set $\Phi=\{V_i\}_{i=1}^{108}$ & Over the set $\Psi=\{V_i\}_{i=1}^{27}$\\
\midrule
Average $||\tilde{u}-u||_V/||u||_V$ & $1.08589\textrm{E}-1$ & $1.18250\textrm{E}-1$ \\
Maximum $||\tilde{u}-u||_V/||u||_V$ & $1.43521\textrm{E}-1$ &  $1.45620\textrm{E}-1$ \\
Minimum $||\tilde{u}-u||_V/||u||_V$ & $7.73181\textrm{E}-2$ &  $8.92147\textrm{E}-2$ \\
\midrule
Average $||\tilde{v}-v||_V/||v||_V$ & $9.06379\textrm{E}-2$ & $1.07225\textrm{E}-1$ \\
Maximum $||\tilde{v}-v||_V/||v||_V$ & $1.27650\textrm{E}-1$ &  $1.27621\textrm{E}-1$ \\
Minimum $||\tilde{v}-v||_V/||v||_V$ & $6.23631\textrm{E}-2$ &  $8.26707\textrm{E}-2$ \\
\midrule
Average $||\tilde{p}-p||_V/||p||_V$ &  $2.89030\textrm{E}-2$ & $2.89858\textrm{E}-2$ \\
Maximum $||\tilde{p}-p||_V/||p||_V$ &  $3.34799\textrm{E}-2$ & $3.28380\textrm{E}-2$ \\
Minimum $||\tilde{p}-p||_V/||p||_V$ &  $2.47451\textrm{E}-2$ & $2.37549\textrm{E}-2$ \\
\midrule
Average $||\widetilde{T}-T||_V/||T||_V$ & $3.66134\textrm{E}-2$ & $3.89187\textrm{E}-2$ \\
Maximum $||\widetilde{T}-T||_V/||T||_V$ &  $4.91492\textrm{E}-2$ & $4.60109\textrm{E}-2$ \\
Minimum $||\widetilde{T}-T||_V/||T||_V$ & $2.57257\textrm{E}-2$ & $2.86674\textrm{E}-2$ \\
\midrule
Average $||\widetilde{T}-T||_{\Gamma_{\textrm{inner}}}/||T||_{\Gamma_{\textrm{inner}}}$ & $2.26461\textrm{E}-2$ & $7.41294\textrm{E}-2$ \\
Maximum $||\widetilde{T}-T||_{\Gamma_{\textrm{inner}}}/||T||_{\Gamma_{\textrm{inner}}}$  & $4.14641\textrm{E}-2$ & $1.12027\textrm{E}-1$ \\
Minimum $||\widetilde{T}-T||_{\Gamma_{\textrm{inner}}}/||T||_{\Gamma_{\textrm{inner}}}$  & $1.19202\textrm{E}-2$ & $2.07173\textrm{E}-2$ \\
\bottomrule
\end{tabular}
\label{Table8}
\end{table}


\subsection{Natural convection in a square enclosure with a cylinder\label{Sect42}}

\citet{cai2021physics} used a physics-informed model to study an inverse problem of two-dimensional forced convection heat transfer (e.g., see Fig. 2 of Ref. \cite{cai2021physics}) as well as steady and unsteady mixed convection heat transfer for flow past a circular cylinder. Moreover, \citet{wang2021reconstruction} employed a physic-informed model for solving an inverse problem of natural convection in a square enclosure but only with a circular inner cylinder. We specifically evaluate the capability of PIPN for an inverse problem of thermally-driven flow in this subsection.

Considering a hot cylinder with the temperature of $T_h$ at its surface located in a cold square enclosure with the temperature of $T_c$ at its surface, the buoyancy force leads to the natural convection; where the solution of the energy equation (Eq. \ref{Eq3}) stands in the source term of the momentum conservation equation (Eq. \ref{Eq2}) using the Boussinesq approximation \cite{lee2010natural}. In this way, the velocity and temperature variables are coupled together in the governing PDEs (Eqs. \ref{Eq1}--\ref{Eq3}). Following the Boussinesq approximation \cite{lee2010natural}, the forcing terms are given by
\begin{eqnarray}
\label{Eq35}
f^x=0,
\end{eqnarray}
\begin{eqnarray}
\label{Eq36}
f^y=\rho G \beta (T-T_{\textrm{ref}}),
\end{eqnarray}
where $G$ is the magnitude of the acceleration due to gravity. The thermal expansion is shown by $\beta$ and $T_{\textrm{ref}}$ indicates the reference temperature. The Rayleigh number ($Ra$) is determined as
\begin{eqnarray}
\label{Eq37}
Ra=\frac{\rho^2 c_p G \beta (T_h - T_c) L^3}{\kappa \mu},
\end{eqnarray}
and the Prandtl number ($Pr$) is expressed as
\begin{eqnarray}
\label{Eq38}
Pr=\frac{c_p \mu}{\kappa}.
\end{eqnarray}
The side length ($L$) of the square enclosure, $H$, is set to 2 m and $H:=[-1 \text{ m}, 1\text{ m}]\times[-1 \text{ m},1 \text{ m}]$ (see Eq. \ref{Eq4}). Zero-velocity Dirichlet condition is enforced at all the boundaries. We set the density ($\rho$), thermal expansion ($\beta$), acceleration due to gravity ($G$), specific heat ($c_p$), and hot temperature ($T_h$) to 1.00, the cold temperature ($T_c$) and reference temperature ($T_{\textrm{ref}}$) to 0.00, the viscosity ($\mu$) and thermal conductivity ($\kappa$) to $2 \sqrt{2} \times 10^{-2.5}$, all in the International Unit System. This set of choice leads to a Rayleigh number of $Ra=10^5$ and a Prandtl number of $Pr=1.0$.

The natural convection problem has been studied for inner cylinders with polygonal shapes (see e.g., Refs. \cite{wang2019natural,savio2022numerical,hussain2019comparison}). In this scenario, we consider three different shapes for $W$ (see Eq. \ref{Eq4}): equilateral heptagon, equilateral octagon, and equilateral nonagon. Particularly, we enlarge the number of geometries by rotating the inner cylinders. Details of the generated geometries can be found in Table \ref{Table7}. According to Table \ref{Table7}, 135 geometries are generated. These geometries are randomly split into two sets of $\Phi=\{V_i\}_{i=1}^{108}$ ($m=108$) and $\Psi=\{V_i\}_{i=1}^{27}$ ($l=27$). In this case, the set $\Psi$ contains unseen geometries from seen categories. In the point clouds, we set $N=5000$, $M_1 = 4340$, $M_2=660$, $M_3=492$, $M_4=105$, and $M_5=130$ (see Sect. \ref{Sect32} for recalling the notations).

We define the inverse problem as follows: given the velocity boundary condition on all the boundaries, the temperature boundary condition only on the outer surface, and a set of sparse observation of the velocity, temperature, and pressure fields at sensor locations; find the full velocity, temperature, and pressure fields at the inquiry points, specifically the temperature distribution on the surface of inner cylinder. Mathematically, this is an ill-posed problem. Accordingly, the loss function of this problem is determined as
\begin{equation}
\label{Eq39}
\begin{split}
\mathcal{J}=\frac{1}{m} \sum_{i=1}^m \big( \lambda_1 r_i^{\text{continuity}} + \lambda_2 r_i^{\text{momentum}_x} + \lambda_3 r_i^{\text{momentum}_y} + \lambda_4 r_i^{\text{velocity}_{\text{BC}}} + 
\lambda_6 r_i^{\text{velocity}_{\text{obs}}}
+ \lambda_7 r_i^{\text{pressure}_{\text{obs}}}
+ \lambda_8 r_i^{\text{energy}}
\\
+ \lambda_9 r_i^{\text{temperature}_{\text{outer-BC}}}
+ \lambda_{10} r_i^{\text{temperature}_{\text{obs}}}\big).
\end{split}
\end{equation}
The associated weights of each residual ($\lambda_i$) are defined similar to those discussed for Eq. \ref{Eq26} and are set to 1.00 for this machine-learning experiment. Note that no boundary condition is specified for the pressure in the loss function (Eq. \ref{Eq39}). Given our computational resources, the batch size ($\mathcal{B}$) can vary from 1 to 4. We obtain the highest performance of PIPN for $\mathcal{B}=2$ in this case study. We select a constant learning rate of $\alpha=0.0005$ and execute the training until PIPN satisfies the following criterion:
\begin{equation}
\label{Eq40}
\mathcal{J}  \leq 7 \times 10^{-4}.
\end{equation}

Generally speaking, the number of sensors and their locations affect the accuracy of outputs of physics-informed neural networks, when an inverse problem is solved (e.g., see Sect. 3.2 of Ref. \cite{wang2021reconstruction}, Sect. 3.1.3 of Ref. \cite{cai2021physics}, and Sect. 4.2 of Ref. \cite{lou2021physics} for a discussion on this matter). For example, 100 sensors were used in Refs. \cite{lou2021physics,wang2021reconstruction}. For this test case, we place 105 sensors inside the domains and 25 sensors on the surface of the inner cylinders, as shown in Fig. \ref{Fig12}. Sensors inside the domains (shown by green triangles in Fig. \ref{Fig12}) measure the velocity, pressure, and temperature values, while the sensors on the inner surfaces (shown by red squares in Fig. \ref{Fig12}) only infer the pressure and temperature values. Location of 25 sensors on the surface of inner cylinders and 25 sensors surrounding the inner cylinders are slightly adjusted based on the shape of their cross sections, while the locations of 80 remaining sensors are fixed over all $V_i \in \Phi$.

For generating the sparse data in the virtual locations of sensors and validation of results obtained by PIPN in the inquiry points, we employ one of our high fidelity finite-element numerical solvers already used in the literature \cite{kashefi2018finite,kashefi2020coarse,kashefi2020coarseb,kashefi2020coarseC,kashefi2021coarseE}. Note that the point clouds are a subset of grid vertices of finite-element meshes used for the numerical solver.

\begin{figure*}[h]
\centering
\includegraphics[width=0.95\textwidth]{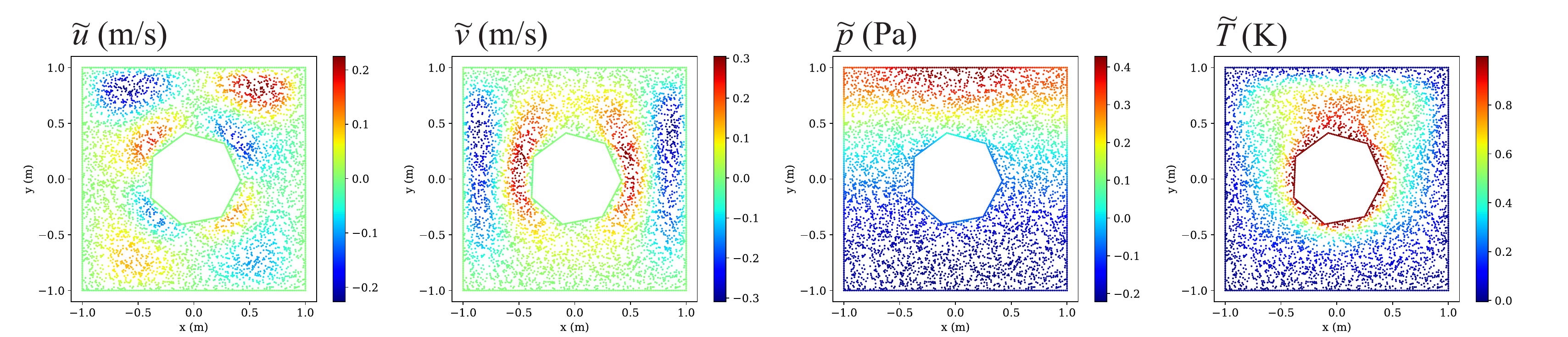}
\includegraphics[width=0.95\textwidth]{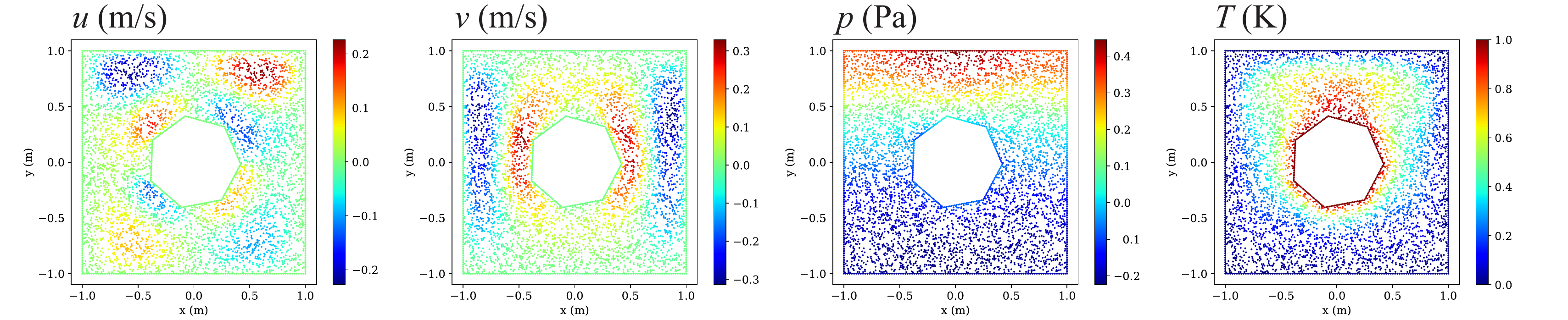}
\includegraphics[width=0.95\textwidth]{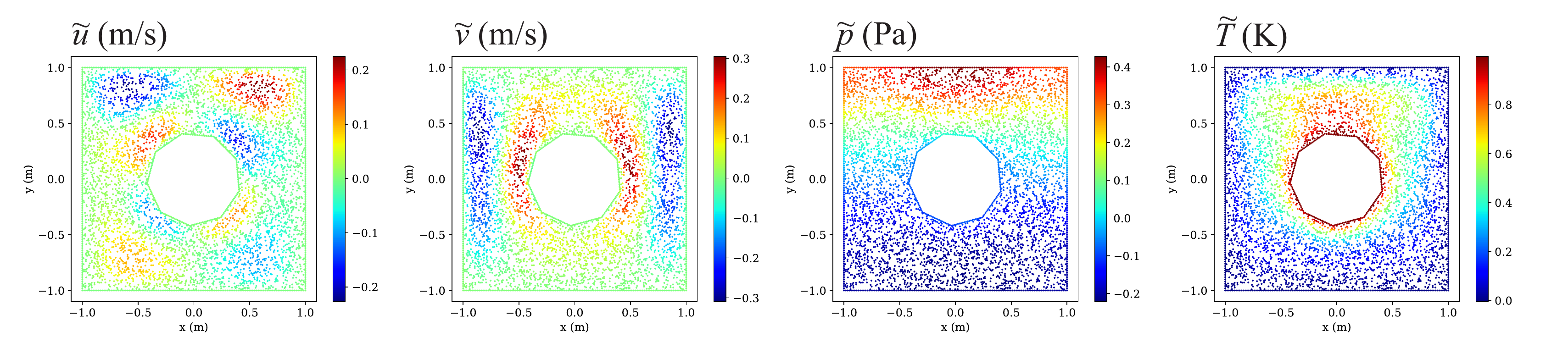}
\includegraphics[width=0.95\textwidth]{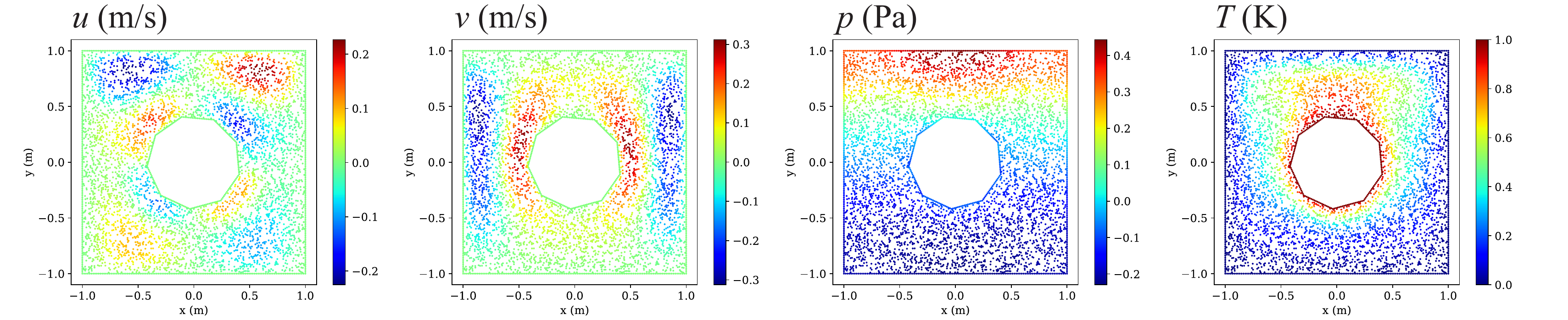}
\caption{Two examples taken from the set $\Phi=\{V_i\}_{i=1}^{108}$ comparing the ground truth solutions to the PIPN solutions for the velocity, pressure, and temperature fields for the thermally-driven flow problem (Sect. \ref{Sect421})}
\label{Fig13}
\end{figure*}



\begin{figure*}[htbp]
\centering
\includegraphics[width=0.95\textwidth]{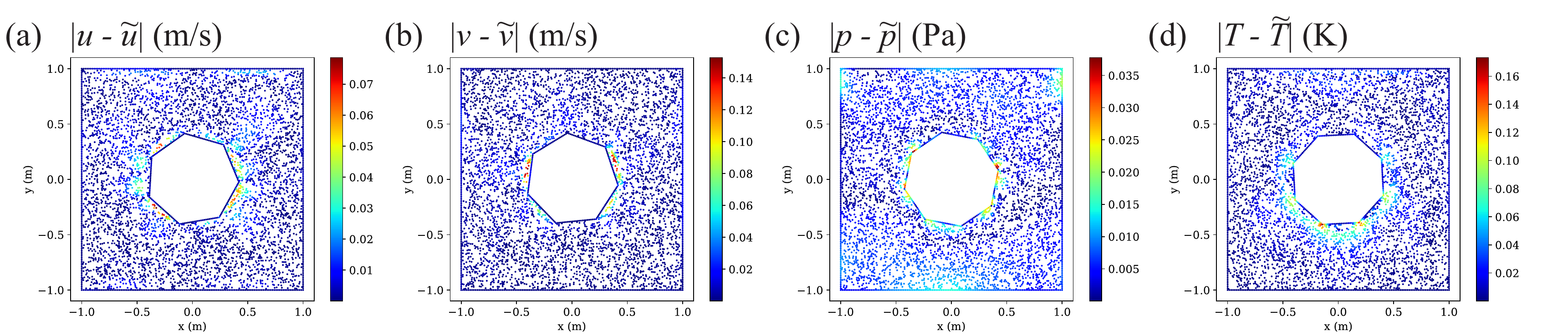}
\includegraphics[width=0.95\textwidth]{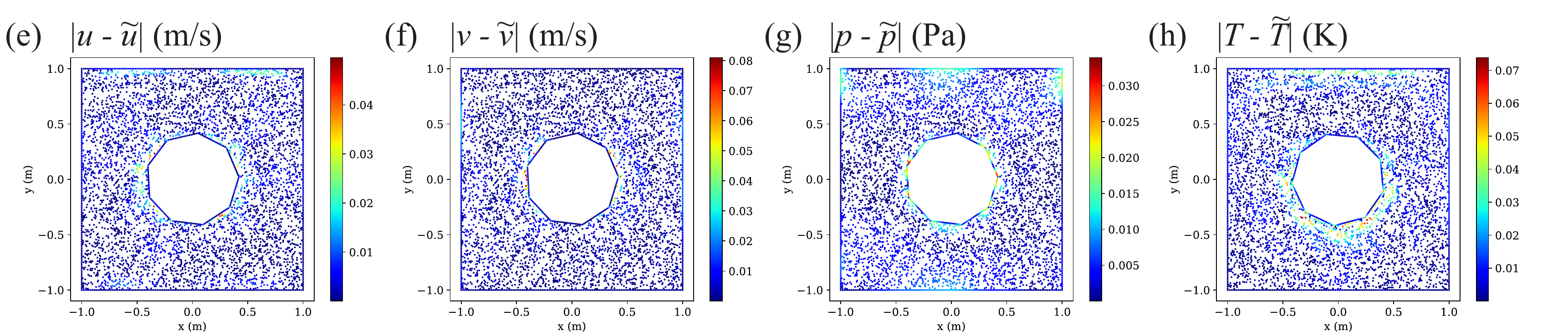}
\caption{Distribution of absolute pointwise error when the relative pointwise error ($L^2$ norm) becomes (\textbf{a}) maximum for $\Tilde{u}$, (\textbf{b}) maximum for $\Tilde{v}$, (\textbf{c}) maximum for $\Tilde{p}$, (\textbf{d}) maximum for $\widetilde{T}$, (\textbf{e}) minimum for $\Tilde{u}$, (\textbf{f}) minimum for $\Tilde{v}$, (\textbf{g}) minimum for $\Tilde{p}$, and (\textbf{h}) minimum for $\widetilde{T}$ for the thermally-driven flow problem in the set $\Phi=\{V_i\}_{i=1}^{108}$ (Sect. \ref{Sect421}).}
\label{Fig14}
\end{figure*}


\subsubsection{General analysis\label{Sect421}}
\begin{figure*}[htbp]
\centering
\includegraphics[width=0.95\textwidth]{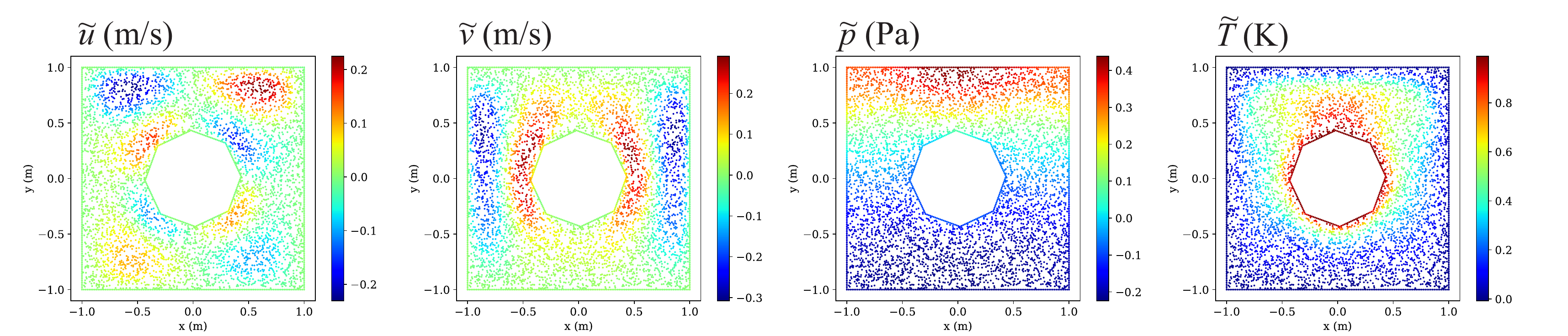}
\includegraphics[width=0.95\textwidth]{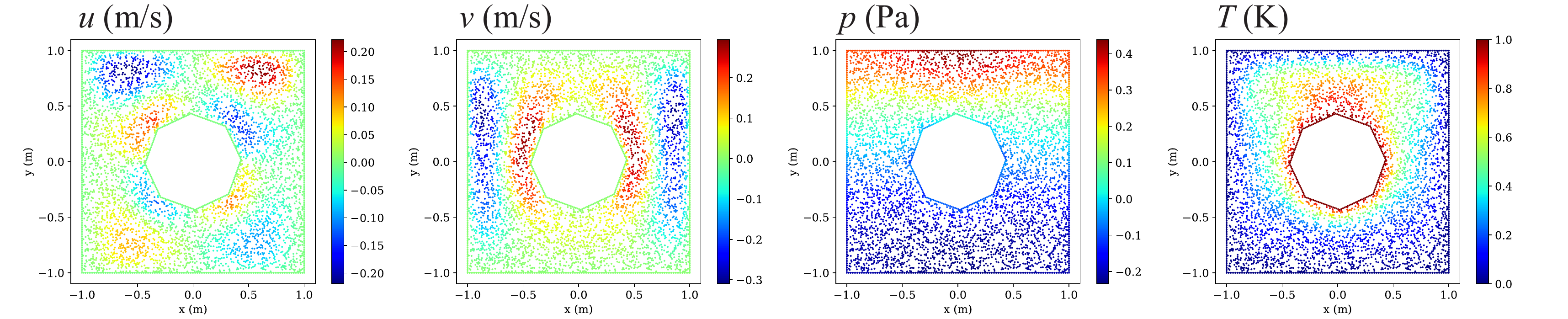}
\includegraphics[width=0.95\textwidth]{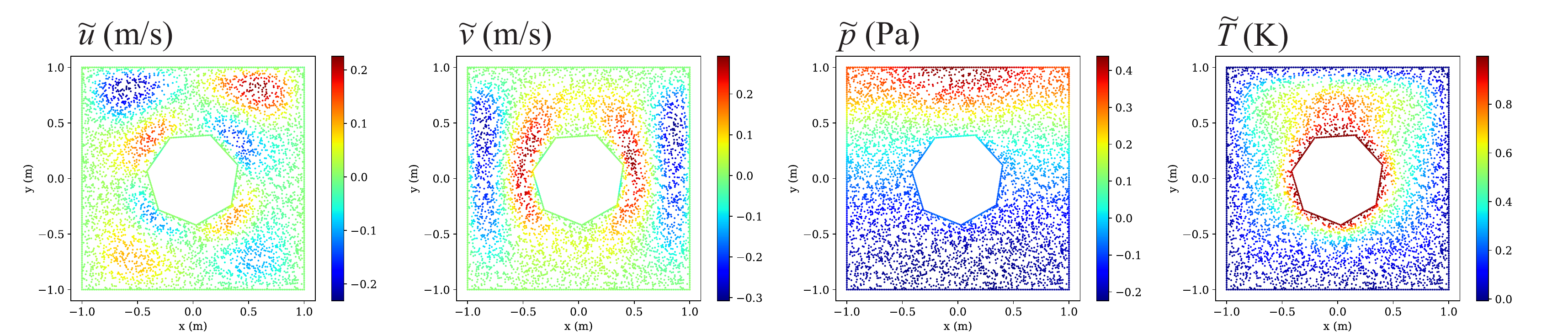}
\includegraphics[width=0.95\textwidth]{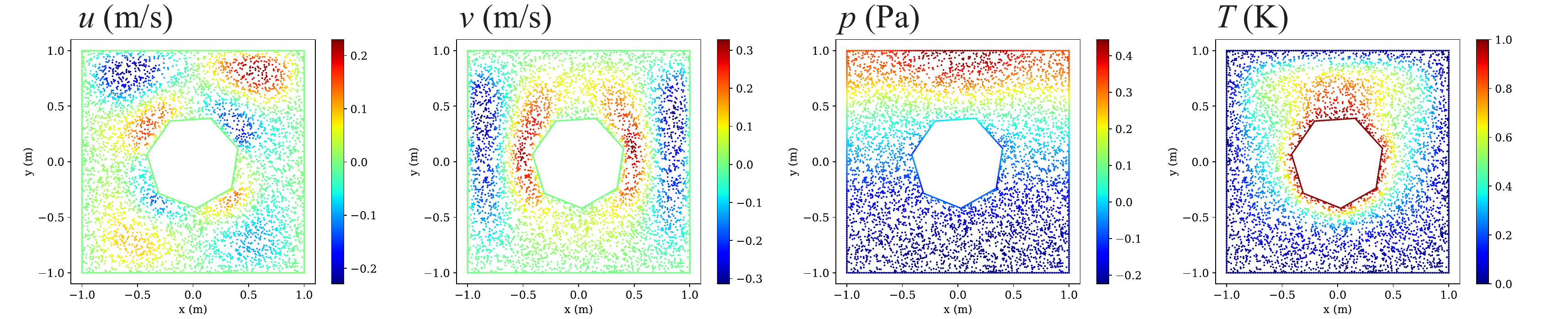}
\caption{Two examples taken from the set $\Psi=\{V_i\}_{i=1}^{27}$ comparing the ground truth solutions to the PIPN predictions for the velocity, pressure, and temperature fields for the thermally-driven flow problem (Sect. \ref{Sect421})}
\label{Fig15}
\end{figure*}

\begin{figure*}
\centering
\includegraphics[width=0.95\textwidth]{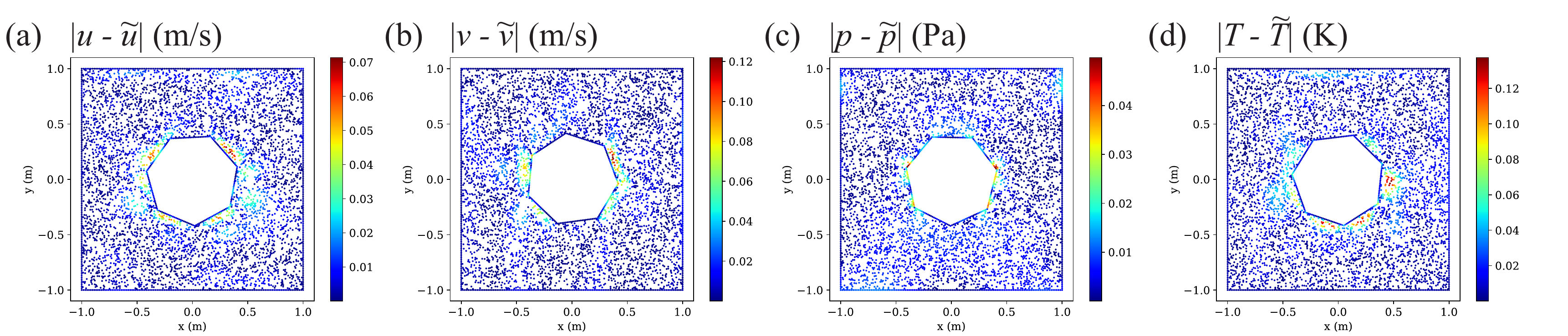}
\includegraphics[width=0.95\textwidth]{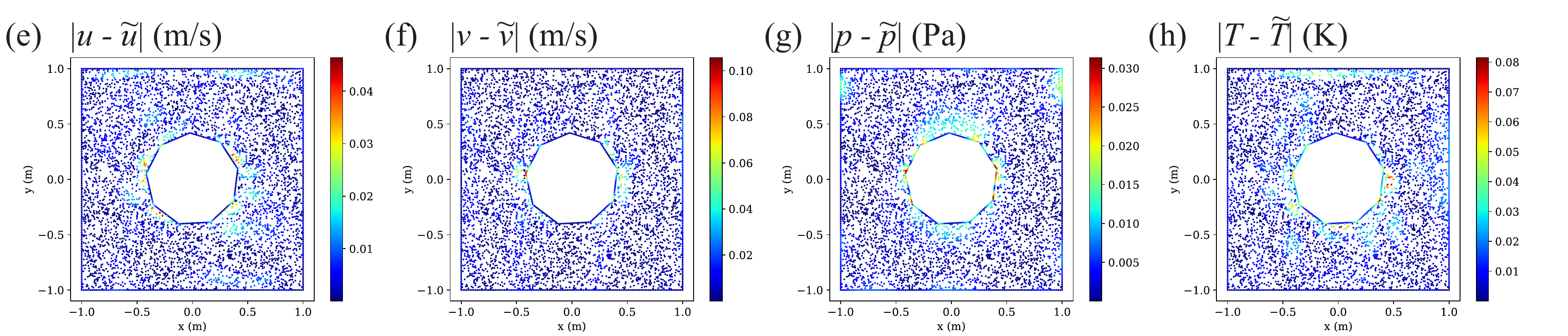}
\caption{Distribution of absolute pointwise error when the relative pointwise error ($L^2$ norm) becomes (\textbf{a}) maximum for $\Tilde{u}$, (\textbf{b}) maximum for $\Tilde{v}$, (\textbf{c}) maximum for $\Tilde{p}$, (\textbf{d}) maximum for $\widetilde{T}$, (\textbf{e}) minimum for $\Tilde{u}$, (\textbf{f}) minimum for $\Tilde{v}$, (\textbf{g}) minimum for $\Tilde{p}$, and (\textbf{h}) minimum for $\widetilde{T}$ for the thermally-driven flow problem in the set $\Psi=\{V_i\}_{i=1}^{27}$ (Sect. \ref{Sect421}).}
\label{Fig16}
\end{figure*}

\begin{figure*}
\centering
\includegraphics[width=0.95\textwidth]{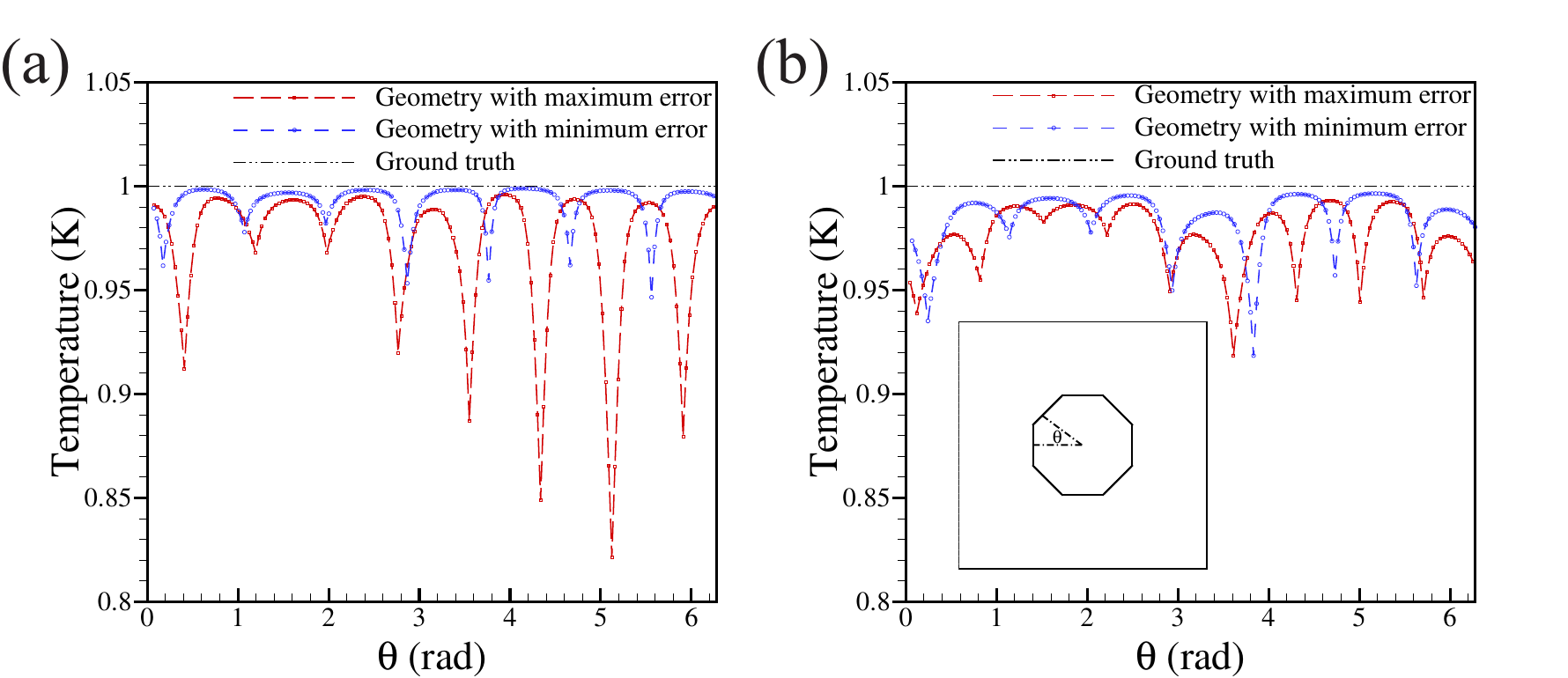}
\caption{Temperature distributions on the surface of the inner cylinder for geometries of the (\textbf{a}) set $\Phi=\{V_i\}_{i=1}^{108}$ and (\textbf{b}) set $\Psi=\{V_i\}_{i=1}^{27}$ when the relative pointwise error ($L^2$ norm) of the temperature over the inner surface becomes maximum and minimum (Sect. \ref{Sect421})}
\label{Fig17}
\end{figure*}

PIPN satisfies the convergence criteria after 2994 iterations, taking 99330 s (approximately 27.6 hours) for computation. Table \ref{Table8} summarizes the error analysis of the PIPN outputs for the both sets of $\Phi=\{V_i\}_{i=1}^{108}$ and $\Psi=\{V_i\}_{i=1}^{27}$. As can be understood from Table \ref{Table8}, the velocity field experiences higher relative pointwise errors ($L^2$ norm) compared to the pressure and temperature fields. Furthermore, the average relative pointwise error ($L^2$ norm) over the set $\Psi=\{V_i\}_{i=1}^{27}$ is higher than the set $\Phi=\{V_i\}_{i=1}^{108}$ for all the fields of interest; specifically this difference is highlighted for the temperature on the inner cylinder surface (2.26461E$-$2 vs. 7.41294E$-$2). This can be explained by two reasons. First, because PIPN parameters (e.g., \textbf{W} and \textbf{b} in Eq. \ref{Eq16}) are primarily trained to achieve the solution for the set $\Phi=\{V_i\}_{i=1}^{108}$ and second because the temperature distribution over the surface of inner cylinders of the set $\Phi=\{V_i\}_{i=1}^{108}$ has been assumed as an unknown variable in the defined inverse problem. These two reasons combined together lead to a higher level of difference in errors for predicting this quantity on the set $\Psi=\{V_i\}_{i=1}^{27}$ compared to the other fields. As a few examples, the velocity, pressure, and temperature fields for two different domains of the set $\Phi=\{V_i\}_{i=1}^{108}$ and set $\Psi=\{V_i\}_{i=1}^{27}$ are visualized in Fig. \ref{Fig13} and Fig. \ref{Fig15}, respectively. A good agreement between the PIPN prediction and the solution obtained by the numerical solver is observed in both Fig. \ref{Fig13} and Fig. \ref{Fig15}. Additionally, Figure \ref{Fig14} and Figure \ref{Fig16} demonstrate the absolute pointwise error distributions over geometries, respectively, from the set $\Phi=\{V_i\}_{i=1}^{108}$ and set $\Psi=\{V_i\}_{i=1}^{27}$, for those where the maximum and minimum relative pointwise errors ($L^2$ norm) of the velocity, pressure, and temperature fields occur. As depicted in Fig. \ref{Fig14} and Fig. \ref{Fig16}, the local maximum errors happen near the boundary of inner cylinders, due to the fact that the highest geometric variation from one domain to another appears in this region. Similarly, \citet{kashefi2021point} observed maximum local errors mostly on boundary points (see Fig. 11 of Ref. \cite{kashefi2021point}), when they \cite{kashefi2021point} employed PointNet \cite{qi2017pointnet} as a supervised computational fluid dynamics model. Furthermore, as can be understood from Fig. \ref{Fig14} and Fig. \ref{Fig16}, the maximum errors belong to domains with the hexagonal or octagonal inner cylinders, whereas the minimum errors belong to domains with the nonagonal inner cylinders, for both the set $\Phi=\{V_i\}_{i=1}^{108}$ and set $\Psi=\{V_i\}_{i=1}^{27}$, demonstrating the challenges of sharper corners. To validate the capability of PIPN for inferring the value of unknown temperature boundary condition, we plot the temperature distribution on the surface of inner cylinders for domains with the maximum and minimum relative pointwise error ($L^2$ norm) for both the set $\Phi=\{V_i\}_{i=1}^{108}$ and set $\Psi=\{V_i\}_{i=1}^{27}$ in Fig. \ref{Fig17}. According to Fig. \ref{Fig17}, the local maxima of the temperature appear at corners of the polygons, indicating that variable prediction on non-smooth boundaries becomes more challenging. Note that because the hyperbolic tangent activation function (Eq. \ref{Eq18}) is utilized in the last layer of PIPN, the predicted temperature has the upper bound of 1.0, as can be seen in Fig. \ref{Fig17}. \textcolor{red}{It is worthwhile to note that points located on the boundary of point-clouds always contribute to the global feature (e.g., see Fig. 12 of Ref. \cite{kashefi2021point} and Fig. 18 of Ref. \cite{qi2017pointnet}). In this way, PointNet \cite{qi2017pointnet} and consequently PIPN identifies boundary points. Hence, PIPN learns boundary values of the domains in the set $\Phi=\{V_i\}_{i=1}^{108}$ during training and afterwards predicts the desired outputs on the boundary of domains in the set $\Psi=\{V_i\}_{i=1}^{27}$ accordingly.}

Next we highlight the speedup factor obtained by the PIPN methodology. PIPN predicts the velocity, pressure, and temperature fields of the set $\Psi=\{V_i\}_{i=1}^{27}$ in approximately 9 s. On the other hand, our numerical solver obtains the solutions on the same set $\Psi=\{V_i\}_{i=1}^{27}$ in approximately 307 s, while each simulation is performed individually on AMD processors with 2.30 GHz clock rate and 24 Gigabytes of RAM. In this sense, the achieved speedup is a factor of approximately 35. Note that the speedup factor reported here is not absolute and strongly depends on the CPU and GPU types as well as coding efficiency of both the finite-element solver and PIPN.

\begin{figure*}
\centering
\includegraphics[width=0.45\textwidth]{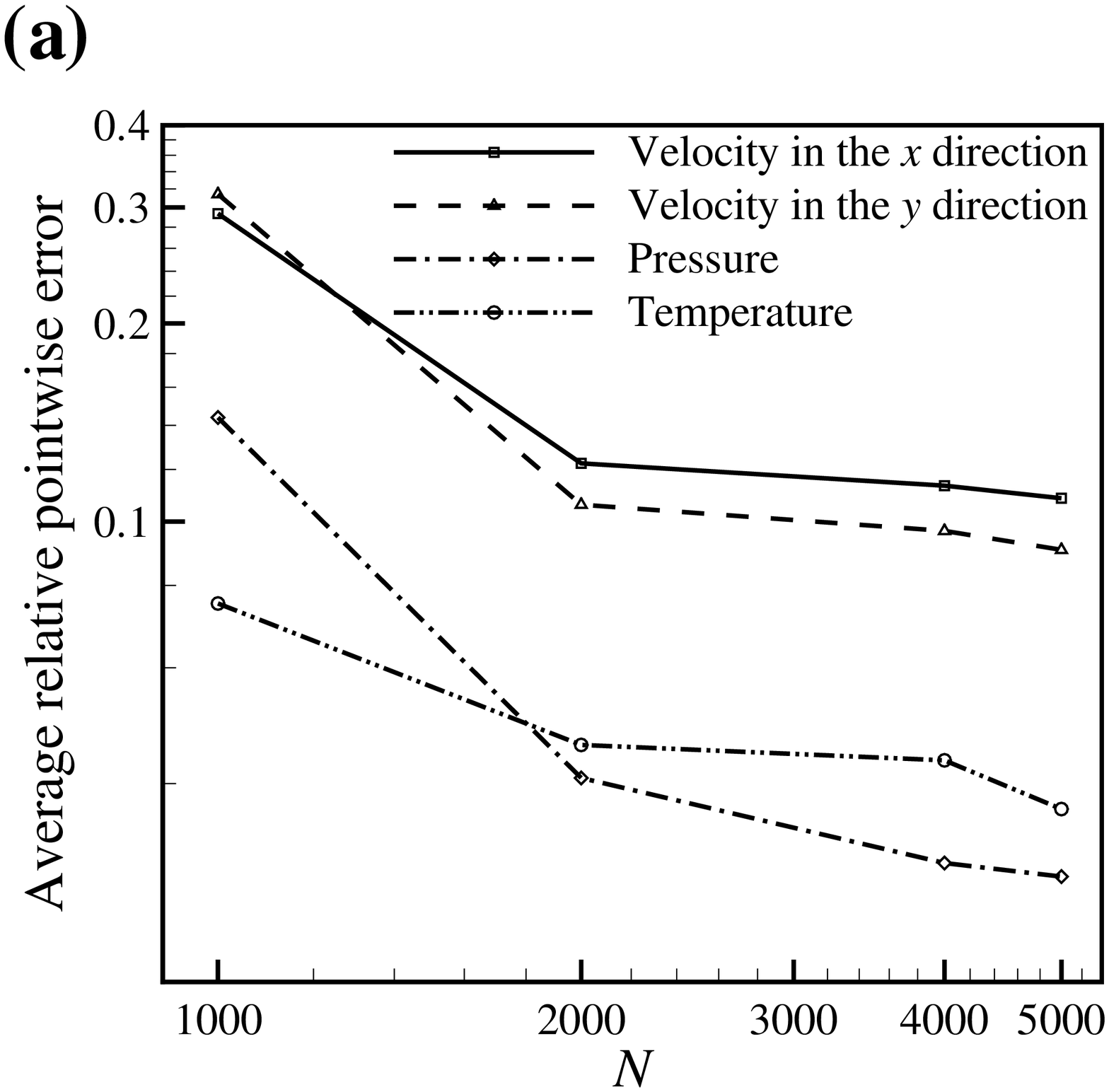}
\includegraphics[width=0.45\textwidth]{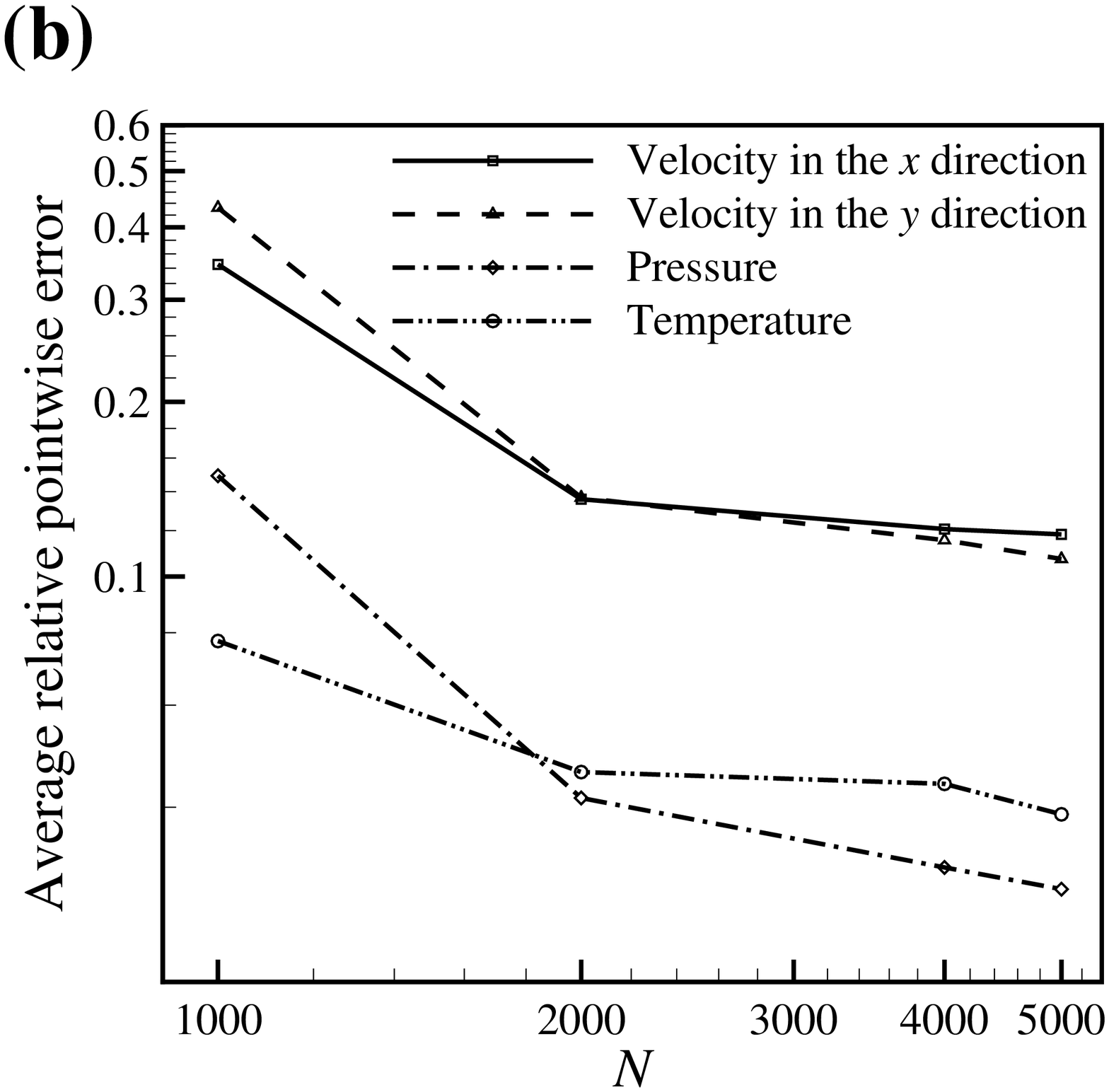}
\caption{
\textcolor{blue}{
Effect of number of grid points ($N$) in point clouds ($\mathcal{X}$) of the (\textbf{a}) set $\Phi=\{V_i\}_{i=1}^{108}$ and (\textbf{b}) set $\Psi=\{V_i\}_{i=1}^{27}$ on the average relative pointwise error ($L^2$ norm) for the buoyancy induced-convection problem discussed in Sect. \ref{Sect421}}}
\label{FigR12}
\end{figure*}

We reevaluate the PIPN performance in the absence of sparse measurements of pressure. Based on this assumption, the loss function is modified as 
\begin{equation}
\label{Eq41}
\begin{split}
\mathcal{J}=\frac{1}{m} \sum_{i=1}^m \big( \lambda_1 r_i^{\text{continuity}} + \lambda_2 r_i^{\text{momentum}_x} + \lambda_3 r_i^{\text{momentum}_y} + \lambda_4 r_i^{\text{velocity}_{\text{BC}}} + 
\lambda_6 r_i^{\text{velocity}_{\text{obs}}}
+ \lambda_8 r_i^{\text{energy}}
\\
+ \lambda_9 r_i^{\text{temperature}_{\text{outer-BC}}}
+ \lambda_{10} r_i^{\text{temperature}_{\text{obs}}}\big).
\end{split}
\end{equation}
The error analysis of this machine-learning experiment is carried out in Table \ref{Table9}. Comparing data of Table \ref{Table9} to Table \ref{Table8}, it is realized that the average relative error of the pressure field for the set $\Phi=\{V_i\}_{i=1}^{108}$ increases by approximately 7218\% (2.89030E$-$2 vs. 2.11516) as a consequence of no pressure observation. Due to an inaccurate reconstruction of the pressure filed on the set $\Phi=\{V_i\}_{i=1}^{108}$, the PIPN prediction for the pressure on the set $\Psi=\{V_i\}_{i=1}^{27}$ introduces an average relative pointwise error ($L^2$ norm) of 2.11388 as well. However, as discussed in Sect. \ref{Sect412}, since the PIPN framework preserves the pressure gradient accuracy, the average relative pointwise error ($L^2$ norm) of the velocity and temperature fields remains roughly unchanged compared to the corresponding values reported in Table \ref{Table8}, for both the set of $\Phi=\{V_i\}_{i=1}^{108}$ and set $\Psi=\{V_i\}_{i=1}^{27}$. Without the pressure term in the loss function (Eq. \ref{Eq41}), PIPN requires 1052 iterations for reaching the convergence criterion (see Eq. \ref{Eq40}), taking 38691 s (approximately 10.8 hours). Analogous to Sect. \ref{Sect412}, removing the pressure term from the loss function leads to a decrease in the computational cost. \textcolor{red}{ Next, we investigate the effect of the observed data and given boundary conditions of the set $\Phi=\{V_i\}_{i=1}^{108}$ polluted by 10\% random Gaussian noise. Table \ref{TableR1} tabulates the average relative pointwise errors ($L^2$ norm) of the velocity, pressure, and temperature fields over the set $\Phi=\{V_i\}_{i=1}^{108}$ and $\Psi=\{V_i\}_{i=1}^{27}$ as a result of noisy data. As can be inferred from Table \ref{TableR1}, the PIPN framework is robust even in the presence of the noisy data.} \textcolor{blue}{Furthermore, we investigate the accuracy of the PIPN predictions for the velocity, pressure, and temperature fields over the set $\Phi=\{V_i\}_{i=1}^{108}$ and set $\Psi=\{V_i\}_{i=1}^{27}$ as a function of number of grid points ($N$) in point clouds ($\mathcal{X}$). As illustrated in Fig. \ref{FigR12}, a coarse representation of point clouds (e.g., $N=$ 1000) causes a considerable increase in the error of predictions. This is mainly because of twofold reasons. First, a poor representation of point clouds negatively affects extraction of the geometric features by PIPN, while the solution of fluid flow fields depend on the geometry of physical domains. Second, low resolutions of point clouds decrease the accuracy of computing nonlinear terms in the momentum and energy equations (see Eqs. \ref{Eq7}--\ref{Eq9}).}


\begin{table}[htbp]
\caption{Error analysis of the velocity, pressure, and temperature fields over the set $\Phi=\{V_i\}_{i=1}^{108}$ and set $\Psi=\{V_i\}_{i=1}^{27}$ in the absence of sparse pressure observations in the loss function (Eq. \ref{Eq41}) for the buoyancy induced-convection problem discussed in Sect. \ref{Sect421}. $||\cdots||_V$ shows the $L^2$ norm over the entire domain ($V$), while $||\cdots||_{\Gamma_{\textrm{inner}}}$ indicates the $L^2$ norm over the inner surface ($\Gamma_{\textrm{inner}}$).}
\centering
\begin{tabular}{l l l}
\toprule
 & Over the set $\Phi=\{V_i\}_{i=1}^{108}$ & Over the set $\Psi=\{V_i\}_{i=1}^{27}$\\
\midrule
Average $||\tilde{u}-u||_V/||u||_V$ & $1.11005\textrm{E}-1$ & $1.12832\textrm{E}-1$  \\
Average $||\tilde{v}-v||_V/||v||_V$ & $1.08131\textrm{E}-1$ & $1.09404\textrm{E}-1$ \\
Average $||\tilde{p}-p||_V/||p||_V$ & $2.11516$ & $2.11388$ \\
Average $||\widetilde{T}-T||_V/||T||_V$ & $3.79894\textrm{E}-2$ & $3.89161\textrm{E}-2$ \\
Average $||\widetilde{T}-T||_{\Gamma_{\textrm{inner}}}/||T||_{\Gamma_{\textrm{inner}}}$ & $2.62480\textrm{E}-2$ & $7.57655\textrm{E}-2$ \\
\bottomrule
\end{tabular}
\label{Table9}
\end{table}



\begin{table}[htbp]
\textcolor{red}{
\caption{Error analysis of the velocity, pressure, and temperature fields over the set $\Phi=\{V_i\}_{i=1}^{108}$ and set $\Psi=\{V_i\}_{i=1}^{27}$ when the observed data is polluted with 10\% Gaussian noise for the buoyancy induced-convection problem discussed in Sect. \ref{Sect421}. $||\cdots||_V$ indicates the $L^2$ norm over the entire domain ($V$), whereas $||\cdots||_{\Gamma_{\textrm{inner}}}$ indicates the $L^2$ norm over the inner surface ($\Gamma_{\textrm{inner}}$).}
\centering
\begin{tabular}{l l l}
\toprule
 & Over the set $\Phi=\{V_i\}_{i=1}^{108}$ & Over the set $\Psi=\{V_i\}_{i=1}^{27}$\\
\midrule
Average $||\tilde{u}-u||_V/||u||_V$ & $1.12197\textrm{E}-1$ & $1.27092\textrm{E}-1$  \\
Average $||\tilde{v}-v||_V/||v||_V$ & $9.74038\textrm{E}-2$ & $1.27536\textrm{E}-1$ \\
Average $||\tilde{p}-p||_V/||p||_V$ & $2.93036\textrm{E}-2$ & $3.65588\textrm{E}-2$  \\
Average $||\widetilde{T}-T||_V/||T||_V$ & $3.84842\textrm{E}-2$ & $4.32023\textrm{E}-2$ \\
Average $||\widetilde{T}-T||_{\Gamma_{\textrm{inner}}}/||T||_{\Gamma_{\textrm{inner}}}$ & $2.52552\textrm{E}-2$ & $7.89100\textrm{E}-2$ \\
\bottomrule
\end{tabular}
\label{TableR1}}
\end{table}


\begin{table}[htbp]
\caption{Error analysis of the velocity and temperature fields for testing the generalizability of PIPN using the buoyancy induced-convection problem illustrated in Sect. \ref{Sect422}. $||\cdots||_V$ indicates the $L^2$ norm over the entire domain ($V$), while $||\cdots||_{\Gamma_{\textrm{inner}}}$ shows the $L^2$ norm over the inner surface ($\Gamma_{\textrm{inner}}$).}
\centering
\begin{tabular}{l l l l l l}
\toprule
Shape of $W$ (see Eq. \ref{Eq4}) & $\frac{||\tilde{u}-u||_V}{||u||_V}$ & $\frac{||\tilde{v}-v||_V}{||v||_V}$ & $\frac{||\tilde{p}-p||_V}{||p||_V}$ & $\frac{||\widetilde{T}-T||_V}{||T||_V}$ & $\frac{||\widetilde{T}-T||_{\Gamma_{\textrm{inner}}}}{||T||_{\Gamma_{\textrm{inner}}}}$ \\
\midrule
Circle & $1.21878\textrm{E}-1$ & $1.08121\textrm{E}-1$ & $2.09506\textrm{E}-2$ & $3.02129\textrm{E}-2$ & $1.15890\textrm{E}-2$ \\
Equilateral hexagon & $1.65543\textrm{E}-1$ & $1.15348\textrm{E}-1$ & $3.32434\textrm{E}-2$ & $4.36705\textrm{E}-2$ & $3.55274\textrm{E}-2$ \\
\bottomrule
\end{tabular}
\label{Table10}
\end{table}


\subsubsection{Generalizability\label{Sect422}}

\begin{figure*}[htbp]
\centering
\includegraphics[width=0.95\textwidth]{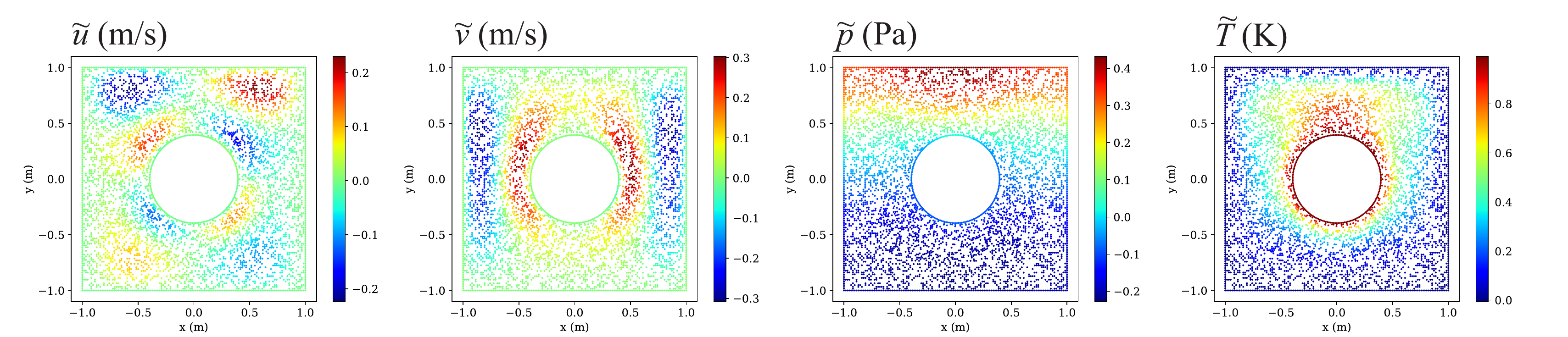}
\includegraphics[width=0.95\textwidth]{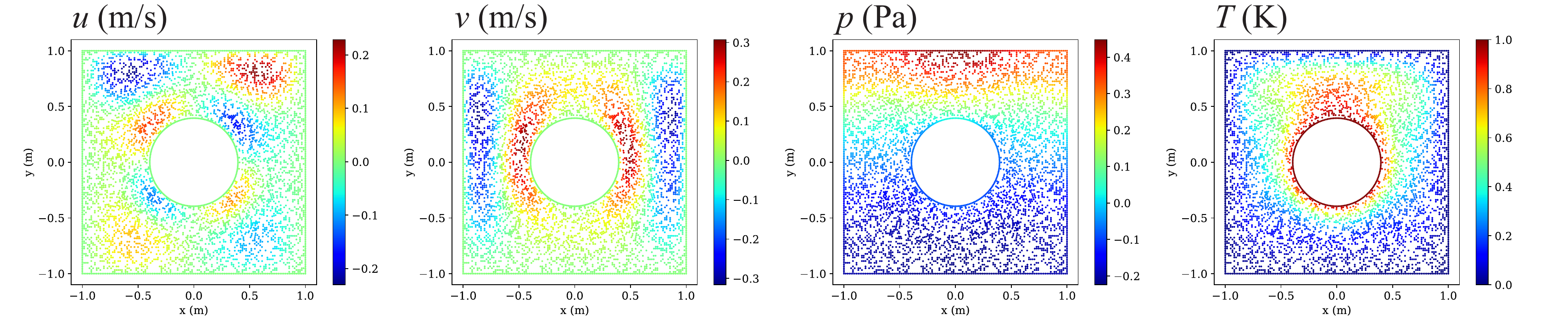}
\includegraphics[width=0.95\textwidth]{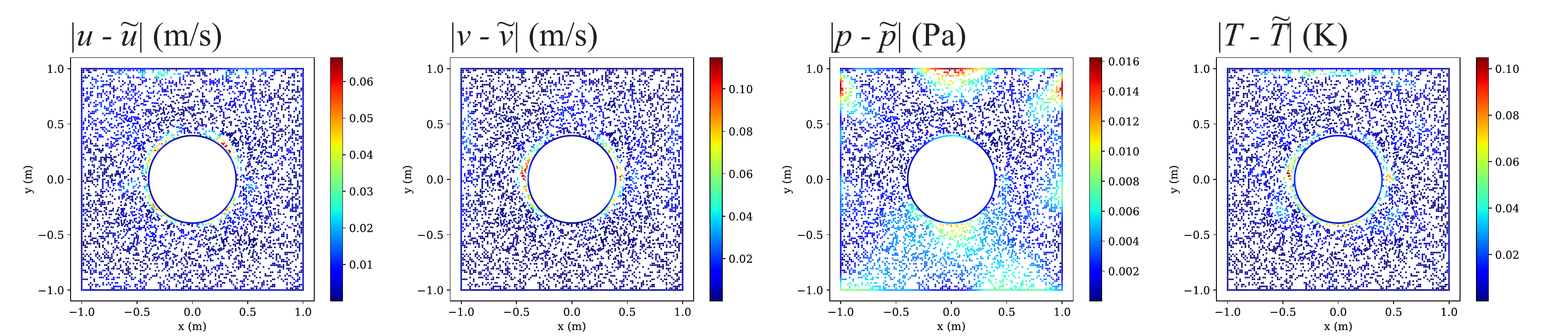}
\textcolor{red}{
\caption{Comparison between the ground truth and PIPN prediction for the velocity, pressure, and temperature fields of the natural convection problem for a squared domain in the set $\Psi=\{V_i\}_{i=1}^{2}$ with a circular inner cylinder for testing the generalizability of PIPN (Sect. \ref{Sect422})}}
\label{Fig18}
\end{figure*}

\begin{figure*}[htbp]
\centering
\includegraphics[width=0.95\textwidth]{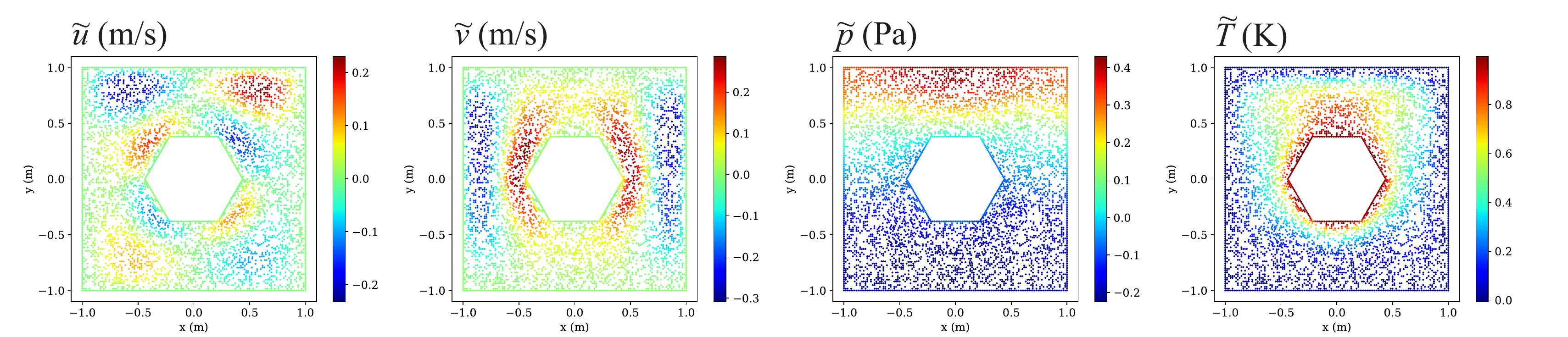}
\includegraphics[width=0.95\textwidth]{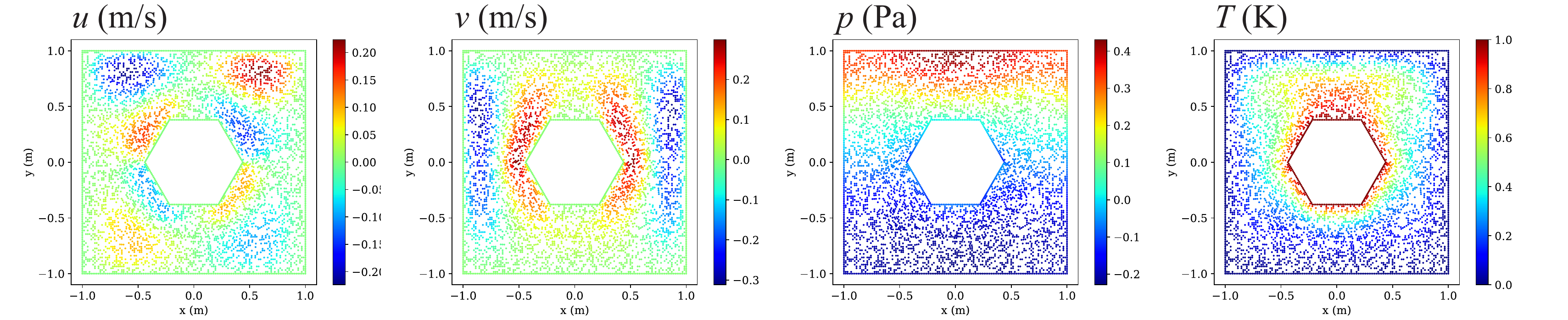}
\includegraphics[width=0.95\textwidth]{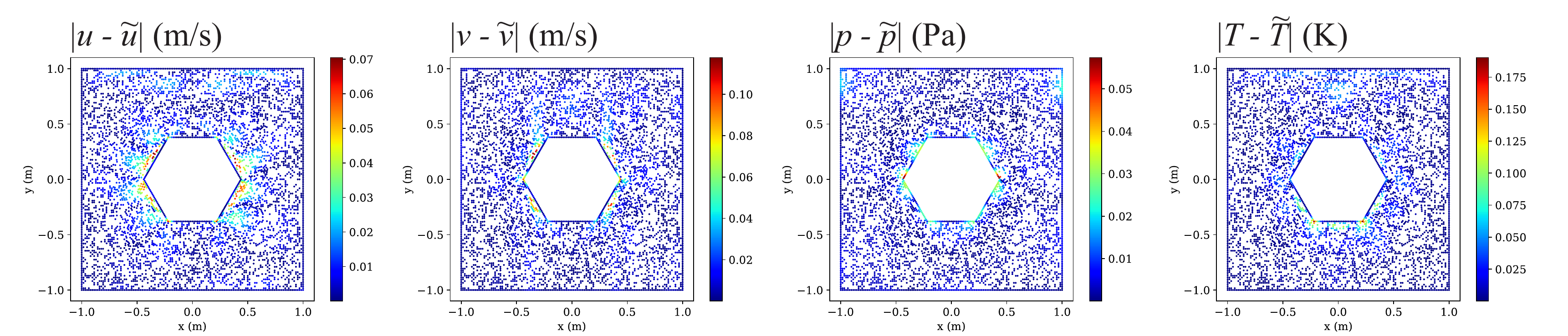}
\textcolor{red}{
\caption{Comparison between the ground truth and PIPN prediction for the velocity, pressure, and temperature fields of the natural convection problem for a squared domain in the set $\Psi=\{V_i\}_{i=1}^{2}$ with a hexagonal inner cylinder for testing the generalizability of PIPN (Sect. \ref{Sect422})}}
\label{Fig19}
\end{figure*}

\begin{figure*}[h]
\centering
\includegraphics[width=0.43\textwidth]{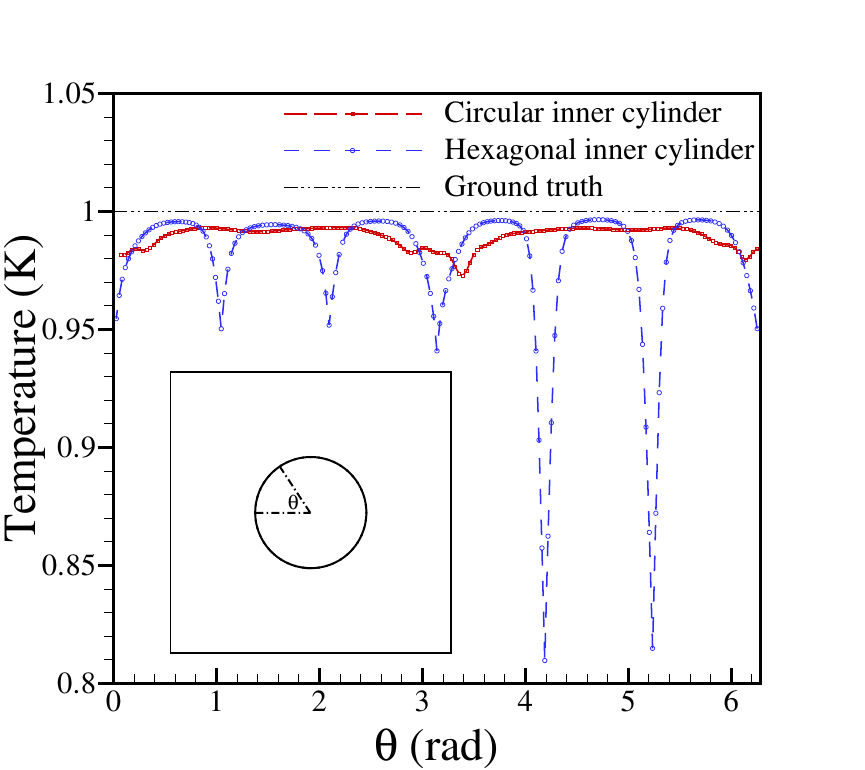}
\textcolor{red}{
\caption{Temperature distributions along the surface of the circular and hexagonal inner cylinders in the set $\Psi=\{V_i\}_{i=1}^{2}$, testing the generalizability of PIPN (Sect. \ref{Sect422})}}
\label{Fig20}
\end{figure*}

Let us recall that the set $\Psi$ was established by unseen geometries but from ``seen categories'' in Sect. \ref{Sect421}. In this subsection, we establish a new set of $\Psi$ consisting of unseen geometries from ``unseen categories'' to investigate generalizability. To this end, the PIPN trained through the inverse problem is now asked to predict the velocity, pressure, and temperature solutions in two computational domains (i.e., $\Psi=\{V_i\}_{i=1}^{2}$), respectively, with a circular inner cylinder with the diameter of $0.395 \times \cos \frac{\pi}{9} \times \csc \frac{\pi}{7}$ m and an equilateral hexagonal inner cylinder with the side length of $\frac{16 \sqrt{3}}{75}$ m, both centered at the origin. Note that PIPN has never seen inner cylinders with circular or hexagonal cross sections during the training. The predictions are graphically presented in Fig. \ref{Fig18} and Fig. \ref{Fig19} respectively for the circular and hexagonal cross sections. Moreover, the temperature distributions along the inner surfaces are plotted in Fig. \ref{Fig20}. In both cases, the PIPN predictions agree well with the ground truth. Similar to the results depicted in Sect. \ref{Sect411}, maximum local errors happens near inner boundaries, except for the pressure field of circular cylinder, as can be seen in Fig. \ref{Fig18}. Additionally, Table \ref{Table10} lists the corresponding error analysis. Based on the relative pointwise error ($L^2$ norm) tabulated in Table \ref{Table10}, the maximum error belongs to the velocity field in the $x$-direction ($\tilde{u}$). Particularly, the domain with a hexagonal inner cylinders experiences 16.554\% relative error for $\tilde{u}$, which is higher than the maximum relative error of $\tilde{u}$ (14.352\%) in the sets considered in Sect. \ref{Sect411} (see Table \ref{Table8}). For all the fields, the hexagonal inner cylinder leads to a higher error compared to the circular inner cylinder, mainly due to the sharpness of the inner boundaries. \textcolor{red}{Additionally, one may define the alternative Rayleigh number ($Ra_R$) such that $Ra_R=Ra\left(\frac{R}{L}\right)^3$, where $R$ is the smallest radial distance between the boundary of inner and outer cylinders. Imposing a large variation of $Ra_R$ for the set $\Psi=\{V_i\}_{i=1}^{l}$ in comparison to $Ra_R$ used in the set $\Phi=\{V_i\}_{i=1}^{m}$, might lead to considerable errors in the prediction of PIPN over the domains of the set $\Psi=\{V_i\}_{i=1}^{l}$. This is mainly because the behavior and pattern of fluid flow fields (e.g., being steady or unsteady) are a function of $Ra_R$, which nominates both physical and geometrical features.} All in all, the outcomes discussed in this subsection are evidences of a successful ability of PIPN to generalize to new and unseen categories of domain geometries.

\section{Summary and future studies \label{Sect5}}

Since the late 2018s, artificial neural networks based on physics-informed models have become popular among the community of computational mathematics and mechanics, mainly due to their abilities to solve PDEs of interest without labeled data for forward problems and with sparse data (observations) for inverse problems. However, recent physics-informed neural networks have been limited to only solve a set of PDEs on a single computational domain with a fixed geometry, fundamentally due to the lack of a mechanism in their architecture to extract geometric features of input computational domains. This makes them practically useless for the purpose of rapid exploration of geometric parameter effects on industrial designs. To overcome the associated challenges, we proposed a novel physics-informed model, called PIPN, for solving time-independent PDEs on sets of domains with complicated geometries. Using PIPN, we first solve a forward (without labeled data) or inverse (with sparse scattered labeled data) problem on multiple geometries. In this procedure, PIPN is also automatically trained over the set of geometries. Then, the trained PIPN is used to find the solution on unseen geometries from seen and unseen categories. The PIPN framework was produced based on the segmentation architecture of PointNet \cite{qi2017pointnet} with necessary adaptations.

In order to examine the performance of the PIPN methodology, we considered two test cases. In the first case, we focused on a forward problem and solved the continuity and Navier-Stokes equations of steady-state incompressible flows for a manufactured solution. Overall, PIPN obtained the velocity field approximately with an average relative pointwise error ($L^2$ norm) of 0.1\% and the pressure field with an average relative pointwise error ($L^2$ norm) of 1.3\% over a set of domains with various geometries. Additionally, it was concluded that although removing the pressure boundary term from the PIPN loss function resulted in a decrease in the accuracy of predicted pressure fields, PIPN maintained the accuracy of the pressure gradient fields, and thus the velocity fields. Furthermore, we realized that by implementing the Navier-Stokes equation in a conservative form, although the required number of spatial derivatives per iteration decreased compared to the non-conservative implementation, it led to a slower convergence and thus higher computational cost. Our machine-learning experiments showed that when body forces varied spatially over a domain, implementing forces as constant tensors led to faster and more accurate outcomes in comparison with symbolic tensor implementations. Finally, the generalizability of the PIPN configuration was evaluated by an error analysis of the velocity and pressure fields predicted on three different unseen circumstances: first, a domain with a cavity as a combination of previously seen cavities inside; second, a domain with distorted outer boundaries, while PIPN had only seen outer boundaries with straight lines; and third, a domain with two circular cavities inside, while PIPN had only seen domains with one circular cavity inside. In all of these cases, the accuracy of computed fields ranged from a reasonable to an excellent level and the relative pointwise error ($L^2$ norm) did not exceed 7.5\%. In the second case, we concentrated on an inverse problem and solved the buoyancy-driven flow in a square enclosure with a cylinder with different cross sections for a Rayleigh number of $Ra=10^5$ and a Prandtl number of $Pr=1.0$. Accordingly, the temperature boundary condition on the inner cylinder surface was unknown, while sparse scattered observations of the flow fields were recorded by virtual sensors. Particularly, the inverse problem was first solved over a set containing 108 domains with equilateral nonagonal, octagonal, and heptagonal inner cylinders using PIPN. Afterwards, the PIPN trained through solving the inverse problem predicted the full velocity, pressure, and temperature fields over a new set containing 27 domains with equilateral nonagonal, octagonal, and heptagonal inner cylinders, but with different orientations compared to the previous set (i.e., unseen geometries from seen categories). In all the cases, the average relative pointwise error ($L^2$ norm) was less than 12\%. Generalizability of PIPN was examined by predicting the flow fields of domains with circular and hexagonal inner cylinders (i.e., unseen geometries from unseen categories). In these cases, PIPN could predict the fluid flow fields with the minimum and maximum relative error of approximately 2\% and 17\%, respectively.  

One of our plans for future researches is the extension of the PIPN framework to solve time dependent incompressible flow problems. To this end, we first temporally integrate the system of equations (Eqs. \ref{Eq1}--\ref{Eq3}) by a forward differentiation formula such as the third-order Runge Kutta scheme (see e.g., Ref. \cite{wazwaz1990modified}), and then we apply a pressure projection method \cite{chorin1968numerical,temam1969approximation,guermond2006overview} for decoupling the solution of pressure and velocity variables (see e.g., an example of these sets of equations in Sect. 3.2 of Ref. \cite{san2013coarse}). Afterwards, we incorporate PIPN with the methodology of neural ordinary differential equations \cite{chen2018neural} to obtain the solutions of evolving PDEs at each time step. A similar approach has been taken in the area of computer vision for evolving rigid objects \cite{rempe2020caspr}.

Graph Neural Networks (GNNs) have been shown as a powerful tool for unstructured grids and irregular geometries for the steady \cite{gao2022physics} and unsteady \cite{pfaff2020learning} problems in computational mechanics. Additionally, \citet{gao2022physics} implemented a physics-informed model embedded in a GNN. However, the authors of these studies \cite{pfaff2020learning,gao2022physics} applied their models to a ``single'' computational domain. For this reason, we plan to design a GNN-based physics-informed model for solving forward and inverse incompressible flow problems on multiple sets of domains with irregular geometries and compare its efficiency with the PIPN framework developed in the current study.

Finally, we plan to extend the PIPN framework to other applications in the area of computational physics such as compressible flows \cite{mao2020physics}, turbulent flows \cite{raissi2019deepTurbulent}, solid mechanics \cite{haghighat2021physics}, fluid flow through porous media \cite{kashefi2021pointPorous}, etc.

\section*{CRediT authorship contribution statement}
\textbf{Ali Kashefi:} Conceptualization, Methodology, Software, Writing - original draft, Writing - review \& editing, Visualization. \textbf{Tapan Mukerji:} Conceptualization, Writing - review \& editing, Project administration, Funding acquisition. 

\section*{Declaration of competing interest}
The authors declare that they have no known competing financial interests or personal relationships that could have appeared to influence the work reported in this paper.

\section*{Acknowledgement}
We acknowledge funding by Shell-Stanford collaborative project on Digital Rock Physics 2.0 for supporting this study. We would also like to thank Steve Graham, the Dean of the School of Earth, Energy, and Environmental Sciences at Stanford University for funding. Additionally, we wish to thank the Stanford Research Computing Center for providing computational resources for this research project. Moreover, A. Kashefi would like to thank Davis Rempe in the department of Computer Science at Stanford University for his helpful guidance on the software engineering aspects of this study. \textcolor{blue}{Additionally, the authors wish to thank the reviewers for their valuable comments.}



\printcredits

\bibliographystyle{cas-model2-names}


\bibliography{cas-refs}


\end{document}